\documentclass[twocolumn,rmp,aps,amssymb]{revtex4}
\usepackage{graphicx}
\usepackage{epsfig}
\usepackage[dvips]{color}
\usepackage{colordvi}
\usepackage{url}

\newcommand{\emap}{EMA$^\prime$}

\newcommand{\et}{{\em et al.}}

\newcommand{\qe}{quasi-elastic}
\newcommand{\qes}{quasi-elastic scattering}
\newcommand{\qep}{quasi-elastic peak}
\newcommand{\sfu}{spectral function}
\newcommand{\cs}{cross section}
\newcommand{\ba}{\begin{eqnarray*}}
\newcommand{\ea}{\end{eqnarray*}}
\newcommand{\beq}{\begin{equation}}
\newcommand{\eeq}{\end{equation}}
\newcommand{\be}{\begin{eqnarray}}
\newcommand{\ee}{\end{eqnarray}}
\newcommand{\magq}{|{\bf q}|}
\newcommand{\magk}{{|{\bf k}|}}
\newcommand{\kv}{\bf k}
\newcommand{\qv}{\bf q}
\newcommand{\hyd}{$^3\!H$}
\newcommand{\nm}{nuclear matter}

\def\Lmunu{L_{\mu\nu}}
\def\Wmunu{W^{\mu\nu}}
\def\lsim{\buildrel < \over {_{\sim}}}
\def\gsim{\buildrel > \over {_{\sim}}}
\newcommand{\qt}{{\widetilde q}}

\newcommand{\gmn}{$G_M^n\,$}
\newcommand{\gmp}{$G_M^p\,$}
\newcommand{\gep}{$G_E^p\,$}

\newcommand{\ha}{$^3\overrightarrow{\mathrm{He}}(\vec{e},e')X\:$}
\newcommand{\han}{$^3\overrightarrow{\mathrm{He}}(\vec{e},e'n)\:$}

\newcommand{\recn}{$^2\textrm{H}(\vec{e},e'\vec{n})p\:$}

\begin{document}
\title{Inclusive quasi-elastic electron-nucleus scattering}

\author{Omar Benhar}
\email{benhar@roma1.infn.it}
\affiliation{INFN, Sezione di Roma, I-00185 Roma, Italy}
\affiliation{Dipartimento di Fisica, Universit\`a ``La Sapienza'', 
I-00185 Roma, Italy}
\author{Donal Day}
\email{dbd@virginia.edu}
\affiliation{Dept. of Physics, University of Virginia, Charlottesville, 
VA 22903, USA}
\author{Ingo Sick}
\email{ingo.sick@unibas.ch}
\affiliation{Dept. f{\"u}r Physik und Astronomie, Universit{\"a}t Basel, 
 CH-4056 Basel, Switzerland}

\begin{abstract}
This article presents a review of the field of inclusive quasi-elastic
electron-nucleus scattering. It discusses the approach used to measure the data 
and includes a compilation of data available in numerical form. The theoretical 
approaches used to interpret the data are presented. A number of results 
obtained from the  comparison between experiment and calculation are then 
reviewed. The analogies and differences to other fields of physics exploiting
quasi-elastic scattering from composite systems are pointed out.    
\end{abstract}
\pacs{25.30.Fj,29.87.+g}
\maketitle
%

\section{Introduction}
%
The energy spectrum of high-energy  leptons (electrons in particular) scattered 
from a nuclear target
displays a number of features. At low energy loss ($\omega$)  peaks due to elastic
scattering and inelastic excitation of discrete nuclear states appear; a
measurement of the corresponding form factors as a function of momentum
transfer $|{\bf q}|$ gives access to the Fourier transform of nuclear (transition)
densities. At larger energy loss, a broad peak due to quasi-elastic
electron-nucleon scattering appears; this peak --- very wide due to nuclear
Fermi motion --- corresponds to processes where
the electron scatters from an individual, moving nucleon, which, after 
interaction with other nucleons, is ejected from the target.  At even larger
$\omega$  peaks that correspond to  excitation of the nucleon to
distinct resonances are visible. At very large $\omega$, a structureless
continuum due to Deep Inelastic Scattering (DIS) on quarks bound in nucleons
appears. A schematic spectrum is shown in Fig.~\ref{ee}.
At momentum transfers above approximately  500 MeV/c the dominant feature of the spectrum
is the quasi-elastic peak.  

A number of questions have been investigated using \qes : 
\begin{itemize}
\item
 The \qe~ cross section integrated over electron energy loss is proportional to
\begin{figure}[hbt]
\begin{center}
\includegraphics[scale=0.42,clip]{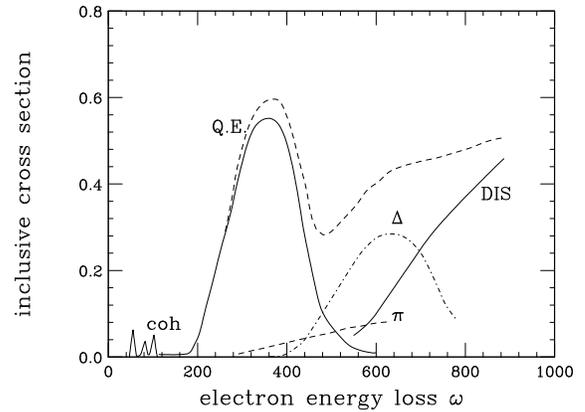}
{\caption[]{Schematic representation of inclusive cross section as 
function of energy loss.
}\label{ee}} 
\end{center} 
\end{figure}
the sum of electron-nucleon cross sections. Historically, this has been
exploited in order to measure the neutron charge and magnetic form factors using
mainly light ($A < 4$) nuclear targets.
Today the emphasis has shifted to exposing possible medium modifications of the 
nucleon form factors. \\
\item Another  integral property of the \qep , its width, provides a direct
measure of the average momentum of nucleons in nuclei, and has been used to
determine nuclear Fermi momenta; contrary to other observables such as densities,
 \qes~ provides a {\em direct} determination via an observable sensitive to the
momenta of nucleons. \\
\item The {\em shape} of the \qep~ depends on the distribution  in energy $E$ and
momentum $\kv$ of the initially bound nucleons. Precise measurements give 
indirect access
to  (integrated properties of) the nuclear \sfu~  $S({\kv} ,E)$ describing this 
distribution. In particular, the tail of the \qep~ at large $|{\bf q}|$ and low $\omega$
is sensitive to the tail  of the \sfu~ at large $\magk$. \\
\item Processes more complicated than one-nucleon knockout, in particular those
 related to  non-nucleonic degrees of freedom and 
Meson Exchange Currents (MEC), also play a role. They can be investigated by
separating the \qe~ response into the longitudinal (charge) and transverse
(magnetic plus convection) pieces, the latter being preferentially affected by MEC.
\\
\item {\em Scaling} is one of the unique features of \qes . This refers to the fact
that the inclusive \cs , which {\em a priori} is a function of two independent
variables, $|{\bf q}|$ and $\omega$,  depends on a {\em single} variable $y(q,\omega)$. 
This scaling property, a consequence
of the   kinematics of the underlying electron-nucleon elastic scattering
process, provides a strong handle on the reaction mechanism. Further, the
scaling violations that are observed reveal how the dynamics go beyond the Impulse 
Approximation (IA) picture of \qes .
\end{itemize}

Inclusive \qe~ electron-nucleus scattering is not the only process of this type.
This \qe~ process occurs in various other areas of physics, and is being
exploited to learn about the dynamics of the underlying composite system: \\
-- Quasi-elastic scattering of keV-energy photons from electrons bound in
atoms provides information about the energy and momentum distribution of bound
electrons. \\
-- Scattering of eV to keV  neutrons from condensed-matter systems such as liquid Helium
provides a measurement of the $He$ momentum distribution and correlations, and
has been exploited for a long time to isolate the effects of the Bose condensate
in superfluid Helium. \\
-- Quasi-elastic scattering of GeV-energy leptons on the quarks bound in
nucleons (DIS) has provided a wealth of information 
on the quark distribution functions; the observation of scaling violations in
DIS  has taught us much about the dynamics of strong interactions. 

Inclusive electron scattering from nuclei is a subclass of \qe~ processes, the
most obvious other representative being $(e,e^\prime p)$. When compared to exclusive
processes where the knocked-out nucleon is detected,  $(e,e')$ corresponds to an
integral over all final states of the nucleon, and consequently provides 
 less specific information. On the
other hand, $(e,e^\prime)$ is  more directly related to the dynamics of the initial hadronic
(nuclear) ground state. The complications of the final hadronic continuum play
a much smaller role.  This is true particularly at large $|{\bf q}|$, as
the electron is sensitive to the fate of the recoiling nucleon in a region of size
$1/|{\bf q}|$; the Final State Interaction (FSI) beyond that region  affects only 
the more exclusive processes.  

Quasi-elastic scattering from nuclei, compared to the other \qe~ processes 
mentioned above, has
one distinct drawback one must deal with: the nucleon the electron scatters from
is not elementary, but can be excited to various states. At large momentum
transfer the result is  that only the low-$\omega$ side of
the \qep~ can be exploited, the large-$\omega$ side  is obscured by the
overlap with $\Delta$ excitation.

During the last decade, the remarkable progress of our understanding 
of \qes~ has allowed us to define a number of features precisely. 
In particular for light nuclei and nuclear matter, \qes~ has provided 
accurate determinations of (integrated properties of) the nuclear 
spectral function. Detailed investigations of the reaction mechanism 
and the conditions necessary/sufficient for scaling 
have clearly shown that the dominant process is elastic
scattering from individual nucleons. The different scale-breaking mechanism 
have been identified. The effects beyond IA, mainly due to 
FSI, have come under much better control. The kinematical region where 
more complicated processes, such as MEC, are important 
has also been identified and the size of two-body current contributions has been 
understood.

In this review we try to give a reasonably comprehensive discussion of the
various aspects of \qes .  The quantitative
understanding of the \cs~ starts from a  description in terms of IA; effects 
beyond IA such as the role of FSI and  MEC are addressed next. We then  
describe some of the experimental aspects important for study of this reaction. 
We then give a compilation of the
experiments that have been performed and point out where the \cs s (most often not 
published in numerical form) can be found. We also briefly describe experiments
that provide the nucleon form factors needed to understand \qes .
 We subsequently discuss
scaling and the related superscaling. For light nuclei and nonrelativistic final
states, exact calculations can be performed. For the lower momentum
transfers, an alternative approach, the use of the Euclidean response, is
available and presented.   We then study the results
obtained after a longitudinal/transverse (L/T) separation of the \cs , and their
impact on the Coulomb sum rule.  A bothersome correction, the effect of
Coulomb distortion on the \cs s, is addressed as well. We also show how data for an important model
system for nuclear theory, infinite nuclear matter, can be obtained. Last, we
address other fields of \qes , and discuss the common aspects.

\section{Electron-nucleus scattering in  impulse approximation 
}\label{sec:IA}
\subsection{Electron-nucleus cross section}
\label{sec:eAxsec}

The differential cross section of the process
\beq
e + A \rightarrow e^\prime + X \ ,
\label{eA:process}
\eeq
in which an electron of initial four-momentum $k_e\equiv(E_e,{\bf k}_e)$ scatters off
a nuclear target to a state of four-momentum
$k^\prime_e\equiv(E_{e^\prime},{\bf k}_{e^\prime})$, the target final state being undetected,
can be written in Born approximation as \cite{Itzykson80}
\beq
\frac{d^2\sigma}{d\Omega_{e^\prime} dE_{e^\prime}} =
\frac{\alpha^2}{Q^4}\frac{E_{e^\prime}}{E_e}\ \Lmunu\Wmunu \ ,
\label{eA:xsec}
\eeq
where $\alpha=1/137$ is the fine structure constant, 
$d\Omega_{e^\prime}$ is the differential solid angle in the direction 
specified by ${\bf k}_{e^\prime}$, $Q^2=-q^2$ and 
$q=k_e-k_{e^\prime} \equiv (\omega,{\bf q})$ is the four momentum transfer.

The tensor $\Lmunu$, that can be written neglecting the lepton mass as
\beq
\Lmunu = 2 \left[ k_e^\mu k_{e^\prime}^\nu + k_e^\nu k_{e^\prime}^\mu  
 - g^{\mu\nu} (k_e k_{e^\prime}) \right] \ ,
\label{lepten}
\eeq
where $g^{\mu \nu} \equiv {\rm diag}(1,-1,-1,-1)$ and $(k_e k_{e^\prime})=E_e
E_{e^\prime}-{\bf k}_e \cdot {\bf k}_{e^\prime}$  is fully specified by the measured electron kinematical
variables. All the information on target structure is contained
in the tensor $\Wmunu$, whose definition involves the initial and final nuclear
states $|0\rangle$ and $|X\rangle$, carrying four-momenta $p_0$ and $p_X$,
as well as the nuclear current operator $J^\mu$: 
\beq
\Wmunu=\sum_{X}  \langle 0 |J^\mu|X \rangle\langle X|J^\nu|0\rangle
\delta^{(4)}(p_0+q-p_X)\ ,
\label{nuclear:tensor}
\eeq
\noindent
where the sum includes all hadronic final states.

The most general expression of the target tensor of Eq.~(\ref{nuclear:tensor}), 
fulfilling the 
requirements of Lorentz covariance, conservation of parity and gauge invariance, can be 
written in terms of two structure functions $W_1$ and $W_2$ as
\be
\nonumber
\Wmunu & = & W_1 \left( -g^{\mu\nu} + \frac{q^\mu q^\nu}{q^2} \right)  \\
 & + & \frac{W_2}{m^2} \left(p_0^\mu - \frac{(p_0 q)}{q^2}q^\mu \right)
                   \left(p_0^\nu - \frac{(p_0 q)}{q^2}q^\nu \right) \ ,
\label{genw}
\ee
where $m$ is the nucleon mass and the structure functions depend on the two scalars $Q^2$ 
and $(p_0 q)$. In the target rest frame $(p_0 q) = m\omega$ and $W_1$ and $W_2$ become 
functions of the measured momentum and energy transfer $|{\bf q}|$ and ${\omega}$.

Substitution of Eq.~(\ref{genw}) into Eq.~(\ref{eA:xsec}) leads to 
\be
\nonumber
\frac{d^2\sigma}{d\Omega_{e^\prime} dE_{e^\prime}} & = &
\left( \frac{d\sigma}{d\Omega_{e^\prime}}\right)_M \\  
 & \times & \left[ W_2(|{\bf q}|,\omega) 
            + 2 W_1(|{\bf q}|,\omega) \tan^2\frac{\theta}{2} \right] \ ,
\label{eA:xsec12}
\ee
where $\theta$ and 
$(d\sigma/d\Omega_{e^\prime})_M= \alpha^2 \cos^2(\theta/2)/4E_e\sin^4(\theta/2)$ 
denote the electron scattering angle and the Mott cross section, respectively. 

The right hand side of Eq.~(\ref{eA:xsec12}) can be further rewritten singling out
the contributions of scattering processes induced by longitudinally (L) and 
transversely (T) polarized virtual photons. The resulting expression is
\be
\nonumber
\frac{d^2\sigma}{d\Omega_{e^\prime} dE_{e^\prime}} & = & 
\left( \frac{d\sigma}{d\Omega_{e^\prime}}\right)_M  \ \left[ 
\frac{Q^4}{\magq^4}\ R_L(|{\bf q}|,\omega)  \right. \\ 
& + & \left. \left( \frac{1}{2} \frac{Q^2}{\magq^2} 
+ \tan^2\frac{\theta}{2} \right)  R_T(|{\bf q}|,\omega) \right] \ ,
\label{eA:xsecLT}
\ee
where the longitudinal and transverse structure functions are trivially related 
to $W_1$ and $W_2$ through
\beq
R_T(|{\bf q}|,\omega) = 2 W_1(|{\bf q}|,\omega)
\eeq
and
\beq
\frac{Q^2}{\magq^2}R_L(|{\bf q}|,\omega) = W_2(|{\bf q}|,\omega) - 
\frac{Q^2}{\magq^2} W_1(|{\bf q}|,\omega) \ .
\eeq

In principle, calculations of $\Wmunu$ of Eq.~(\ref{nuclear:tensor}) at moderate momentum 
transfer $( {\bf |q|} < 0.5\, {\rm GeV/c})$
can be carried out within nuclear many-body theory (NMBT), using
nonrelativistic wave functions to describe the initial and final
states and expanding the current operator in powers of ${\bf |q|}/m$
\cite{Carlson98}, where $m$ is the nucleon mass. 
 The available results for medium-heavy targets have been mostly obtained
using the mean field approach, supplemented by the inclusion of
model residual interactions to take into account long range correlations 
\cite{Dellafiore85}. 

On the other hand, at higher values of
${\bf |q|}$, corresponding to beam energies larger than $\sim$1 GeV,
 describing the final states $|X\rangle$
in terms of nonrelativistic nucleons is no longer possible.
Due to the prohibitive difficulties involved in a fully consistent treatment of the 
relativistic nuclear many-body problem, calculations of $\Wmunu$ in this regime require 
a set of simplifying assumptions, allowing one to take into account the relativistic motion of
final state particles carrying momenta $\sim {\bf q}$, as well as 
inelastic processes leading to the production  of hadrons other than protons and neutrons.

\begin{figure}[htb]
\includegraphics[scale=0.8,clip]{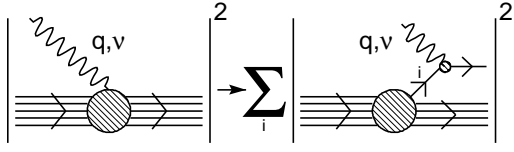} 
\caption{Schematic representation of the IA regime, in which
the nuclear cross section is replaced by the incoherent sum of
cross sections describing scattering off individual nucleons, the
recoiling $({\rm A}-1)$-nucleon system acting as a spectator.}
\label{IA:cartoon}
\end{figure}

The IA scheme is based on the assumptions that i) as the space 
resolution of the electron probe is $\sim 1/|{\bf q}|$, at large momentum
transfer scattering off a nuclear target reduces to the incoherent sum of
elementary scattering processes involving individual {\it bound} nucleons 
(see Fig.~\ref{IA:cartoon}) and  
ii) there is no FSI between the 
hadrons produced in electron-nucleon scattering and the recoiling nucleus 
\footnote{
Coherent effects, not included in the impulse approximation picture, appear 
in DIS even at large $|{\bf q}|$ for values of the Bjorken scaling variable 
$x =Q^2/2m \omega \lsim 0.2$, corresponding to very large electron energy loss.}. Under these assumptions, 
a relativistic particle in the final state is
completely decoupled from the spectator system, and the description of its
motion becomes a simple kinematical problem.

Within the IA picture the nuclear current appearing in Eq.~(\ref{nuclear:tensor}) 
is written as a sum of one-body currents
\beq
J^\mu \rightarrow \sum_i j_i^\mu \ ,
\label{currIA}
\eeq
while $|X\rangle$ reduces to the direct product of the hadronic state produced at the 
electromagnetic vertex, carrying momentum ${\bf p}_x$, and the state describing the 
$(A-1)$-nucleon residual system, carrying momentum 
${\bf p}_{\cal R}= {\bf q}-{\bf p}_x$ (in order to simplify the notation, spin 
indices will be omitted) 
\beq
|X\rangle \rightarrow |x,{\bf p}_x\rangle
\otimes |{\cal R},{\bf p_{\cal R}}\rangle \ .
\label{resIA}
\eeq
Using Eq.~(\ref{resIA}) we can replace
\be
\nonumber
\sum_X | X \rangle \langle X |  & \rightarrow & \sum_{x}
\int d^3p_x  | x,{\bf p}_x \rangle \langle {\bf p}_x,x | \\
&  \times  & \sum_{{\cal R}} d^3p_{{\cal R}}
| {\cal R}, {\bf p}_{{\cal R}} \rangle \langle {\bf p}_{{\cal R}}, {\cal R} | \
.
\label{sumn}
\ee
Substitution of Eqs.~(\ref{currIA})-(\ref{sumn}) into 
Eq.~(\ref{nuclear:tensor}) and
insertion of a complete set of free nucleon states, satisfying
\beq
\int d^3k\  | {\rm N},  {\bf k}\rangle\langle {\bf k},  {\rm N} |=1 \ ,
\eeq
then leads to the factorization of the nuclear current matrix element according to
\be
\nonumber
\langle 0 | J^\mu | X\rangle & = & 
\left( \frac{m}{\sqrt{|{\bf p}_{{\cal R}}|^2 + m^2}} \right)^{1/2}
 \langle 0 | {\cal R}, {\bf p}_{{\cal R}} ; {\rm N},-{\bf p}_{{\cal R}} \rangle \\
 & \times & \sum_i \langle -{\bf p}_{\cal R},N | j^\mu_i | x,{\bf p}_x \rangle \ ,
\ee
where the factor $(m/\sqrt{|{\bf p}_{{\cal R}}|^2 + m^2})^{1/2}$ takes into 
account the 
implicit covariant normalization of $\langle -{\bf p}_{\cal R},N |$ in the matrix
element of $j_i^\mu$.  

As a result, the incoherent contribution to Eq.~(\ref{nuclear:tensor}) can be rewritten in the form
\be
\nonumber
\Wmunu & = & \sum_{x,{\cal R}} \int d^3p_{{\cal R}}\ d^3p_x
| \langle 0 | {\cal R},{\bf p}_{{\cal R}};{\rm N},-{\bf p}_{{\cal R}} \rangle |^2 \\
\nonumber
& \times &  \frac{m}{E_{\cal R}}\ \sum_i
\langle -{\bf p}_{{\cal R}},{\rm N}| j^\mu_i | x,{\bf p}_x \rangle
\langle {\bf p}_x ,x | j^\nu_i | {\rm N},-{\bf p}_{{\cal R}} \rangle \\
& \times & \delta^{(3)}({\bf q}-{\bf p}_{{\cal R}}-{\bf p}_x)
\delta(\omega+E_0-E_{\cal R}-E_x),
\label{hadrtenIA}
\ee
where $E_0$ is the target ground state energy and 
$E_{\cal R} = \sqrt{|{\bf p}_{{\cal R}}|^2 + M_{\cal R}^2}$, $M_{\cal
R}$ being the mass of the recoiling system and $E_x$ the energy of the final
state X.

Finally, using the identity
\be
\nonumber
\delta(\omega+E_0-E_{\cal R}-E_x) & = & \int
dE \ \delta(E-m+E_0-E_{\cal R}) \\ 
& \times &  \delta (\omega-E+m-E_x) \ ,
\label{deltaIA}
\ee
and defining the nucleon spectral function as 
\be
\nonumber
S_N({\bf k}, E) & = & \sum_{\cal R}
|\langle 0|{\cal R},-{\bf k};{\rm N},{\bf k} \rangle |^2  \\
& \times & \delta(E-m+E_0-E_{\cal R}) \ ,
\label{specfunIA}
\ee
where the index $N = p,n $ labels either a proton or a neutron, we can cast
Eq.~(\ref{nuclear:tensor}) in the form
\be
\nonumber
\Wmunu({\bf q},\omega) & = & \int d^3k \ dE \  
\left(\frac{m}{E_{{\bf k}}}\right) \left[ Z S_p({\bf k}, E) 
w_p^{\mu\nu} \right. \\
& + & \left. (A-Z) S_n({\bf k}, E) w_n^{\mu\nu} \right] \ ,
\label{hadrten2}
\ee
$A$ and $Z$ being the target mass number and number of protons, respectively. In 
Eq.~(\ref{hadrten2}), $E_{{\bf k}} = \sqrt{|{\bf k}^2|+m^2}$ and
\be
\nonumber
w_N^{\mu\nu} & = & \sum_x \langle {\bf k},{\rm N}| j^\mu_N | x,{\bf k}+{\bf q} \rangle 
\langle {\bf k}+{\bf q},x | j^\nu_N | {\rm N},{\bf k} \rangle  \\
& \times & \delta({\widetilde \omega} + E_{{\bf k}} - E_x ) \  .
\label{nucleon:tensor}
\ee
    with (see Eqs.~(\ref{hadrtenIA}) and (\ref{specfunIA}))
     \beq
     {\widetilde \omega} = E_x - E_{{\bf k}} = E_0 + \omega - E_{{\cal R}}
     - E_{{\bf k}} = \omega - E + m - E_{{\bf k}} \ .
     \label{omega:tilde}
     \eeq
The above equations show that within the IA scheme the definition of the 
electron-nucleus cross section involves two important elements: i) the tensor 
$w_N^{\mu\nu}$, defined by Eq.~(\ref{nucleon:tensor}), describing the electromagnetic 
interactions of a {\it bound} nucleon carrying momentum ${\bf k}$ and ii) the spectral 
function, defined by Eq.~(\ref{specfunIA}), yielding its momentum and removal energy 
distribution. These quantities will be further discussed in the following sections. 

\subsection{Electron scattering off a bound nucleon}
\label{sec:eNxsec}

While in electron-nucleon scattering in free space the struck particle is given 
the entire four momentum transfer $q\equiv(\omega,{\bf q})$, in a scattering process 
involving a bound nucleon a fraction $\delta \omega$ of the energy loss goes
into the spectator system. This mechanism emerges in a most natural fashion from 
the IA formalism.  

Assuming that the current operators are not modified by the nuclear environment,
the quantity defined by Eq.~(\ref{nucleon:tensor}) can be identified with 
the tensor describing electron scattering off a {\it free} nucleon at four 
momentum transfer $\tilde{q}\equiv({\bf q},{\widetilde \omega})$. Hence, Eq.~(\ref{nucleon:tensor}) 
shows that within IA binding is taken into account through the replacement
\beq
q\equiv(\omega,{\bf q}) \rightarrow {\widetilde q}\equiv({\widetilde \omega},
{\bf q}) \ .
\label{qtilde}
\eeq

The interpretation of $\delta \omega = \omega - {\widetilde \omega}$ as the amount of 
energy going into the recoiling spectator system becomes particularly transparent 
in the limit $|{\bf k}|/m \ll 1$, in which Eq.~(\ref{omega:tilde}) yields 
$\delta \omega = E$. 

In the case of quasielastic scattering, to be discussed in this review, $w_N^{\mu\nu}$
of Eq.~(\ref{nucleon:tensor}) can be obtained from the general expression 
(compare to Eq.~(\ref{genw}))
\be
\nonumber
w_N^{\mu\nu}  & = & w^N_1 \left( -g^{\mu\nu} + \frac{\qt^\mu \qt^\nu}{\qt^2} \right) \\
 & + & \frac{w^N_2}{m^2} \left(k^\mu - \frac{(k \qt)}{\qt^2}\qt^\mu \right)
                   \left(k^\nu - \frac{(k \qt)}{\qt^2}\qt^\nu \right) \ ,
\label{genw2}
\ee
where $k\equiv (E_{\bf k},{\bf k})$ and the two structure functions $w_1$ and $w_2$ are
simply related to the measured electric and magnetic nucleon form factors, 
$G_{E_N}$ and $G_{M_N}$, through
\beq
w^N_1 = -\frac{\qt^2}{4m^2}\ \delta\left({\widetilde \omega} + \frac{\qt^2}{2m} \right)
 \ G_{M_N}^2 \ ,
\label{qe1}
\eeq
\be
\nonumber
w^N_2 & = & \frac{1}{1 - \qt^2/4 m^2} \ \delta\left({\widetilde \omega} + 
\frac{\qt^2}{2m} \right) \\
 & \times & \left( G_{E_N}^2 - \frac{\qt^2}{4m^2} G_{M_N}^2 \right) \ .
\label{qe2}
\ee

While the replacement of $\omega$ with $\tilde{\omega}$ is reasonable on physics
grounds and in fact quite natural in the context of the PWIA analysis presented
above,   it poses a
considerable conceptual problem in that it leads to a violation of current 
conservation, which requires 
\beq
q_\mu w_N^{\mu\nu} = 0 \ .
\label{gauge:inv}
\eeq
A widely used prescription to overcome this difficulty has been proposed  
in the early 1980's \cite{Forest83}. It amounts to using a 
tensor ${\widetilde w}_N^{\mu\nu}$ whose
time components are given by Eq.~(\ref{nucleon:tensor}), while the longitudinal ones
are obtained from the time components requiring that the continuity equation 
(\ref{gauge:inv}) be satisfied. Taking the $z$-axis along the direction of ${\bf q}$
one obtains 
\be
\nonumber
{\widetilde w}_N^{\mu\nu} & = & w_N^{\mu\nu}({\widetilde q}) \ , 
\ {\rm for} \ \mu \ {\rm and/or} \ \nu = 0 \ , \\
{\widetilde w}_N^{3\nu} & = & \frac{\omega}{|{\bf q}|} \ w_N^{0\nu}({\widetilde q}) \ ,
\label{defo:cc1}
\ee
with ${\widetilde q}$ given by Eq.~(\ref{qtilde}). 

The above prescription is manifestly non-unique. For example, one might have just as 
well chosen to use Eq.~(\ref{nucleon:tensor}) to obtain the longitudinal components of 
${\widetilde w}^{\mu\nu}$ and the continuity equation to obtain the time components. 
However, differences between these two procedures to restore gauge invariance only 
affect the longitudinal response $R_L$. As a consequence, they are expected to become 
less and less important as the momentum transfer increases, electron 
scattering at large $\magq$ being largely dominated by transverse contributions. 
Numerical calculations of the quasielastic cross section off $^3\!He$ have confirmed that 
these differences
become indeed vanishingly small already in the pion production region {\cite{Meier89}.

The uncertainty associated with the implementation of current conservation has been
recently analyzed by Kim, who has compared the inclusive cross sections computed using the
Coulomb gauge (equivalent to the deForest prescription) to those obtained from
the Landau and Weyl gauges for a variety of targets and kinematical condition
\cite{Kim04}. The results corresponding to the different choices agree with one another
to better than $\sim$ 1 \% for backward angle kinematics, thus confirming that the
transverse response is unaffected by gauge ambiguities. For the longitudinal
response, Kim finds more substantial ambiguities, as a consequence of the huge
scalar/vector-potentials present in the $\sigma-\omega$ model; these large
potentials, however, only occur when trying to describe  nuclei --- strongly 
correlated systems --- using a mean-field basis. 

Other prescriptions based on the use of free nucleon currents \cite{Forest83}
 have been recently employed in the analysis of $(e,e^\prime p)$ data in the region of 
large missing momentum and energy, where the effects of using different off-shell 
extrapolations of electron-nucleon scattering become sizable \cite{Rohe04a}. 
More fundamental approaches,
involving the explicit calculation of the off-shell form factors, necessarily rely on 
oversimplified dynamical models \cite{Naus87,Naus90}.

In conclusion the violation of gauge invariance in the IA scheme, while  
in principle an intricate one, turns out to be only marginally relevant to 
inclusive electron scattering at large momentum transfer. The main
 effect of nuclear binding can be easily accounted for with the replacement 
$\omega \rightarrow {\widetilde \omega}$.

\subsection{The nuclear spectral function}
\label{sec:Ske}

The spectral function, defined by Eq.~(\ref{specfunIA}), gives the probability of removing a 
nucleon from the target nucleus leaving the residual system with energy $E_{\cal R} = E_0 - m + E$.

Within the shell model picture, based on the assumption that nucleons in a nucleus
behave as independent particles moving in a mean field, the spectral function reduces to
\beq
S_{SM}({\bf k},E) = \sum_{n\epsilon\{F\}} |\phi_n({\bf k})|^2 \delta(E-E_n) \ ,
\label{S:SM}
\eeq
where $\phi_{n}({\bf k})$ is the momentum-space wave function associated with the single particle 
shell model state $n$, $E_n$ is the corresponding energy eigenvalue and the sum is extended to all 
occupied states belonging to the Fermi sea $\{F\}$. 

The results of electron- and hadron-induced nucleon knock-out experiments have provided 
overwhelming evidence of the inadequacy of the independent particle model to describe the 
full complexity of nuclear dynamics. While the peaks corresponding to knock-out 
from shell model orbits can be clearly identified in the measured energy spectra, the 
corresponding strengths turn out to be 
consistently and sizably lower than expected, independent of the nuclear mass number. 

This discrepancy is mainly due to the effect of dynamical correlations induced 
by the nucleon-nucleon (NN) force, whose effect is not taken into account in the independent 
particle model. Correlations give rise to scattering processes, leading to the 
virtual 
excitation of the participating nucleons to states of energy larger than the 
Fermi energy, 
thus depleting the shell model states within the Fermi sea. As a result, the spectral 
function acquires tails extending to the region of large energy and momentum, where 
$S_{SM}({\bf k},E)$ of Eq.~(\ref{S:SM}) vanishes.

The typical energy scale associated with NN correlations can be estimated considering
a pair of correlated nucleons carrying momenta ${\bf k}_1$ and ${\bf k}_2$ much
larger than the Fermi momentum ($\sim 250$ MeV/c). In the nucleus rest frame, where 
the remaining $A-2$ particles carry low momenta, ${\bf k}_1 \approx -{\bf k}_2 = {\bf k}$. 
Hence, knock-out of a nucleon of large momentum ${\bf k}$ leaves the residual system with a 
particle in the continuum and requires an energy
\beq
E \approx E_{thr} + \frac{{\bf k}^2}{2m}\ ,
\label{corr:en}
\eeq
much larger than the typical energies of shell model states ($\sim 30$ MeV). 
The above equation, where  $E_{thr}$ denotes 
the threshold for two-nucleon removal, shows that large removal energy and large nucleon 
momentum are strongly correlated.

Realistic theoretical calculations of the spectral function have been carried out within 
NMBT, according to which the nucleus consists of a collection of $A$ 
nucleons whose dynamics are described by the nonrelativistic Hamiltonian  
\beq
H = \sum_{i=1}^{A} \frac{{\bf k}_i^2}{2m} + \sum_{j>i=1}^{A} v_{ij}
 + \sum_{k>j>i=1}^A V_{ijk} \ .
\label{H:A}
\eeq
In the above equation,  ${\bf k}_i$ is the momentum of the $i$-th constituent and $v_{ij}$ and $V_{ijk}$
describe two- and three-nucleon interactions, respectively. The two-nucleon potential,
that reduces to the Yukawa one-pion-exchange potential at large internucleon distance, is
obtained from an accurate fit to the available data on the
two-nucleon system, {\em i.e.} deuteron properties and $\sim$ 4000 NN
scattering data  \cite{Wiringa95}. The additional three-body term
$V_{ijk}$ has to be included in order to account for the binding energies of the
three-nucleon bound states \cite{Pudliner95b} and the empirical saturation properties 
of uniform nuclear matter \cite{Akmal97}; this term results from the fact that
non-nucleonic constituents (such as $\Delta$'s) have been excluded.

The many-body Schr\"odinger equation associated with the Hamiltonian
of Eq.~(\ref{H:A}) can be solved exactly, using stochastic methods,
for nuclei with mass number $A \leq 12$. The resulting energies of the ground and low-lying
excited states are in excellent agreement with the experimental
data \cite{Pieper01}. Accurate calculations can also be carried out for uniform
nuclear matter, exploiting translational invariance and using either
a variational approach based on cluster expansion and chain summation
techniques \cite{Akmal97}, or G-matrix perturbation theory \cite{Baldo00}.

Nonrelativistic NMBT has been employed to obtain the spectral functions of
the three-nucleon systems \cite{Dieperink76,Ciofi80,Meier83}, Oxygen \cite{Geurts96,Polls97} 
and symmetric nuclear matter, having $ A \rightarrow \infty$ and $Z=A/2$ 
\cite{Benhar89,Ramos89}. Calculations based on NMBT but involving some simplifying 
assumptions have been also carried out for $^4\!He$ \cite{Ciofi90b,Morita91,Benhar93}.

\begin{figure}[htb]
\includegraphics[scale=1.0,clip]{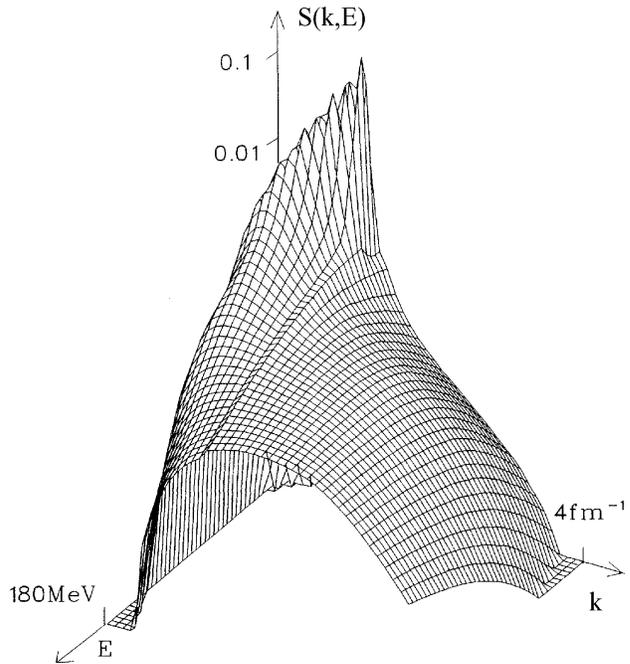}
\caption{Nuclear matter spectral function calculated using correlated
basis function  perturbation theory \protect\cite{Benhar89}.}
\label{nm:Pke}
\end{figure}

As an example, Fig.~\ref{nm:Pke} shows the results of a nuclear matter 
calculation\footnote{As in 
symmetric nuclear matter $S_p({\bf k},E)=S_n({\bf k},E)$, the spectral function shown 
in the figure corresponds to an isoscalar nucleon.
}  carried out using correlated basis function (CBF) perturbation theory
 \cite{Benhar89}. In addition to the peaks corresponding to single particle states, 
{\em i.e.} to 
bound one-hole states of the (A--1)-nucleon system, the resulting $S_N({\bf k},E)$ 
exhibits a broad background, extending 
up to $E \sim 200$ MeV and $|{\bf k}| \sim 800$ MeV/c, associated 
with $n$-hole ($n$--1)-particle (A--1)-nucleon states  
in which at least one nucleon is excited to the continuum. The correlation ridge at 
$E \sim {\bf k}^2/2m$ (see Eq.~(\ref{corr:en})) is 
clearly visible. Note that, in absence of interactions, the surface shown 
in Fig.~\ref{nm:Pke} collapses to a collection of $\delta$-function peaks distributed along 
the line $|E| = {\bf k}^2/2m$, with $|{\bf k}| < k_F \approx 250$ MeV/c.

The proton spectral functions of nuclei with A $>$ 4 have been modeled using the
Local Density Approximation (LDA) \cite{Benhar94b}, in which the experimental
information obtained from nucleon knock-out measurements is combined
with the results of theoretical calculations of the nuclear matter
$S({\bf k},E)$ at different densities.

The kinematical region corresponding to low missing energy and momentum, where shell model 
dynamics dominates, has been extensively
studied by coincidence $(e,e^\prime p)$ experiments. The spectral function extracted from the data 
is usually written in the factorized form (compare to Eq.~(\ref{S:SM}))
\beq
S_{MF}({\bf k},E) = \sum_{n\epsilon\{F\}} Z_n\ |\phi_n({\bf k})|^2 F_n(E-E_n) \ ,
\label{S:MF}
\eeq
where the {\it spectroscopic factor} $Z_n < 1$ and the function $F_n(E-E_n)$, describing 
the energy width of the $n$-th state, account for the effects of residual 
interactions not included in the mean field picture. In the 
$Z_n \rightarrow 1$ and $F_n(E-E_n) \rightarrow \delta(E-E_n)$ limit 
Eq.~(\ref{S:MF}) reduces to Eq.~(\ref{S:SM}).

The correlation contribution to the nuclear matter spectral function
has been calculated using CBF perturbation theory for a wide range of density 
values \cite{Benhar94b}. Within the LDA scheme, these results can be used to 
obtain the corresponding quantity for a finite nucleus of mass number $A$ from
\beq
S_{corr}({\bf k},E) = \int d^3r\ \rho_A({\bf r})
S^{NM}_{corr}({\bf k},E;\rho = \rho_A({\bf r})) \ ,
\label{S:corr}
\eeq
where $\rho_A({\bf r})$ is the nuclear density distribution and
$S^{NM}_{corr}({\bf k},E;\rho)$ is the correlation part of 
the spectral function of uniform nuclear matter at density $\rho$.
    The correlation part of the nuclear matter spectral function can be 
     easily singled out at zero-th order of CBF, being associated to 
     two hole-one particle intermediate states. At higher orders, however, one hole 
     and two hole-one particle states are coupled, and the identification 
     of the correlation contributions becomes more involved.
     A full account of the calculation of $S^{NM}_{corr}({\bf k},E)$ can be found
     in \cite{Benhar94b}.

The full LDA spectral function is written in the form
\beq
S_{LDA}({\bf k},E) = S_{MF}({\bf k},E) + S_{corr}({\bf k},E) \ ,
\label{S:LDA}
\eeq
the spectroscopic factors $Z_n$ of Eq.~(\ref{S:MF}) being constrained
by the normalization requirement
\beq
\int d^3k\ dE\ S_{LDA}({\bf k},E) = 1 \ .
\label{S:norm}
\eeq

A somewhat different implementation of LDA has also been proposed
\cite{Neck95}. Within this approach the nuclear
matter spectral function is only used at $k>k_F(r)$, $k_F(r)$ being the
local Fermi momentum, whereas the correlation background at $k<k_F(r)$
is incorporated in the generalized mean field contribution. Comparison
between the resulting Oxygen momentum distribution and that obtained
by Benhar {\it et al.}  shows that they are in almost
perfect agreement.  

The LDA scheme is based on the premise that short range nuclear dynamics are 
unaffected by surface and shell effects. The validity of this assumption
is   supported by the results of theoretical calculations
of the nucleon momentum distribution
\beq
n({\bf k}) = \int dE\ [Z S_p({\bf k},E) + (A-Z) S_n({\bf k},E) ]\ ,
\label{momdis}
\eeq
showing that for A$\ge$4 the quantity $n({\bf k})/A$ becomes nearly independent of
$A$ at large $|{\bf k}|$ ($\gsim 300$ MeV/c). This feature, illustrated in 
Fig.~\ref{scal:nk}, suggests that the correlation part of the spectral function 
also scales with the target mass number, so that $S^{NM}_{corr}({\bf k},E)$
can be used to approximate $S_{corr} (k,E)$ at finite $A$.

\begin{figure}[ht]
\includegraphics[scale=0.95,clip]{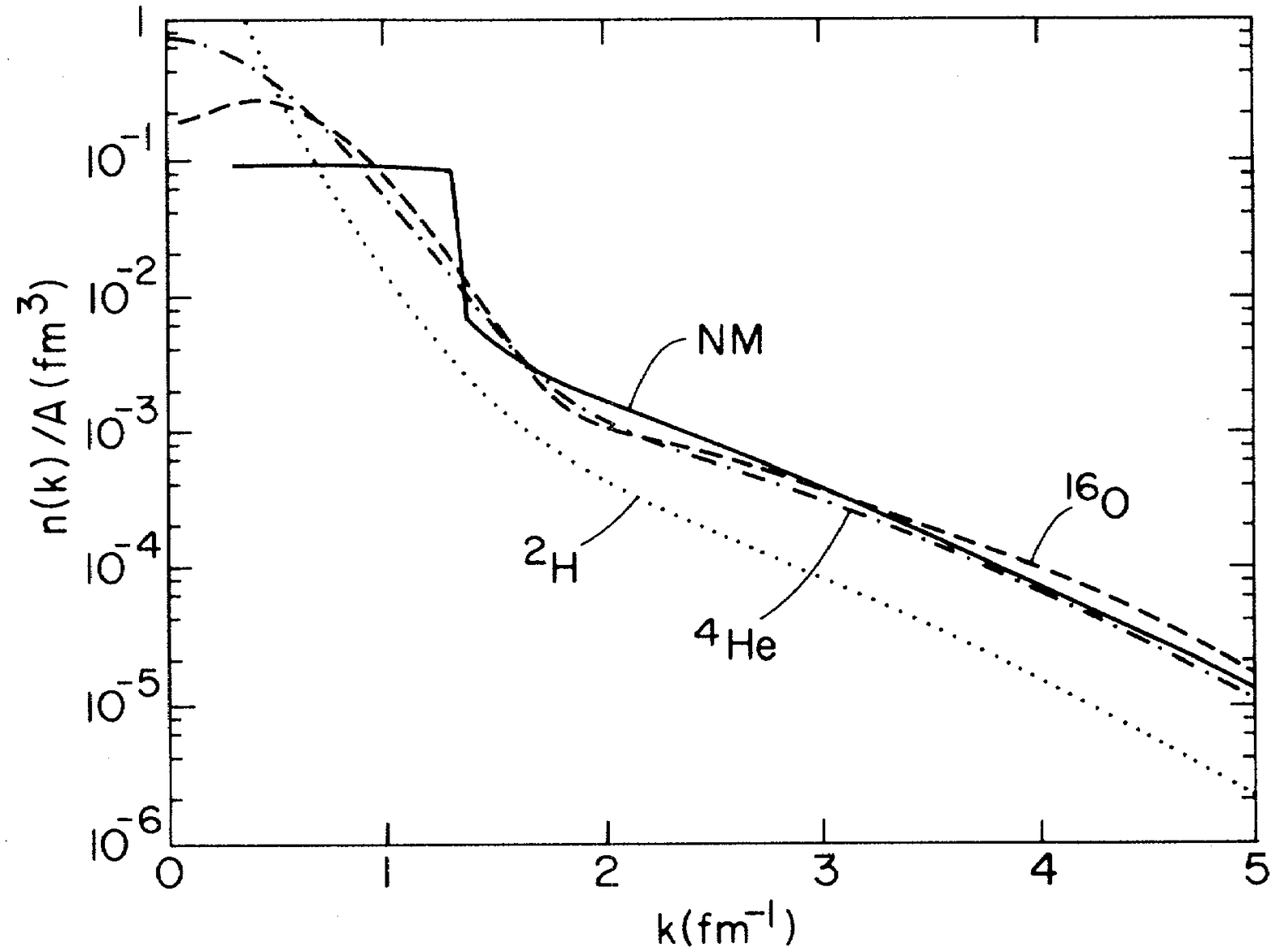}
\caption{Calculated momentum distribution per nucleon in $^2\!H$, $^4\!He$,
$^{16}\!O$ and uniform nuclear matter \protect\cite{Schiavilla86,Benhar93a}.}
\label{scal:nk}
\end{figure}

A direct measurement of the correlation component of the
spectral function of $^{12}C$, from the 
$(e,e^\prime p)$ cross section at missing momentum and energy up to $\sim$ 800
MeV/c
and $\sim 200$ MeV, respectively, has been recently carried out  by the JLab
E97-006 Collaboration \cite{Rohe04a}. The data  from the preliminary
analysis appear to be consistent with the theoretical predictions based
on LDA.

\subsection{Contribution of inelastic processes}
\label{inelastic}

The approach described in the previous sections is not limited to quasielastic 
processes. The tensor defined in Eqs.~(\ref{hadrten2}) and (\ref{nucleon:tensor}) 
describes electromagnetic transitions of the struck nucleon to any hadronic 
final state. 

To take into account the possible production of hadrons other than 
protons and neutrons one has to replace $w^N_1$ and $w^N_2$ given by 
Eqs.~(\ref{qe1}) and (\ref{qe2}) with the inelastic nucleon structure functions extracted 
from the analysis of electron-proton and electron-deuteron scattering data 
\cite{Bodek81}. The resulting IA cross section can be written as in Eq.~(\ref{eA:xsec12}), 
the two nuclear structure functions $W_1$ and $W_2$ being given by \cite{Benhar97}
\be
\nonumber
& & W_1(|{\bf q}|,\omega) = \int d^3k\ dE\  \left\{ Z S_p({\bf k},E) 
\left( \frac{m}{E_k} \right)\right. \\
& & \ \times \left[ w^p_1(|{\bf q}|,{\widetilde \omega}) 
\left.  + \frac{1}{2}\frac{w^p_2(|{\bf q}|,{\widetilde \omega})}{m^2}
 \frac{|{\bf k} \times {\bf q}|^2}{|{\bf q}|^2}  \right] + \ldots \ \right\}
\label{W1A}
\ee
and
\be
\nonumber
&  & W_2(|{\bf q}|,\omega) = \int d^3k\ dE \left\{ ZS_p({\bf k},E) 
\left( \frac{m}{E_{\bf k}} \right) \right. \\
\nonumber
& &  \ \ \ \times \left[ w^p_1(|{\bf q}|,{\widetilde \omega}) \frac{q^2}{|{\bf q}|^2}
            \left( \frac{q^2}{\qt^2} - 1 \right) \right. \\
\nonumber
     &  & \ \ \ + \frac{ w^p_2(|{\bf q}|,{\widetilde \omega})}{m^2}
 \left( \frac{ q^4 }{ |{\bf q}|^4 }
 \left( E_{\bf k} -{\widetilde \omega}\ 
\frac{ E_{\bf k}{\widetilde \omega} - {\bf k}\cdot{\bf q} }{ \qt^2 } \right)^2 \right. \\
&  & \ \ \ - \left. \left.  \left.
\frac{1}{2}\frac{q^2}{\magq^2}\frac{|{\bf k}\times{\bf q}|^2}{|{\bf q}|^2}
 \right)  \right] + \ldots \ \right\} \ ,
\label{W2A}
\ee
where the dots denote the neutron contributions. Eqs.~(\ref{W1A}) and (\ref{W2A}) 
are obtained using the prescription of Eq.~(\ref{defo:cc1}) \cite{Forest83} to preserve 
gauge invariance. Note that the standard 
expression \cite{Atwood73}, widely used in studies of nuclear effects in 
deep inelastic scattering, can be recovered from the above equations 
replacing ${\widetilde \omega} \rightarrow \omega$ and 
$E_{\bf k} \rightarrow M_A - E_{\cal R}$.

\begin{figure}[ht]
\includegraphics[scale=0.5,clip]{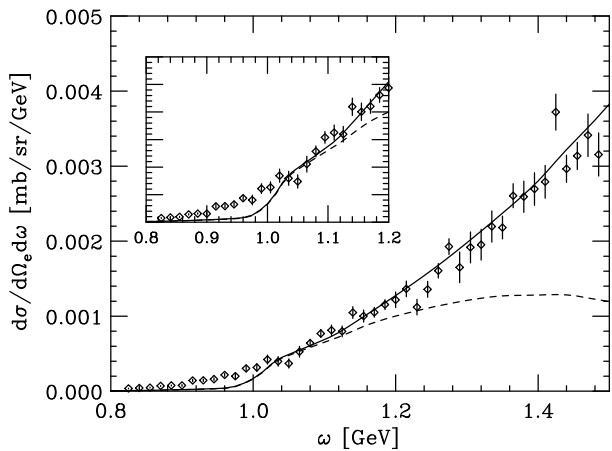}
\caption{Inclusive electron scattering cross section at $E_e = 3.595$GeV and
$\theta = 30^\circ$. The data points represent the extrapolated nuclear
matter cross section \protect\cite{Day89} while the solid and dashed lines show the
results of IA calculations carried out with and without inclusion of the inelastic 
contributions, respectively \protect\cite{Benhar91}.}
\label{inelxs}
\end{figure}

As an example, Fig.~\ref{inelxs} shows the quasi-elastic (dashed line) and total 
 (solid line) inclusive cross sections of uniform nuclear matter, at beam energy 
$E_e = 3.595$ GeV and scattering
angle $\theta = 30^\circ$, evaluated using 
a phenomenological fit of the nucleon structure functions $w^N_1$ and 
$w^N_2$ \cite{Bodek81} and the above mentioned spectral function \cite{Benhar89}.

The data show that the transition from the quasi elastic to the inelastic 
regime, including resonant and nonresonant pion production as well as deep inelastic 
processes, is a smooth one, thus suggesting the possibility of a unified representation.

The approach based on NMBT and the IA yields a good description of the measured cross section 
at energy loss $\omega \gsim$ 1 GeV, corresponding to $x \lsim 1.3$ (note that in the 
kinematics of Fig.~\ref{inelxs} the top of the quasi free bump corresponds to
$\omega = \omega_{QE} \sim 1.4$ GeV). On the other hand, the data at lower energy loss
are largely underestimated.

The failure of IA calculations to explain the measured cross sections at 
$\omega \ll \omega_{QE}$
has long been recognized, and confirmed by a number of theoretical studies, 
carried out using highly realistic spectral functions 
\cite{Benhar89,Meier83,Ciofi92}, see {\em e.g.} Fig.~\ref{meier1}. It has to be 
ascribed to 
FSI between the struck nucleon and the spectator particles, that move strength 
from the  region of the quasi free bump to the low $\omega$ tail. 
This mechanism will be analyzed in the next Section. 

\begin{figure}[ht]
\includegraphics[scale=0.4,clip]{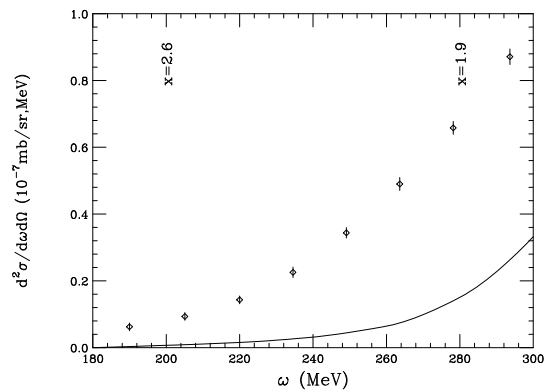}
\caption{Inclusive electron scattering cross section at $E_e = 7.26$GeV and
$\theta = 8^\circ$ for $^3$He. The data points are from \cite{Day79}, 
the solid line shows the IA calculation based on the $^3$He spectral function 
\cite{Meier83}. Approximate values for the scaling variable $x$ are indicated on
top. \label{meier1}}
\end{figure}

In conclusion, NMBT and the IA provide a consistent and computationally 
viable approach, yielding a quantitative description of the data in both the quasi elastic 
and inelastic regime, with the only exception of the region of very low energy loss.
Theoretical studies in which nuclear binding effects are included                              
using realistic spectral functions also provide a quantitative                                      
account of the size and density dependence of the European Muon 
Collaboration (EMC) effect \cite{Benhar97,Benhar99b}.  


\subsection{Different implementations of the IA scheme}
\label{ambiguities}

In spite of the fact that the basic assumptions underlying IA can
be unambiguously stated, in the literature one finds two different
definitions of the IA inclusive cross section: the one discussed in 
Sec.~\ref{sec:eAxsec}, involving the target spectral function, and another
one, written in terms of the target momentum distribution of Eq.~(\ref{momdis}) 
\cite{Rinat96}.

It has been shown \cite{Benhar01b} that the definition in
terms of the spectral function follows from minimal use of the
assumptions involved in the IA scheme, and
correctly takes into account the correlation between momentum and
removal energy of the participating constituent. On the other hand, a more
extended use of the same assumptions leads to a definition of the IA cross 
section in which the nucleon spectral function is written in the 
approximated form 
\beq
S({\bf k},E) = \frac{1}{A}~ n({\bf k})~ 
\delta\left( E+\frac{ |{\bf k}|^2 }{ 2m }\right) \ ,
\label{rinat1}
\eeq
so that the information on the target removal energy distribution 
is  lost.

Figure \ref{ia:amb} shows a comparison between the response 
function of a nonrelativistic model of uniform nuclear matter\footnote
{The response function shown in Fig.~\protect\ref{ia:amb} is proportional to 
the inclusive cross section in the case of scattering of a scalar probe. The
generalization to the electromagnetic target tensor of 
Eq.~(\protect\ref{nuclear:tensor}) is straightforward. 
}
obtained from the full spectral function (solid line) and the approximation 
of Eq.~(\ref{rinat1}) (dashed line).

\begin{figure}[ht]
\epsfysize=2.0in
\epsfbox{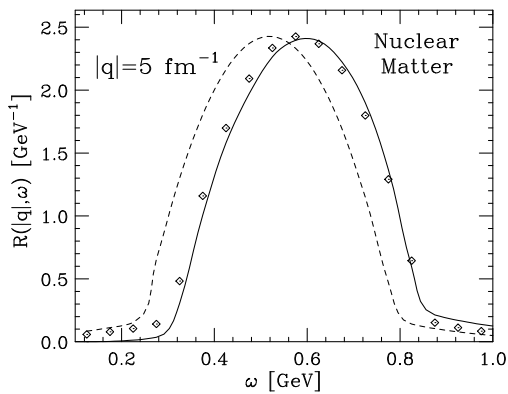}
\caption{
Response of  uniform nuclear matter at equilibrium 
density and $\magq$=5 fm$^{-1}$.
The solid and dashed lines have been obtained using the full spectral function
 \protect\cite{Benhar89} and the approximation of
Eq.~(\protect\ref{rinat1}), respectively. 
The diamonds represent the results
obtained when shifting the dashed line by ${\bar E}= 62$ MeV
\cite{Benhar01b}.
}
\label{ia:amb}
\end{figure}

The two responses have similar shape, their width being dictated by the
momentum distribution. However, they are shifted with respect to one another.
The peak of the dashed curve is located at
 energy loss $\omega \sim |{\bf q}|^2/2m$, corresponding to 
elastic scattering off a free stationary nucleon, whereas 
the solid line, due to the removal energy distribution described by the spectral 
function, peaks at significantly larger energy. The shift is roughly given by
the average nucleon removal energy \footnote{Note that for spectral functions describing 
{\em correlated} nucleons ${\bar E}$ is much larger than the average of 
single-particle energies.} 
\beq
{\bar E} = \int d^3k dE\  E S({\bf k},E) = 62\ {\rm MeV}\ .
\eeq
This feature is illustrated by the diamonds of Fig.~\ref{ia:amb}, that show the 
response obtained replacing
$|{\bf k}|^2/2m \rightarrow (|{\bf k}|^2/2m - {\bar E})$ in the argument of the energy 
conserving $\delta$-function of Eq.~(\ref{rinat1}). The results of this calculation 
turn out to be much closer to those derived in the previous sections.

In addition to the shift in the position of the peak, 
the dashed and solid curves sizably differ at low energy transfer, where
the response obtained using the momentum distribution is much larger.
Obviously, to identify corrections to the response arising from mechanisms 
not included in the IA picture, one has to start from the definition
involving the minimal set of approximations. 
The results of Fig.~\ref{ia:amb} show that a quantitative understanding 
of the effects of FSI, which are known to dominate the inclusive cross section at 
low $\omega$, requires the use of the spectral function in the calculation of the IA
cross section.
\section{Final state interactions}
\label{sec:FSI}
The existence of strong FSI in quasi-elastic  scattering has long been
experimentally established. The results of a number of $(e,e^\prime p)$ measurements,
covering the kinematical domain corresponding to $0.5 \lsim Q^2 \lsim 8.0$ (GeV/c)$^2$
\cite{Garino92,O'Neill95,Abbott98,Garrow02,Rohe05}, clearly show that the flux
of outgoing protons is strongly suppressed with respect to the IA predictions.
The observed attenuation ranges from 20-40 \% in Carbon to 50-70 \% in Gold.

Being only sensitive to rescattering processes taking place within a distance 
$\sim 1/|{\bf q}|$ of the electromagnetic vertex, the inclusive cross section at 
high momentum transfer in general is largely unaffected by FSI.
However, the systematic discrepancies between data and the results of highly
accurate IA calculations indicate that the effects of FSI can become 
appreciable, indeed dominant, in the low $\omega$
region, where the cross sections produced by IA calculations become very small. 
As  IA cross sections in this region are most sensitive to the high momentum
and high removal energy tails of the nuclear spectral function 
\footnote{When ignoring the E-distribution of $S(k,E)$ the cross section at low
$\omega$ is related to the strength at large $k$. However, the spectral function
exhibits a strong correlation between {\em large} $|{\bf k}|$ and {\em large}
$E$. For example, in nuclear matter at equilibrium density, more than 50\% of
the strength at $|{\bf k}| = 1.5fm^{-1}$ (just above the Fermi surface) is
located at $E >$ 80MeV.}
a quantitative understanding of FSI is required to unambiguously 
identify correlation effects.

In inclusive processes FSI has two effects: i) an energy shift of the cross section, 
due to the fact that the struck nucleon moves in the average potential generated by the 
spectator particles and ii) a redistribution of the strength, leading to the quenching
of the quasielastic peak and an enhancement of the tails, as a consequence 
of NN scattering processes coupling the one particle-one hole final state to more
complex n-particle n-hole configurations.

Early attempts to include FSI effects were based on the Green's function formalism                                        
and multiple scattering theory, leading to a description of the dynamics in terms                                       
of a complex optical potential \cite{Horikawa80}.                                                                       
However, while providing a computationally practical scheme to account for the loss of                                  
flux in the one-nucleon removal channel, the optical potential model employed                                                  
relies on the mean field picture of the nucleus, and does not                                    
include the effect of dynamical NN correlations.                                                                        
                                                                                                                        
A similar approach has been  adopted \cite{Chinn89} to analyze  the                                         
longitudinal and transverse responses of Eq.~(\ref{eA:xsecLT}) and                                                      
investigate the possible importance of relativistic effects using Dirac 
bound-state                                    
wave functions and optical potentials. Although the authors  suggest                              
that relativity may play an important role in suppressing the response functions,                                       
particularly $R_L$, the interpretation of their results is hindered by                                                 
the large uncertainty associated with relativistic descriptions of nuclear 
dynamics and the appearance of the deep potentials, driven by the attempt to
describe correlated systems using an  independent particle model. 
Systematic studies of relativistic effects                                     
carried out within NMBT show that they are indeed rather small. For example,                                            
when  using the relativistic kinetic energy operator and                                          
when including boost corrections to the NN potential,  a change of                                                    
the binding energy of nuclear matter at equilibrium density of                                                          
$\sim$10\%  only has been found \cite{Akmal98}.

A different approach, based on NMBT and a generalization of Glauber theory 
of high energy proton scattering \cite{Glauber59} was  proposed  
 in the early 90's \cite{Benhar91}. This treatment of FSI, generally 
referred to as the 
Correlated Glauber Approximation (CGA), rests on the premises that
i) the struck nucleon moves along a straight trajectory with constant velocity
(eikonal approximation), and ii) the spectator nucleons are seen by the
struck particle as a collection of fixed scattering centers
(frozen approximation).

Under the  assumptions given above the expectation value of the propagator of the
struck nucleon in the target ground state can be written in the factorized form
\cite{Petraki03}
\beq
U_{{\bf k}+{\bf q}}(t) = U^0_{{\bf k}+{\bf q}}(t)
{\bar U}^{FSI}_{{\bf k}+{\bf q}}(t)\ ,
\eeq
where $U^0_{{\bf k}+{\bf q}}(t)$ is the free space propagator, while FSI
effects are described by the quantity $(R\equiv({\bf r}_1,\ldots,{\bf r}_A))$
\beq
{\bar U}^{FSI}_{{\bf k}+{\bf q}}(t) = \langle 0 |
U^{FSI}_{{\bf k}+{\bf q}}(R;t) | 0 \rangle \ ,
\label{eik:prop0}
\eeq
with
\beq
U^{FSI}_{{\bf k}+{\bf q}}(R;t) = \frac{1}{A} \sum_{i=1}^A
{\rm e}^{-i \sum_{j > i} \int_0^t dt^\prime
\Gamma_{{\bf k}+{\bf q}}(|{\bf r}_{ij} + {\bf v}t^\prime |) } \ .
\label{eik:prop}
\eeq

In Eq.~(\ref{eik:prop}) ${\bf r}_{ij}={\bf r}_{i}-{\bf r}_{j}$ and 
$\Gamma_{{\bf k}+{\bf q}}(|{\bf r}|)$ is the coordinate-space 
t-matrix, related to the NN scattering amplitude at incident momentum 
${\bf k}+{\bf q}$ and momentum transfer ${\bf p}$, $A_{{\bf q}}(p)$ by 
\beq
\Gamma_{{\bf k}+{\bf q}}({\bf r}) = - \frac{2 \pi}{m} \int \frac{d^3p}{(2 \pi)^3}
e^{i{\bf p}\cdot{\bf r}} A_{{\bf k}+{\bf q}}({\bf p})\ .
\label{def:Gamma}
\eeq 
At large $|{\bf q}|$, ${\bf k}+{\bf q} \approx {\bf q}$ and the eikonal
propagator of Eq.(\ref{eik:prop}) becomes a function
of $t$ and the momentum transfer only. 

The scattering amplitude extracted from the measured NN cross sections is generally
parameterized in the form 
\beq
A_{{\bf q}}(p) = i \frac{|{\bf q}|}{4 \pi}\ \sigma_{NN}(1 - i\alpha_{NN})
e^{ -(\beta_{NN}|{\bf p}|)^2 }\ ,
\label{para:ampl}
\eeq
where $\sigma_{NN}$ and $\alpha_{NN}$ denote the total cross section and
the ratio between the real and the imaginary part of the amplitude, respectively,
while the slope parameter $\beta_{NN}$ is related to the range of the interaction. 

The quantity
\beq
P_{\bf q}(t) = \langle 0 | |U^{FSI}_{{\bf q}}(R;t) |^2 | 0 \rangle \ 
\eeq
measures the probability that the struck nucleon does not undergo rescattering 
processes during a time $t$ after the electromagnetic interaction 
\cite{Rohe05}. In absence of FSI, 
i.e. for vanishing $\Gamma_{{\bf q}}$, $ P_{\bf q}(t) \equiv 1$.

Note that $P(t)$ is trivially related to the nuclear
transparency $T_{{\bf q}}$, measured in coincidence $(e,e^\prime p)$ 
experiments \cite{Garino92,O'Neill95,Abbott98,Garrow02,Rohe05}, through
\beq
T_{{\bf q}} = \lim_{t \rightarrow \infty} P_{{\bf q}}(t) .
\eeq

It is very important to realize that, as shown by Eqs.~(\ref{eik:prop0}) and 
(\ref{eik:prop}), the probability that a 
rescattering process occurs is not simply dictated by 
the nuclear density distribution $\rho_A({\bf r}_j)$,  the probability 
of finding a spectator
at position ${\bf r}_j$. The rescattering probability depends upon the {\it joint} probability 
of finding the
 struck particle at position ${\bf r}_i$ {\it and} a spectator at position 
${\bf r}_j$, that can be written in the form
\beq
\rho^{(2)}({\bf r}_i,{\bf r}_j) = 
\rho_A({\bf r}_i)\rho_A({\bf r}_j) g({\bf r}_i,{\bf r}_j) \ .
\label{def:rho2}
\eeq

Due to the strongly repulsive nature of nuclear interactions at short range, 
$\rho^{(2)}({\bf r}_i,{\bf r}_j)$ is  dominated by NN correlations, 
 whose behavior is described by the correlation function $g({\bf r}_i,{\bf r}_j)$.
The results of numerical calculations carried out within NMBT yield $g({\bf r}_i,{\bf r}_j) \ll 1$
at $|{\bf r}_{ij}| < 1$ fm. This feature is illustrated in Fig.~\ref{corr:hole}, showing 
the average ${\overline \rho}^{(2)}({\bf r}_i,{\bf r}_j)$, defined as
\beq
{\overline \rho}^{(2)}({\bf r}_{ij}) = \frac{1}{A} \int d^3R_{ij}\ 
 \rho^{(2)}({\bf r}_i,{\bf r}_j) \ ,
\label{average:rho2}
\eeq
with ${\bf R}_{ij} = ({\bf r}_i+{\bf r}_j)/2$, evaluated for both NN and $pp$ pairs
with and without inclusion of dynamical correlation effects. 

\begin{figure}[ht]
\includegraphics[scale=0.46,clip]{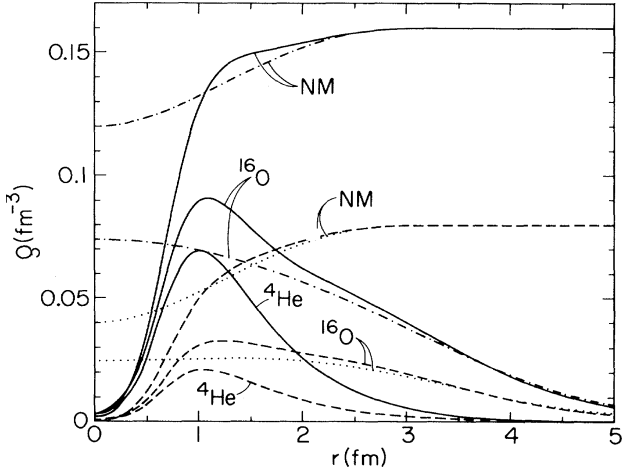}
\caption{Averaged distribution functions, defined by
Eq.~(\protect\ref{average:rho2}), of NN (solid lines) and $pp$
 (dashed lines) pairs in $^4\!He$, $^{16}\!O$ and uniform nuclear matter 
at equilibrium density (NM).
The corresponding quantities obtained neglecting correlation effects
are shown by the dot-dash and dotted lines, respectively \cite{Benhar93a}.}
\label{corr:hole}
\end{figure}

From Eq.~(\ref{eik:prop}) it follows that within the CGA the 
energy shift and the redistribution of the inclusive strength are driven by the real and
the imaginary part of the NN scattering amplitude, respectively. However, at large
${\bf q}$  the imaginary part of $\Gamma_{{\bf q}}$, which gives rise to the real part of
${\bar U}^{FSI}_{{\bf q}}$, is dominant.

Neglecting the contribution of the real part of $A_{\bf q}$ altogether, the CGA inclusive cross 
section can be written as a convolution integral, involving the cross section evaluated 
within the IA, {\em i.e.} using Eqs.~(\ref{eA:xsec}), (\ref{lepten}) and (\ref{hadrten2}), and 
a folding function embodying FSI effects:  
\beq
\frac{d\sigma}{d\Omega_{e^\prime} d\omega} = \int d\omega^\prime \
\left( \frac{d\sigma}{d\Omega_{e^\prime} d\omega^\prime} \right)_{IA} \
 f_{{\bf q}}(\omega - \omega^\prime)\ ,
\label{sigma:FSI}
\eeq
$f_{{\bf q}}(\omega)$ being defined as
\be
\nonumber
f_{{\bf q}}(\omega) & = & \delta(\omega) \sqrt{ T_{{\bf q}} } 
 + \int \frac{dt}{2 \pi}\ {\rm e}^{i \omega t}
\left[ {\bar U}^{FSI}_{{\bf q}}(t) - \sqrt{ T_{{\bf q}} } \right] \\
 & = & \delta(\omega) \sqrt{ T_{{\bf q}} } + F_{{\bf q}}(\omega) \ ,
\label{ff}
\ee
and normalized according to
\beq
\int_{-\infty}^{+\infty} d\omega f_{{\bf q}}(\omega) = 1 \ .
\eeq
The preceeding  equations show that the strength of FSI is governed by
both $T_{{\bf q}}$ and the width of $F_{{\bf q}}(\omega)$. In absence
of FSI $ {\bar U}^{FSI}_{{\bf q}}(t) \equiv 1$, implying in turn
$T_{{\bf q}}=1$ and $f_{{\bf q}}(\omega) \rightarrow \delta(\omega)$.

In principle, the real part of the NN scattering amplitude
can be explicitly included in Eq.~(\ref{eik:prop}) and treated on the same footing 
as the imaginary part. However, its effect turns out to be appreciable only at
$t \sim 0$, when the attenuation
produced by the imaginary part is weak. The results of numerical calculations
show that an approximate treatment, based on the use of a time independent
optical potential, is adequate to describe the energy shift produced
by the real part of $\Gamma_{{\bf q}}$ \cite{Benhar94b}, whose size of 
$\sim$10 MeV
is to be compared to a typical electron energy loss of a few hundred MeV.

The shape of the folding function is mainly determined by the total NN cross
section $\sigma_{NN}$. In the energy region relevant to scattering of few GeV
electrons, $\sigma_{NN}$ is dominated by the contribution of inelastic processes
and nearly independent of energy. As a consequence, the scattering amplitude 
of Eq.~(\ref{para:ampl}) grows linearly with ${\bf q}$ and both 
${\bar U}^{FSI}_{{\bf q}}(t)$ and the folding
function $f_{{\bf q}}$ become independent of ${\bf q}$.

\begin{figure}[ht]
\includegraphics[scale=0.5,clip]{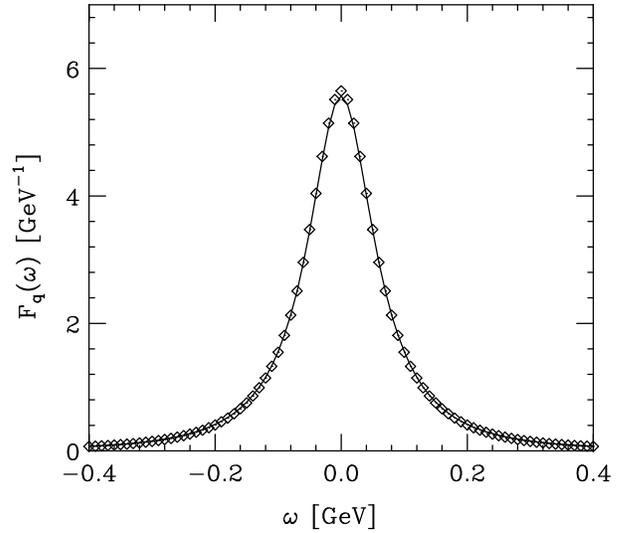}
\caption{
Folding functions $F_{{\bf q}}$ of Eq.~(\protect\ref{ff}), calculated in 
uniform nuclear matter at equilibrium density. 
The solid line and the diamonds correspond to momentum transfer $|\bf{q}|$ = 2.2 
and 3.4 GeV/c, respectively \protect\cite{Benhar99}.
}
\label{ff:shape}
\end{figure}

This feature is illustrated in Fig.~\ref{ff:shape}, showing that the $F_{{\bf q}}$
calculated in uniform nuclear matter at equilibrium density
at momentum transfers $|{\bf q}|$ = 2.2 and 3.4 GeV/c are nearly identical. 

Dynamical NN correlations also affect the shape of the folding function. Inclusion 
of correlations through the distribution function $g({\bf r}_i,{\bf r}_j)$ 
results in a strong quenching of the tails and an enhancement of the peak 
of $F_{{\bf q}}$, leading to a significant suppression of FSI effects. 

The effect of FSI is illustrated in Fig.~\ref{FSI:1}, showing the inclusive cross 
section of uniform nuclear matter at a beam energy $E_e = 3.595$ GeV and a scattering 
angle $\theta = 30^\circ$, corresponding to momentum transfer 
$|{\bf q}| \sim 2$ GeV/c.

Comparison between theory and the  data in Fig.~\ref{FSI:1} clearly shows that
at  $\omega < 1.1$ GeV, 
where quasielastic scattering dominates
\footnote 
{In the kinematics of Fig.~\ref{FSI:1}, inelastic processes only contribute 
$\sim$ 5\%  
of the inclusive cross section at $\omega = 1.1$ GeV, and become negligibly small at 
lower $\omega$.} 
and which correspond to $x > 1$,  $x = Q^2/2m\omega$ 
being the Bjorken scaling variable,  FSI effects are large and must be taken into account. 
The results obtained within the CGA are in good agreement with the data in the region 
$\omega > 800$ MeV, i.e. for $x \lsim 1.8$, while at higher $x$ the experimental 
cross section is largely overestimated. The dashed line has been obtained neglecting 
the effect of dynamical correlations on the distribution function 
$g({\bf r}_i,{\bf r}_j)$.
Comparison between the solid and dashed lines provides a measure of the quenching 
of FSI due to NN correlations.

\begin{figure}[htb]
\includegraphics[scale=0.5,clip]{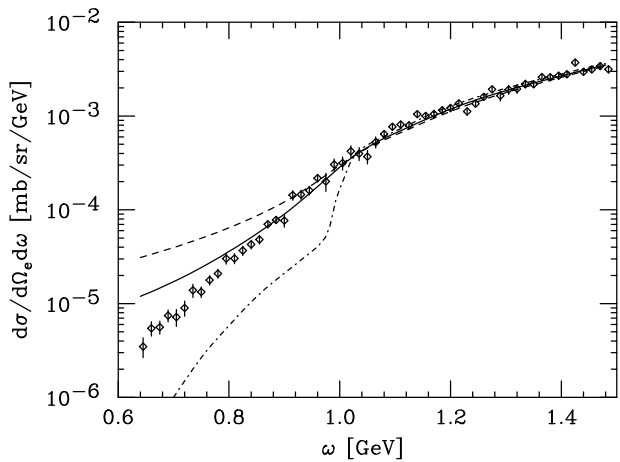}
\caption{
Inclusive electron scattering cross section at $E_e = 3.595$GeV and 
$\theta = 30^\circ$. The data points represent the extrapolated nuclear 
matter cross section \protect\cite{Day89}, while the solid and dashed lines show the 
results obtained in CGA including FSI effects, with and without taking into account 
correlation effects. For comparison, the IA cross section is also shown 
by the dot-dash line \protect\cite{Benhar91}.
}
\label{FSI:1}
\end{figure}
%

\begin{figure}[tbh]
\includegraphics[scale=0.5,clip]{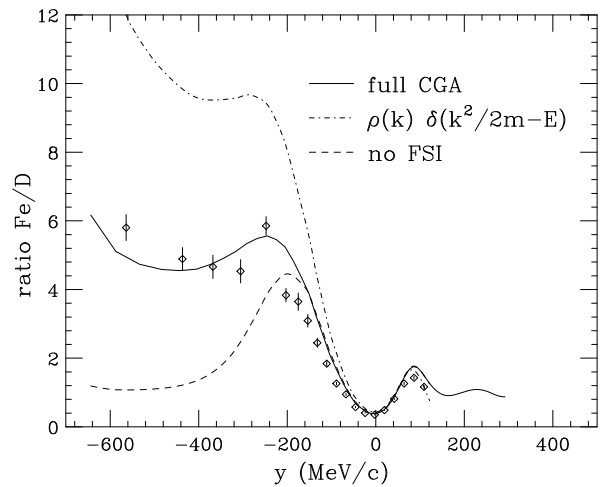}
\caption{
Ratios of inclusive cross sections of iron and deuterium at $E_e = 3.595$GeV and 
$\theta = 25^\circ$. Solid line: full calculation; dashed line: IA calculation, 
neglecting FSI in both iron and deuteron; dot-dash line: calculation carried 
out using the approximate spectral function of Eq.~(\ref{rinat1}) 
\cite{Benhar95e}.  }
\label{FSI:2}
\end{figure}

The ability of the CGA to provide a quantitative understanding of FSI in the region 
$x < 2$ is further illustrated in Fig.~\ref{FSI:2}, showing the cross section ratio
\beq
R = \frac{d\sigma(e + ^{56}\!Fe \rightarrow e^\prime + X)}
{d\sigma( e + ^2\!H \rightarrow e^\prime + X)} \frac{2}{56}  \ ,
\label{ratio}
\eeq
at $E_e = 3.595$GeV and $\theta = 25^\circ$. Note that $R$ of Eq.~(\ref{ratio})
 is only defined up to $y\sim$--700 MeV/c, corresponding to $x=2$, 
the kinematical limit for inclusive scattering  off an A=2 target (for the
definition of $y$ see Sec.~\ref{scale}). 

The solid line in Fig.~\ref{FSI:2} corresponds to the full CGA calculation, 
providing a good description of the experiments over the whole range of $y$, whereas 
the IA results, represented by the dashed line, lie well below the data at 
$y<-200$ MeV/c ($x> 1.5$). 
For comparison, Fig.~\ref{FSI:2} also shows
the results obtained using the approximate spectral function of Eq.~(\ref{rinat1}), which 
turn out to largely overestimate the data at negative $y$.

The results of Figs.~\ref{FSI:2} and \ref{meier1} clearly  rule out the 
interpretation of the 
behavior of the cross section ratio at $y \lsim -200$ MeV/c in terms of scattering
off strongly correlated nucleon pairs \cite{Egiyan03}. This interpretation in 
fact assumes the validity of the IA 
picture, which is known to fail at large negative $y$, and does not take into account 
the large effect of FSI.

Notwithstanding its success in describing the existing inclusive data at large
negative $y$, the CGA appears 
to consistently overestimate FSI effects at larger $-y$. As the validity of the eikonal 
approximation
is well established in the kinematical region apposite to scattering of few GeV electrons, 
possible corrections to the CGA scheme are likely to be ascribable either to modifications 
of the NN scattering amplitude or to the inadequacy of the approximations leading 
to the convolution expression for the cross section.

It has been pointed out  \cite{Benhar91} that the  use of the free-space 
amplitude  to describe
NN scattering in the nuclear medium may be questionable. Pauli blocking
and dispersive corrections are known to be important at moderate 
energies \cite{Pandharipande92}.
However, their effects on the calculated inclusive cross section have been found to 
be small in the kinematical region corresponding to $|{\bf q}| \gsim$ 2 GeV/c, and
decrease as $|{\bf q}|$ increases \cite{Benhar95f}. Corrections to the
amplitude associated with its extrapolation to off-shell energies are also
expected to be small at $|{\bf q}| >$ 2 GeV/c \cite{Benhar96b}.

Modifications of the free-space NN cross section may also originate from the
internal structure of the nucleon. It has been suggested \cite{Brodsky82,Mueller82} that
elastic scattering on a nucleon at high momentum transfer can only occur
if the nucleon is found in the Fock state having the fewest number of constituents,
so that the momentum transfer can be most effectively shared among them.
This state is very compact, its size being proportional to 1/$|{\bf q}|$, and therefore
interacts weakly with the nuclear medium. Within this picture a nucleon,
 after absorbing a large momentum, travels through nuclear matter experiencing
very little FSI, {\em i.e.} exhibits {\it color transparency} (CT), before it 
 evolves back to its standard configuration with a characteristic timescale.

 CT may be particularly relevant to the analysis of inclusive 
electron-nucleus scattering at $x >$ 1, where elastic scattering is the dominant reaction
mechanism, since it leads to a significant quenching of FSI.
In fact, the influence of CT is expected to be much larger for $(e,e^\prime)$ 
than for $(e,e^\prime p)$; in $(e,e^\prime)$ FSI occur mainly very close to the electromagnetic
vertex, at a distance of less than   $\sim 1/\magq$, where the compact configuration has 
not yet evolved back to the ordinary proton. It has been suggested that the 
modification of the NN scattering amplitude
due to onset of CT may explain the failure of CGA to reproduce the data in 
the region of very low $\omega$ \cite{Benhar91,Benhar93,Benhar94b}.

However, the recent measurements  show no enhancement of 
the nuclear transparency up to $Q^2 \sim 8$ (GeV/c)$^2$ \cite{Garrow02,Dutta03} and
seem to rule out observable CT effects in (e,e'p) at beam energies of few GeV.
Therefore, the excellent agreement between the measured inclusive cross sections
and the results of theoretical calculation \cite{Benhar91,Benhar93,Benhar94b} 
carried out using CGA and the quantum 
diffusion model of CT \cite{Farrar88} may be accidental.

An improved version of CGA has been recently proposed  
\cite{Petraki03}. Within this approach the initial momentum of the struck nucleon, 
which is averaged over in CGA, is explicitly taken into account. 
As a result, one goes beyond the simple convolution form of the 
inclusive cross section and gets a generalized folding function, 
depending on both the momentum transfer ${\bf q}$ and the initial momentum ${\bf k}$.
Numerical calculations
of the nonrelativistic response of uniform nuclear matter at $1 \lsim |{\bf q}| \lsim 2$ GeV/c
show that the inclusion of this additional momentum dependence leads to a sizable quenching 
of the low energy loss tail of the inclusive cross section, with respect to the predictions of 
CGA.

Even though they cover a limited kinematical range and have been obtained 
using a somewhat oversimplified model, the available results suggest that a better 
treatment of the momentum distribution of the struck nucleon may improve
the agreement between theory and data in the region of $x>2$, where CGA begins to 
fail. A systematic study of the dependence of the rescattering probability on the 
initial momentum, based on a relativistically consistent formalism, is 
presently being carried out \cite{Benhar06}.

A different approach to describe FSI in inclusive electron-nucleus scattering, based
on the relativistic generalization of the Gersch-Rodriguez-Smith (GRS) $ 1/|{\bf q|}$ expansion
of the response of many-fermion systems \cite{Gersch73a}, has been proposed
 \cite{Gurvitz02}. Numerical studies of the cross section of
$^4\!He$ \cite{Viviani03}, carried out using  realistic wave functions to compute 
the $\left< ^4 \!He|^3 \!He\right>$ and
$\left<^4 \!He|^3 \!H\right>$ transition matrix elements and a model spectral function 
\cite{Ciofi94}
to describe the many-body breakup channels, show that the generalized GRS expansion
provides a fairly good overall description of the SLAC data 
\cite{Day93} at beam energy of 3.6 GeV. 

\section{Experiments} \label{exp}                                                                                       
Studies of inclusive electron-nucleus scattering have been performed at a number                                        
of facilities. The bulk of the currently available data (at momentum transfers
$\magq$ on the order of 500 MeV/c and above upon which we  focus here) has been
generated at Saclay (France), Bates-MIT (Boston), Jefferson Lab (Newport News)
and SLAC (Stanford). Isolated data sets are                                             
available from other facilities (see next section).                                                                     
                                                                                                                        
Without exception, magnetic spectrometers have been used to momentum-analyze the                                          
scattered electrons. The magnet arrangements are varied, and at the four facilities
mentioned above  were D(ipole), DD, Q(uadrupole)QQD and QQDDQ systems,
respectively. These spectrometers had a maximum bending momentum of                                          
0.6, 0.9, 7 and 8 GeV/c, respectively and solid angles between 4 and 6.8 msr                                            
\cite{Leconte79,Bertozzi79,Yan97,Mo65}. As an example, we                                                               
show in Figs.~\ref{HMS},\ref{fpdet} the JLab High Momentum Spectrometer (HMS)
 of Hall C and its associated  focal-plane detector setup.                                                                                             
                                                                                                                        
\begin{figure}[hbt]                                                                                                     
\begin{center}                                                                                                          
\includegraphics[scale=0.65,clip]{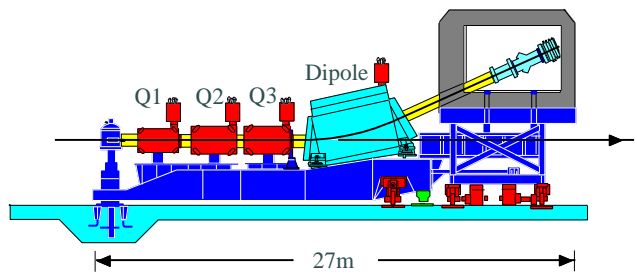}                                                                          
{\caption[]{HMS spectrometer in hall C at Jefferson Lab with 4 magnets arranged
as QQQD and the heavily shielded detector hut.                                                                  
}\label{HMS}}                                                                                                           
\end{center}                                                                                                            
\end{figure}                                                                                                            
                                                                                                                        
In most cases the electron track in the focal plane was reconstructed using                                               
2 to 4 planes of multiwire chambers, and the trigger involved the use of                                                
fast scintillator detectors, often 2 planes of segmented paddles. Typically                                           
the focal plane detector also included a Cerenkov detector needed to separate                                           
electrons from slower particles such as pions. At the high-energy                                               
facilities a  total absorption counter provided an additional, cruder,                                           
energy measurement to assist in the discrimination of electrons from other,
often more numerous, charged particles.                                                                                                     
                                                                                                                        
\begin{figure}[hbt]                                                                                                     
\begin{center}                                                                                                          
\includegraphics[scale=0.35,angle=-90,clip]{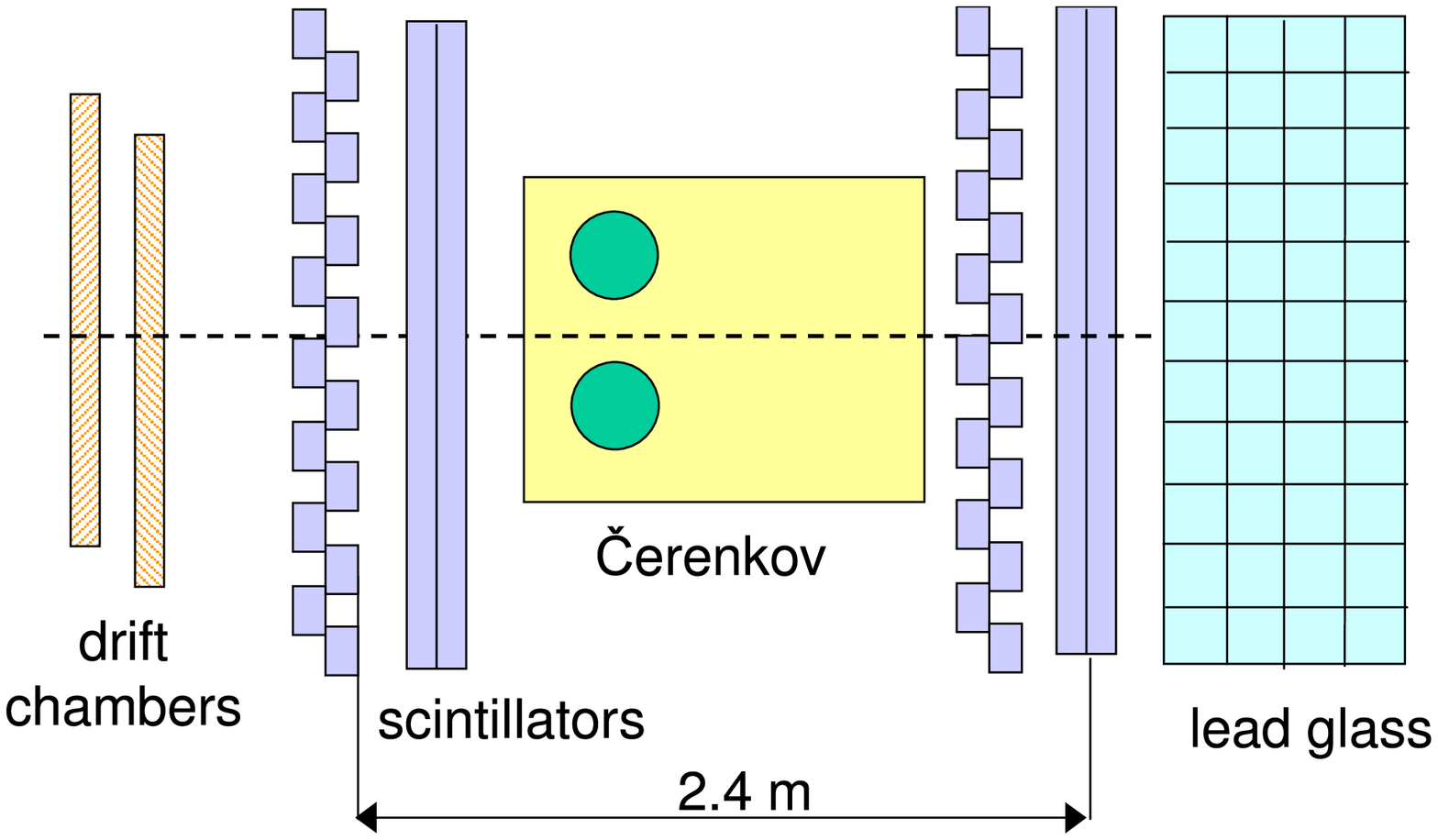}                                                                     
{\caption[]{HMS detector package, consisting of multiple planes of multiwire
drift chambers, fast scintillators, a sub-atmospheric gas Cerenkov counter  and 
 a lead glass total absorption counter.                                                                                               
\label{fpdet}}}                                                                                                         
\end{center}                                                                                                            
\end{figure}                                                                                                            
                                                                                                                        
The data acquisition systems of the earlier, lower-energy facilities were 
rather
restricted because of computer speed, memory and high data rates. Consequently
much of the data was pre-processed online and stored in spectra; this  allowed 
little  off-line analysis of the
data. The data acquisition at the higher energy facilities included event-mode 
acquisition                                        
where all the coordinates, times  and amplitudes are registered for further off-line                                    
analysis. Event-mode acquisition  has important                                                  
advantages in that during the course of the off-line analysis shortcomings of 
the  acquisition hardware can be discovered and (often) corrected. This typically leads to much                                        
more reliable cross sections.                                                                                           
                                                                                                                        
In many cases the product of effective solid angle and detector efficiency is                                                  
not easily established. While the efficiency of the detector system can be                                              
determined by exploiting the redundancy of the detector elements, the effective solid                                   
angle of spectrometers (not using a solid-angle defining slit of well-known                                              
geometry) is difficult to establish. For that reason many of the experiments, in particular                                              
the ones at the                                                                                                         
higher-energy facilities,  use elastic scattering from a liquid hydrogen                                       
target as check; the e-p cross sections are well known. The data on hydrogen                                            
often are also used  to determine offsets in the scattered electron momenta and 
angles as reconstructed from positions and angles of the electrons in the
detectors.
                                                                                                                        
                                                                                                                        
The inclusive quasielastic spectrum extends over a momentum range that exceeds
(in most cases) the momentum acceptance of the spectrometers. Consequently data
taking entails a sequence of runs at stepped values of the central spectrometer
momentum.  The resulting set of spectra, if taken with                                                
sufficient overlap, can then be used to determine both the dependence in $\delta p
\equiv \Delta p/p_{central}$ due to detector acceptance and inefficiencies 
and the true                                        
$\omega$-dependence of the \cs~\cite{Crannell66}.                                                                       
                                                                                                                        
An experimental aspect peculiar to inclusive scattering is the need to                                            
accurately know the dispersion of the spectrometer. As the cross section is                                           
ultimately determined for bins in energy loss, the accuracy of the differential                                              
momentum dispersion enters directly in the accuracy of the cross sections. The                                          
dispersion can be determined using focal-plane scans with an elastic peak,                                              
provided that the relation between the integral over $B dl$ and the quantity                                            
measured --- typically the $B$-field at one point --- is linear and has no
offset at zero momentum. For spectrometers with high resolution the dispersion
can be determined using elastic scattering off nuclei with different mass.  This                                                                                                            
determination of a precise dispersion often has not been given the appropriate                                          
attention.                                                                                                              
                                                                                                                        
After correcting the data for the various shortcomings of the spectrometers and                                         
detector packages (energy- and angle-offsets, detector inefficiencies, ...)  one                                        
major correction is the removal of radiative effects from external and                                           
internal Bremsstrahlung. While the theory for radiative effects is well under                                           
control \cite{Mo69},                                                                                                    
the specifics of \qes~ lead to the need for extensive modeling. In                                                  
\qes , the energy loss can be very large as  can be the energy loss due to                                          
radiative effects. Since radiation can be emitted before or after the \qes~                                             
process, the calculation of radiative effects requires knowledge of the \qe~                                            
structure function for all incident electron beam energies down to                                                      
$E_e-\omega_{max}$, where $E_e$ is the incident energy and $\omega_{max}$ is                                            
the largest energy loss of interest. Collecting the required quasi-elastic data 
at the corresponding lower incident electron                                                 
energies is often too time consuming. As a consequence, only few such data or                                           
none at all are taken.  In order to obtain the \qe~ cross section at all                                                
energies extensive modeling of the data in the region covered by experiment is                                          
needed, such that a reliable extrapolation to the other regions is feasible.                                            
With the discovery of $y$-scaling (see Sec.\ref{scale}) this modeling has                                              
been greatly facilitated,                                                                                               
as the scaling function $F(y)$  represents a quantity that is only very weakly                                          
dependent on energy and angle and can be interpolated over a large range. This                                          
unfortunately does not apply to the $\Delta$-region, which at large $\omega$ overlaps                                          
with the \qep . Modeling of the $\Delta$ contribution can  be facilitated                                               
when using the approach based on superscaling \cite{Amaro05}.                                                                          
                                                                                                                        
Despite the difficulties introduced by the radiative corrections, the impact on                                         
the overall accuracy is relatively small. For cross sections that do not depend                                               
steeply on $\omega$ the effect of the radiative corrections is small;                                          
the uncertainty introduced is also small if the model cross section is not
grossly deficient.                                               
                                                                                                                        
There is another correction --- not always considered --- that is unique to cross
section measurements of continuum spectra at large energy losses. One must
account for the  contributions of  electrons with  momenta outside the                                                                                
momentum acceptance of the spectrometer which interact with   part of the                                             
spectrometer  yoke or vacuum chamber, and are rescattered into the detector.                                            
Usually the trajectory of these electrons cannot be traced back through the                                              
spectrometer as their momentum is                                                                                       
not known. This process can lead to background, which in general is in the                                              
\%-region, but can become problematic when {\em e.g.} an L/T-separation is                                              
performed. In this case the effect  is greatly enhanced in magnitude  if the L- or                                            
T-contribution is                                                                                                       
small as compared to the other one, as happens at the larger momentum transfers                                         
where the longitudinal response gives a small contribution to the cross section.                                        
 Such rescattering events can become                                                                                    
especially important if the experiment uses long (liquid or gas) targets that                                           
extend beyond the acceptance of the spectrometer; in this case electrons can                                            
enter the spectrometer through the solid-angle defining collimator                                                      
 and scatter from the vacuum chamber or pole pieces.                                                                    
                                                                                                                        
Experimentally, provisions against scattering from the yoke (but not the pole                                           
pieces) include the use                                                                                                 
of baffles on the high/low momentum wall of the vacuum chamber \cite{Leconte79}.                                        
In some cases these backgrounds also have been subtracted by Monte-Carlo                                                 
simulations of the paths of electrons through the spectrometer \cite{Danel90}.                                          
They have been measured using a solid target displaced along                                          
the beam direction, and corrections up to 10\% have been found \cite{Danel90}.                                          
 The most reliable method to remove these contributions has been accomplished
via a secondary measurement of the energy of the electron in the                                           
focal plane, using some total-absorption counter \cite{Jourdan96a}. Once one                                            
knows with reasonable                                                                                                   
accuracy the energy of the detected particles, they can be traced back  and one                                         
can find out whether they  come directly from the target or not. For the                                             
QQDDQ 8GeV spectrometer at SLAC an acceptance-reducing collimator                                                
placed in between the two dipoles  allowed an experimental determination of
the fractional contribution of rescattered events. At large energy loss this
fraction has been found to be quite substantial \cite{Jourdan96a}.                                                                                    
                                                                                                                        
On the whole, this correction for rescattering has been given too little                                                
attention, particularly at the lower energies where their contribution is more                                          
important and where secondary energy-measurements using total-absorption                                                
counters are not very effective.                                                                                        
                                                                                                                        
One further correction that has to be applied to the data is the subtraction                                          
of electrons resulting from $e^+e^-$ pair production in the target. The                                                 
 $e^-$ from this charge-symmetric process give a  significant contribution to 
the inclusive cross                                           
section, particularly at large scattering angles and small energy loss $\omega$ 
where the  \qe~  \cs~ is very small. The effect of this contribution is usually                                             
handled by subtracting the spectrum, measured under identical kinematic
conditions, of the pair produced  $e^+$.                                                                        
                                                                                                                        
\section{Data} 
The modern era of experiments dedicated to quasielastic electron-nucleus 
scattering began in 1969 with a series of data sets from Kharkov
\cite{Dementii69,Titov69,Titov71,Titov72,Titov74}; 
 a first systematic survey of quasielastic electron scattering from a range 
of nuclei ($He$ to $Pb$) \cite{Moniz71,Whitney74,McCarthy76} was 
made at Stanford.    The first attempt to separate the response functions  
 was performed at Stanford when data for $^{40}Ca$ and $^{48}Ca$  were 
collected at constant values of the three-momentum transfer 
\cite{Zimmerman69,Zimmerman76}.

At about the same time quasielastic electron scattering from $^{12}C$ was 
being studied at the Harvard CEA at energies ranging from 1 to 4 GeV and angles 
from 8.5$^\circ$ to 18$^\circ$  using a half-quadrupole spectrometer 
\cite{Stanfield71}. At DESY inclusive electron scattering data in the 
quasielastic region was studied  
\cite{Heimlich74,Zeller73,Heimlich73} from $^6\!Li$ and $^{12}C$  
at energies up to 2.7~GeV and angles as large as 15 degrees.

The high energy beam at SLAC was exploited in the mid-1970's to measure 
quasielastic cross sections at very  high momentum transfer  
($\simeq 4$ (GeV/c)$^2$) on $^3\!He$ \cite{Day79}. A considerable body of 
quasielastic data was  also measured at SLAC  in the mid-1980's 
\cite{Day87,Day93,Potterveld89,Baran88,Baran89,Chen90,Meziani92,Sealock89} 
using the high-intensity, low energy ($\leq$5GeV) beam from the NPAS injector.

Starting in the late 1970's  a series experiments began 
 to measure data for a wide range of nuclei ($^2\!H$ to $U$), often as part of 
a program to separate the structure  functions, both at Bates   
\cite{Altemus80,Deady83,Deady86,Dytman88,Connell87,Zimmerman78,Blatchley86,Reden90,Quinn88,Dow88,Dow87,Hotta84,Yates93} 
and  at Saclay
\cite{Mougey78,Marchand85,Zghiche94,Meziani84,Meziani85,Barreau83,Gueye99}

In 1996 an experiment at JLab, intended to extend the range in four-momentum transfer 
$Q^2$ and Bjorken scaling variable $x$, 
produced data in the quasielastic region out to 
5 (GeV/c)$^2$ \cite{Arrington99,Arrington98}.  In the last year a new experiment 
using the full energy of Jlab (6 GeV)   extended these measurements even 
further using $^2\!H$, $^3\!He$, $^4\!He$, $Be$,  $C$, $Cu$ and $Au$ targets.

As most of the data is not published in numerical form, we have prepared, as 
part of this review, a website {\em Quasielastic Electron Nucleus Scattering 
Archive} \url{http://faculty.virginia.edu/qes-archive}. In this archive we have 
placed all radiatively unfolded 
cross section we could locate in tabular form. The site also gives the 
references and some details on the experiments.   

\begin{figure}[htb]
\begin{center}
\includegraphics[scale=0.49,clip]{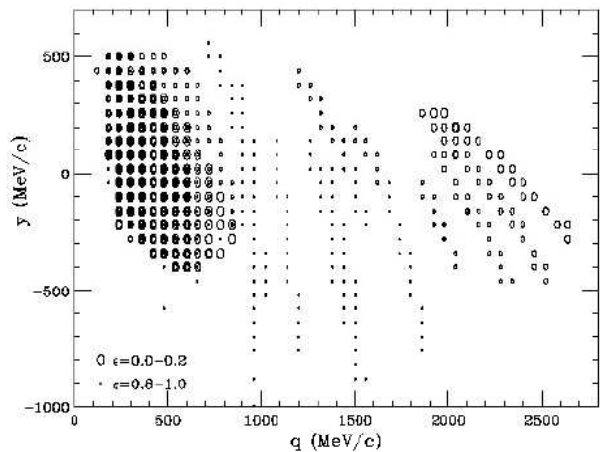}
{\caption[]{Kinematic range covered by the (e,e') data for $^{12}C$, plotted as
function of $\magq$ and $y$ (for definition see Sec.~\ref{scale}). The symbols
encode the range of $\epsilon$ covered by the data (bins of 0.2 in $\epsilon$).
 }\label{overqy}} 
\end{center} 
\end{figure}

For some nuclei, quite a comprehensive set of cross sections is available.
Fig.~\ref{overqy} shows the range in $\magq$ and $y$ covered by the data for
$^{12}C$. The symbols incorporate information on
 $\epsilon = 1./(1.+2 {\bf q}^2 / Q^2 ~tan^2 \theta/2)$, indicating that
only for the lower momentum transfers a large enough $\epsilon$-range is covered
for reliable L/T-separations.

\section{Extraction and use of nucleon form factors}

We began our discussion of the quasielastic scattering process by describing it 
as the incoherent scattering of the high energy electrons from the bound and 
quasi-free nucleons. Any quantitative description of this process then depends 
on an accurate description of the elementary electron-nucleon (proton and 
neutron) cross section.

In this section we will briefly review the status of the form factors, with 
particular attention to the most recent developments. A comprehensive review is 
not called for and the interested reader should consult 
Refs.~\cite{Gao:2003ag,Hyde-Wright:2004gh}
 or the recent 
conference proceedings \cite{Day:2005rs,deJager:2004br}.

 The form factor data, until recently, have been obtained, in the case of 
the proton, through a Rosenbluth separation which exploits of the linear 
dependence of the cross section on the polarization $\epsilon$ of the virtual 
photon. 
In terms of exchange of a single virtual photon the cross section depends on the
Sachs form factors  $G_E$ and $G_M$ via
%
%
%
%
%
\begin{equation}
\frac{d \sigma}{d \Omega} = \sigma_{\rm{NS}}
\left[\frac{G_E^2 + \tau G_M^2}{1+ \tau} +  2\tau G_M^2 \tan^2(\theta/2)\right],
\end{equation} 
with  $\tau=Q^2/4M^2$ and $\sigma_{\rm{NS}}= 
\sigma _{\rm{Mott}}E'/E_{0}$. Rearranging this expression with 
 $\epsilon^{-1} = 
1 + 2(1+\tau)\tan^2(\theta/2)$  yields
\begin{equation} \sigma_R \equiv \frac{d \sigma}{d \Omega} 
\frac{\epsilon(1+\tau)}{\sigma_{\rm{NS}}} = \tau G^2_M(Q^2)+\epsilon G_E^2(Q^2).
\label{rosen1}
\end{equation} 
By making measurements at a fixed $Q^2$ and variable $\epsilon(\theta, E_0)$,
the reduced cross section $\sigma_{\rm{R}}$ can be fit with a straight line with 
slope $G_E^2$ and intercept $\tau G^2_M$. The Rosenbluth technique, because of 
the dominance of \gmp over \gep at large $Q^2$, demands strict control over the 
kinematics, acceptances and radiative corrections over a large range of both 
incident and final electron energies and angles. 

The analysis of quasielastic scattering has utilized  widely accepted 
parameterizations or  models of the nucleon form factors, based on data 
collected, starting in the 1950's, through the 1990's. The data has been 
described by simple fits as well as sophisticated models based on the Vector
Dominance Model VDM, 
dispersion relations and quark models, 
see \cite{Gao:2003ag,Hyde-Wright:2004gh,Perdrisat:2006hj} for a review. 

The lack of a free neutron target with which to study the neutron properties 
forces the experimental effort to turn to investigate the neutron bound in a 
nucleus. The deuteron, weakly bound and with a ground state that, in  principle, 
is calculable, has been for unpolarized studies the nucleus of choice. Until 
the early 1990's the extraction of $G_E^n$ was done most successfully through
either small angle elastic electron scattering from the
deuteron
\cite{PhysRevLett.9.521,PhysRevLett.13.353,PhysRevLett.25.1774,Galster:1971fk,Simon:lr,Platchkov:1989ch} 
or by quasielastic $e$-D scattering
\cite{PhysRev.139.B458,PhysRev.146.973,Bartel:1973qy,Lung:1992bu}.

In  the  IA the elastic electron-deuteron cross section 
is the sum of proton
and neutron responses with deuteron wave function weighting.
The coherent nature of elastic scattering
gives rise to an interference term between the neutron and proton response 
which
allows the smaller $G_E^n$ contribution to be extracted. Still,  the large 
proton
contribution must be removed. 
 Experiments have been able to achieve small
statistical errors but remain very sensitive to deuteron wave function model 
leaving a significant residual dependence on the NN potential.

Quasielastic $e$-D scattering  provides a complementary approach to the
extraction of $G_E^n$. In the IA model  pioneered by  Durand and McGee
\cite{PhysRevLett.6.631,PhysRev.158.1500} the cross section is incoherent sum 
of p and n cross section
The extraction of $G_E^n$ requires both a Rosenbluth separation and the 
subtraction
of the sizeable proton contribution. It suffers, unfortunately, from unfavorable
error propagation and a sensitivity to the deuteron structure.

$G_M^n$ has been extracted from both inclusive and exclusive quasielastic 
scattering from the deuteron.  $G_M^n$ when determined from the ratio, 
$^2\textrm{H}(e,e' p)/^2\textrm{H}(e,e'n)$, always working near 
the top of the quasielastic peak, has the smallest uncertainties arising 
from  FSI, MEC and details of the ground state wavefunction. For both the 
ratio technique and $^2\textrm{H}(e,e'n)p$ the absolute efficiency of the 
neutron detector must be determined. The most precise data are from MAMI 
\cite{Kubon02}, recently $G_M^n$ was measured out to 5 (GeV/c)$^2$ 
\cite{Lachniet:2005qr,Brooks:2005gt} at JLab.


It has been known for many years that  the ''small'' nucleon electromagnetic form 
factors, $G_E^n$ and $G_E^p$ at large  $Q^2$ 
could be measured through   spin-dependent elastic scattering from the nucleon, 
accomplished either through a measurement of the
scattering asymmetry of polarized electrons from a polarized nucleon 
target~\cite{Dombey:1969wk,Donnelly:1985ry,Raskin:1988kc}
 or equivalently by measuring the polarization transferred to the 
nucleon~\cite{Akhiezer:1974em,Arnold:1980zj}.

	In the scattering of polarized electrons from a polarized target, 
an asymmetry appears in the elastic scattering cross section when the beam 
helicity is  reversed
	  due to the presence of a polarized piece, $\Delta$, in addition to 
the unpolarized piece, $\Sigma$. The total cross section is 
\begin{equation}
\sigma(h)  = \Sigma + h\Delta;\hspace{0.5cm}
 h = \pm 
p_{beam}.
\end{equation}
The asymmetry 
\begin{equation}
A ={\sigma_+ - \sigma_-\over{\sigma_+ + \sigma_-}} = 
{\Delta\over\Sigma}.
\end{equation}
can be written schematically ($a$, $b$, $c$, and $d$ are known kinematic factors)
 as    
\begin{equation}
A=\frac{ a  \cos\Theta^{\star} \left(G_{M}\right)^2  + 
          b  \sin\Theta^{\star} \cos\Phi^{\star} G_EG_M}
        {c \left(G_{M}\right)^2 + d \left(G_{E}\right)^2}
\end{equation} where $\Theta^{\star}$ and $\Phi^{\star}$ fix the target 
polarization axis.
With the target polarization axis in the scattering plane and perpendicular 
to $\vec{q}$, ($\Theta^{\star}, \Phi^{\star}=90^{\circ},0^{\circ}$) the 
asymmetry $A_{TL}$ is proportional to $G_EG_M$. With the polarization axis 
in the scattering plane and parallel to 
{\bf q} ($\Theta^{\star}, \Phi^{\star}=0^{\circ},0^{\circ}$), one measures 
the transverse asymmetry $A_T$, which in the case of a free nucleon would be 
insensitive to \gmn (depend simply on the kinematic factors). In \ha scattering 
the denominator contains contributions arising from both the protons and the 
neutrons and at modest $Q^2$ is sensitive to \gmn\cite{Gao:2003ag}.

In elastic scattering of polarized electrons from a nucleon, the nucleon 
obtains (is transferred) a polarization whose components, $P_l$ (along the 
direction of the nucleon momentum)  and $P_t$ (perpendicular to the nucleon 
momentum) are proportional to $G_M^2$ and $G_EG_M$ respectively.
 Polarimeters are sensitive only to the perpendicular polarization components 
so precession of the nucleon spin before the polarimeter in the magnetic field
 of the spectrometer (for the proton) or a dipole (inserted in the path of 
neutron)  allows a measurement of the ratio $P_t/P_l$ and the form factor ratio:
 $G_E/G_M = - \frac{P_t}{P_l}\frac{(E_0+E')}{2M_N}\tan(\theta/2).$ 
The recoil polarization technique has allowed precision measurements of  
$G_E^p$ to nearly 6 (GeV/c)$^2$ \cite{Jones:1999rz, Gayou:2001qt,Gayou:2001qd}. 

These data on the proton have generated a great deal of activity as they have
revealed a major discrepancy between the Rosenbluth data and the polarization 
transfer data for $G_E^p$ at $Q^2 > 2 GeV^2/c^2$. 
A detailed discussion of these topics  is beyond 
the scope of this review; it is discussed in the recent  
papers \cite{Gao:2003ag,Hyde-Wright:2004gh,Perdrisat:2006hj}
%


Extraction of the neutron form factors  using polarization observables is 
complicated by the need to account for Fermi motion, MEC, and FSI, 
complications that are absent when scattering from a proton target. 
Fortunately it has been found for the deuteron that  in kinematics that 
emphasize quasi-free neutron knockout both the  transfer polarization 
$P_t$~\cite{Arenhovel:1987zm} and the beam-target asymmetry 
$A_V^{eD}$~\cite{Arenhovel:1988qh} are especially
sensitive to $G_E^n$ and  relatively  insensitive to the NN
potential describing the ground state of the deuteron and other reaction 
details. See Figure~\ref{fig:aren1}. The utility of $^3$He as a polarized 
neutron arises from  the fact that in the ground state (dominantly a 
spatially symmetric S wave) the proton spins cancel and the $^3$He spin  
is carried by the unpaired neutron.  Calculations~
\cite{Golak:2000nt,Ishikawa:1998gb} of the beam-target asymmetry from a 
polarized $^3$He target have shown it to have only modest model dependence.

Extraction of the neutron form factors from beam-target asymmetry measurements 
in {\it inclusive} scattering are hindered by the dominant contribution of the 
proton, even with a polarized $^3$He where the protons are responsible for just 
10\% of the $^3$He polarization. This is especially true in the case of the 
neutron charge form factor  \cite{Jones:1993hg,Jones-Woodward:1991ih} 
though methods to minimize the role of the proton have been 
proposed \cite{CiofidegliAtti:1994cm}. $G_M^n$ has been extracted from 
inclusive polarized electron-polarized $^3$He scattering at Bates and Jefferson 
Lab out to $Q^2$=0.6(GeV/c)$^2$ \cite{Gao:1994ud,Xu:2000xw,Xu:2002xc}. Inclusive asymmetry 
measurements on $^3$He in the threshold region \cite{Xiong:2001vb} have been 
used to successfully test nonrelativistic Faddeev calculations which include 
MEC and FSI. 

Fortunately the development of polarized beams and targets have been able to 
leverage the utility of {\it exclusive} quasielastic scattering from bound 
neutrons in both $^3$He and $^2$H providing precision data on  the 
electric  form factors of the neutron.  Coincidence
measurements allow one to avoid completely the subtraction of the dominant 
proton.


Since the early work on the neutron at
Bates \cite{Jones:1993hg,Eden:1994ji} through  \ha and 
$D(\vec{e},e' \vec{n})p$ respectively, the further development of high 
polarization beams and targets,
together with high duty factor accelerators, has improved the data set 
(and outlook)  for $G_E^n$ 
\cite{Meyerhoff:1994ev,Herberg:1999ud,Passchier:1999cj,Ostrick:1999xa,Becker:1999tw,Rohe:1999sh,Zhu:2001md,Warren:2003ma} through either \( \vec{D}(\vec{e},e'n)p \), \han or  \recn and $G_M^n$ through \ha \cite{Gao:1994ud,Xu:2000xw,Xu:2002xc}.

\begin{figure}[htp]
\centerline{ 
\includegraphics[width=3.2in]{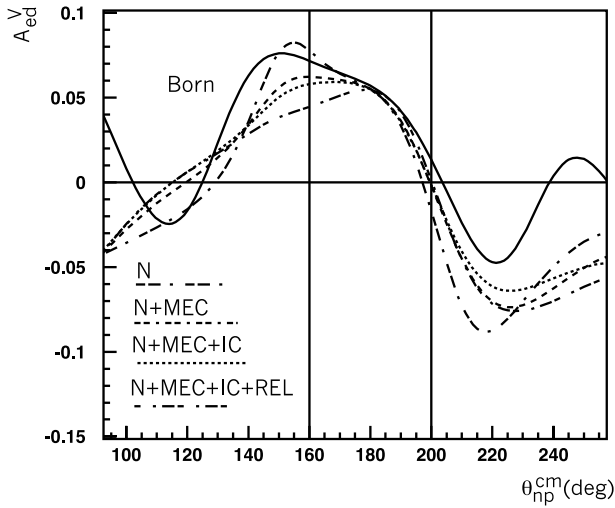}}
\caption{A calculation
\cite{Arenhoevel:1992xu, Arenhovel:1988qh,Arenhovel:1988qhe}
of the  electron-deuteron vector   asymmetry $A_{ed}^V$ at  $Q^2= 0.5$ 
(GeV/c)$^2$ which demonstrates
the insensitivity of $A_{ed}^V$ to the reaction model and subnuclear degrees 
of freedom at angles near 180 deg.
}\label{fig:aren1}
\end{figure}


In order to understand the discrepancy between polarization transfer and
Rosenbluth results for $G_E^p$ several investigators~
\cite{Blunden:2003sp,Rekalo:2003xa,Guichon:2003qm,Chen:2004tw} have explored 
the possibility of two-photon exchange corrections.  
While only incomplete calculations exist, the results of 
Ref.~\cite{Blunden:2003sp,Chen:2004tw} account for part of the difference.

 The most recent work by Chen {\it{et al.}}~\cite{Chen:2004tw}  
describes the process in terms of hard scattering from a quark and use 
Generalized Parton Distributions  to 
describe the quark emission and absorption. Chen {\em et al.} argue that when 
taking the 
recoil polarization form factors as input, the addition of the two-photon 
corrections reproduces the Rosenbluth data. However, 
Arrington~\cite{Arrington:2004ae} has shown that when the corrections of  
Chen {\it{et al.}} are applied to the new Jefferson Lab Rosenbluth data,  
which have small errors, (see below and Figure~\ref{fig:super2}) only 
one-half of the discrepancy is explained.



\begin{figure}[!ht]\centerline{\resizebox{0.45\textwidth}{!}{
\includegraphics{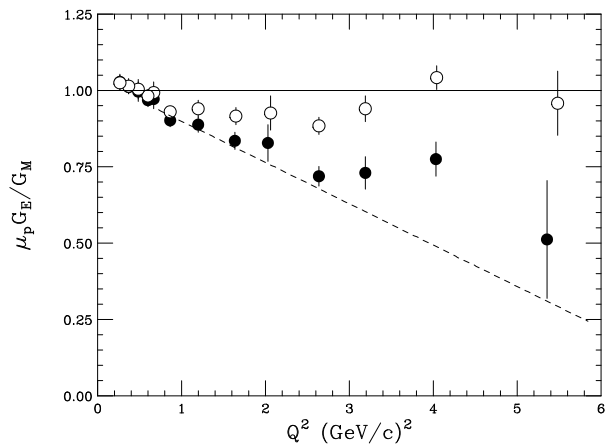}}}
\caption{ Result from the global analysis of
the L-T (Rosenbluth) data \cite{Arrington:2006zm}. Open (full) points are
obtained before (after) two-photon exchange corrections. The dashed line is a
fit to the polarization-transfer data.
}\label{fig:super2}
\end{figure}

Direct  tests for the existence of two-photon exchange include  measurements 
of the ratio $\sigma(e^+p)/\sigma(e^-p)$, where the real part of the 
two-photon exchange amplitude leads to an enhancement, and in  Rosenbluth 
data where it can lead to non-linearities in $\epsilon$. There is no 
experimental evidence of non-linearities in the Rosenbluth data and the 
$e^+/e^-$ ratio data~\cite{Mar:1968qd} are of only modest precision, making 
it difficult to absolutely confirm the presence of two-photon effects in these 
processes.

It is the imaginary part of the two-photon amplitude that can lead to single 
spin asymmetries but again the  existing 
data~\cite{Kirkman:1970er,Powell:1970qt} are of insufficient precision to 
allow one to make a statement.   There is, however, one observable that has 
provided unambiguous evidence for a two-photon effect in $ep$ elastic 
scattering. Groups in both the US~\cite{Wells:2000rx} and 
Europe~\cite{Maas:2004pd}
have measured the transverse polarized beam asymmetry. These measurements 
are significant but have limited utility in solving the $G^p_E$ discrepancy. 
The reader interested in more detail about the existence of two-photon effects 
and their role on the form factor measurements should refer to 
Ref.~\cite{Arrington:2004ae}.




For quasi-elastic electron-nucleus scattering, the focus of this review paper,
one therefore must ask the question: which form factors should one use in the
interpretation of the data? What one needs is a parameterization  of the
electron-nucleon cross section for the various kinematics, and this
parameterization is provided by the form factors $G_E$ and $G_M$ determined via
Rosenbluth separation of the e-N cross sections. Whether the $G's$ from
Rosenbluth separations or polarization transfer, with or without two-photon
corrections, are the {\em true} (one-photon exchange) form factors is largely 
irrelevant.

In any case, the discrepancy is largely confined to the proton 
electric form factor $G_E^p$. At large momentum transfers (above 1 (GeV/c)$^2$) 
the quasi-elastic cross section is weakly dependent on $G_E^p$. For example at 
forward angles at 1 (GeV/c)$^2$ the electric part of the the proton cross 
section is approximately 30\%, decreases to less than 20\% at 2 (GeV/c)$^2$ 
and to less than 10\% at 4 (GeV/c)$^2$. The electric proton contribution to 
quasielastic electron-nucleus scattering is almost a factor of two smaller 
due to the  magnetic contribution of the neutron cross section. 

We also note that another potential problem recently brought up
\cite{Bystritskiy07}, the difference between radiative corrections calculated by
different authors, hardly influences the interpretation of quasi-elastic
electron-nucleus scattering. For the time being, most of the data on both the 
nucleon
and nuclei have been radiatively corrected using the approach of Mo+Tsai
(including various improvements made over the years), and eventual shortcomings
drop out when using nucleon form factors in predicting quasi-elastic cross
sections.

\section{Scaling} \label{scale}
In general, the inclusive cross section is a function of two independent 
variables, the momentum transfer ${\bf q}$ and the energy transfer $\omega$ of the
electron. (The L- or T-nature of the scattering might be considered as a third
independent --- discrete --- variable). 
{\em Scaling} refers to the dependence of the cross section on a single  
 variable $y({\bf q},\omega)$, itself dependent on 
${\bf q}$ and $\omega$. This scaling property is basically a consequence of momentum-
and energy-conservation in the quasi-free scattering process \cite{West75,Sick80}. 

Inclusive scattering by a ''weakly" interacting probe such as the electron  
can often be interpreted in terms of IA (see Sect. \ref{sec:IA}). 
Quantitative derivations of scaling in IA have been given in several places, 
see {\em e.g.}  \cite{Day90}. Here, we first
look at a {\em qualitative} consideration, which however contains much of 
the basic physics.   Energy and momentum conservation for quasi-free scattering 
 off an initially bound  nucleon with momentum $\kv$  yields
\begin{equation}
\omega = [{(\kv+\qv)^2}+m^2]^{1/2} -m + E + E_{recoil} \ ,  
\end{equation}
with $E_{recoil}=\kv^2$/$(2m(A-1))$.
Splitting $\kv$ into its components $k_\parallel$ and
$k_\perp$ parallel and perpendicular to $\qv$ and assuming $\magq, \omega 
\rightarrow \infty$, such that the $k_\perp^2$ and recoil- and removal-energy 
 terms can be neglected, yields
$(\omega + m)^2 = k_\parallel^2 + 2 k_\parallel \magq + \magq^2 - m^2$. 
 This equation reveals that $k_\parallel = y(\qv,\omega)$, $\qv$ and $\omega$ no 
longer are independent variables. The same value of $y$ ($\equiv k_\parallel$)
can result from different combinations  of $\magq$ and $\omega$.
 The cross section 
$\sigma(\magq,\omega)$ divided by the electron-nucleon cross section 
$\sigma_{eN}(\magq,\omega)$ and a kinematic factor gives a function $F(y)$ that only depends on 
$y$. This function $F$ has an easy approximate interpretation: it represents the 
probability to find
in the nucleus a nucleon of momentum component $y$ parallel to
$\qv$.

The quantitative derivation of $y$-scaling is more involved  
\cite{Day90,Pace82}. In the limit of
very large momentum transfer one also finds scaling; the main {\em quantitative}
difference  concerns the restriction of the region ${\magk},E$  which  
contributes to $F(y)$; Fig.~\ref{fig2rev} shows an example. In the limit
of very large $\magq$, the scaling function is given by 
\beq
F(y)=2 \pi \int_y^\infty k~dk~ \tilde{n}(y;\magk )
\eeq with 
\beq
\tilde{n}(y;\magk) = \int_0^{\varepsilon} dE~S(\magk,E) \label{ntilde}.
\eeq
Allowing the upper integration limit  to $\infty$,
then $\tilde{n}$ would correspond to 
\beq
n(k)= \int_0^\infty dE~S(\magk,E)
\eeq and $F(y)$ would become the probability $n(k_\parallel)$ to find a nucleon
with momentum component $k_\parallel$ in the nucleus. This approximation is
quite reasonable; due to the rapid fall-off of $S({\kv},E)$ with increasing $\magk$ or
$E$, the integral is dominated by the region near $(\magk=y,~ E=0)$  

%
\begin{figure}[htb]
\begin{center}
\includegraphics[scale=0.48,clip]{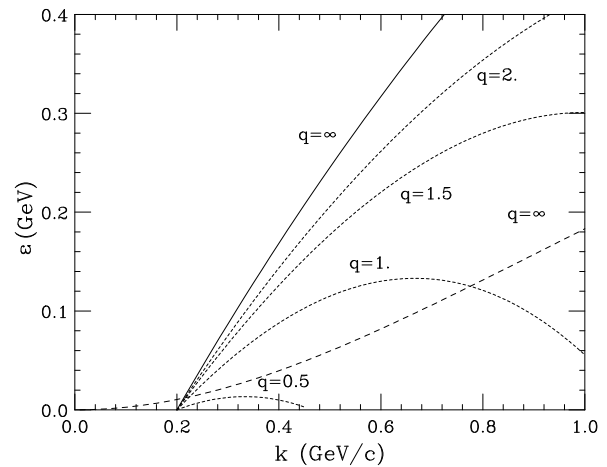}
{ \caption{ 
\label{fig2rev} Integration region of $S({\kv},E)$ in Eq.~\ref{ntilde} 
(below corresponding curves) 
contributing to $F(y)$  at $y$=--0.2GeV/c and given $\magq$ \protect{\cite{Day90}}.
The long-dashed line shows the location of the ''ridge'' of $S({\kv},E)$ where, at large
$\magk$, most of the strength is expected to occur.  } }  
\end{center}
\end{figure}

As an illustration  of scaling in quasi-elastic electron scattering, 
we show in Fig.~\ref{over1dhe3} 
some of the available inclusive  scattering data for
%
\begin{figure}[htb]
\begin{center}
\includegraphics[scale=0.48,clip]{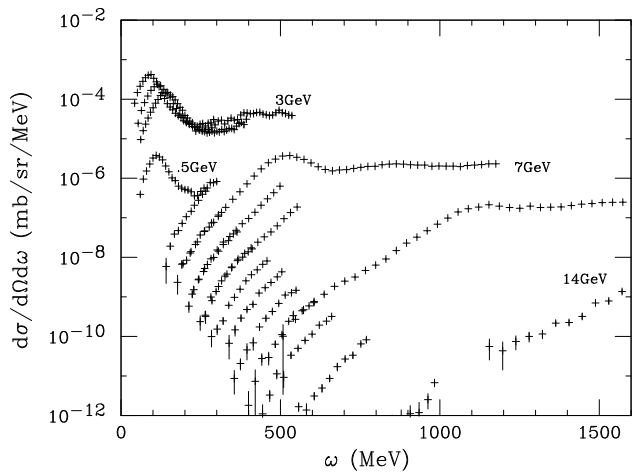}
{ \caption{ 
\label{over1dhe3} Cross sections for $^3\!He (e,e^\prime)$ as function of incident 
electron energy and energy loss, for different incident electron energies.
 } }  
\end{center}
\end{figure}
$^3\!He$. The data cover a large kinematic range  and 
extend over many orders of magnitude in cross section, and the
quasi-elastic peak shifts over a large range of $\omega$ with increasing $\magq$. 
The same data, now shown in terms of the scaling function and plotted
as a function of the scaling variable $y$ 
(Fig.~\ref{dfyhe31})
show an impressive scaling behavior for $y<0$, {\em i.e.} for the
low-$\omega$ side of the quasi-elastic peak. The cross section that differs by 
several orders of magnitude define the {\em same} function $F(y)$.
For \mbox{$y>0$} the values strongly diverge, primarily  due to the more involved
kinematics and the different $q$-dependence of the inelastic e-N cross section
($\Delta$-excitation in particular) which contributes to the inclusive cross
section at large $\omega$ and $q$. 
\begin{figure}[htb]
\begin{center}
\includegraphics[scale=0.48,clip]{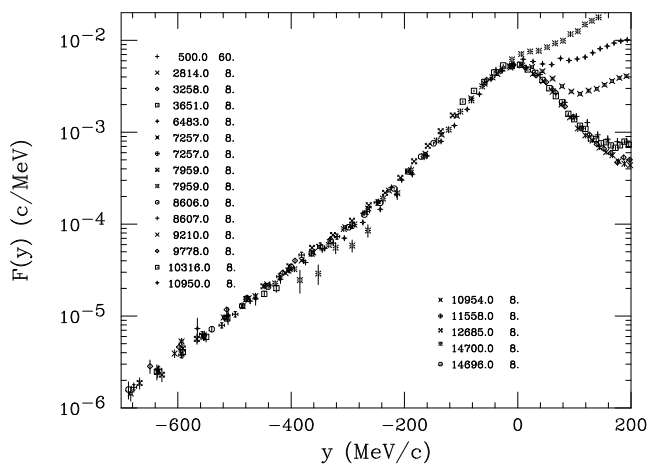}
{ \caption{ 
\label{dfyhe31} Scaling function for $^3\!He (e,e^\prime)$. The various data sets are
labeled by electron energy (MeV) and angle (deg). 
  } }  
\end{center}
\end{figure}

In the IA, because of the distribution of strength in $E$ of the spectral
function, we expect that the asymptotic limit of the scaling function would be
approached from below with increasing  $\magq$. 

This scaling property allows  the data to be exploited in several ways: 
The presence or absence of scaling tells us something about the reaction 
mechanism (we assumed IA without FSI to derive scaling), a residual 
$\magq$-dependence of $F(y)$ can tell us something about the $\magq$-dependence of the 
in-medium nucleon form factor and FSI, and the functional form of the
experimental scaling function provides some insight to the 
 nuclear spectral function. We  address  some of these points in more detail
below.

{\bf Reaction mechanism.} The data show a marked deviation from scaling 
behavior for $y>0$, indicating that in this region processes other than 
quasi-free scattering, such as MEC, pion production, $\Delta$-excitation and
DIS contribute.  This is confirmed by substituting MEC or $\Delta$ cross
sections for the electron-nucleon cross section in our development of $F(y)$. 
In this instance the data profoundly fails to scale for $y<0$ as well.
 We thus may conclude in particular that, at places where the data approximately
scale, the contributions of reaction mechanisms other than quasi-free 
scattering are smaller than the residual non-scaling of the data. Thus, 
scaling gives us direct information on the reaction mechanism, a question we
must understand before we can use the data for quantitative interpretation.

{\bf Bound-nucleon form factor.} The scaling function is secured by dividing the 
experimental cross sections by the electron-nucleon cross section. If the
$\magq$-dependence of the electron-nucleon cross section is not correct, the
data  will not scale, at least as long as the
range of $\magq$ covered is large enough to lead to a large variation of the 
elementary cross section. In order to exploit this idea 
quantitatively, 
one can compute the scaling function using a modified nucleon form factor
describing the assumed revision of the  bound-nucleon form factor, depending on
one parameter,
and then fit $F(y)$ using a flexible parameterization.
The minimum of  $\chi^2$ of this fit gives the best value for the parameter
modifying the nucleon form factor. 
%
%
It has been found for Iron, for example \cite{Sick86b}, that within the systematic
error of 3\% no change of the nucleon radius (more precisely  the size
parameter in the Dipole formula) can be supported by the data. As the cross
sections receive a $\sim$70\% contribution from  magnetic scattering, 
this result mainly applies to the magnetic radius; the limit on any 
medium modification  of the charge
radius is twice as large. Given the number of models that  predict a 
sizable influence of the nuclear medium on the nucleon form factors, the 
information provided by scaling behavior is quite constraining.

{\bf Constituent mass.} In the calculation of the scaling variable $y$, the mass
$m_c$ 
of the constituent the electron scatters from plays an important role, at least
as long as the recoiling constituent is not ultra-relativistic (recoil energy
$>$3 GeV). The dependence of the scaling function on $m_c$ can be exploited 
to learn about the
nature of the constituent. This is important for  small
$\omega$ and large $q$, where it has been suggested that scattering from 
quark-clusters or individual
quarks plays a role \cite{Kumano88,Pirner81}.  For the
kinematical region explored, the data are best explained by scattering from {\em
nucleons}.

It is only possible to identify the participating constituent (its mass and 
form factor) if the data covers  a large range of momentum transfer. If
only a limited range is considered, accidental compensations can occur which
then obscure an interpretation. For instance, the simultaneous observation of
$y$-scaling (nucleon mass and form factor) and $\xi$-scaling (pointlike,
massless constituents) was rather confusing until it could be shown to result
from an accidental cancellation of $q$-dependencies \cite{Day04,Benhar95c}.
    
{\bf Nucleon FSI.} The final state interaction of the knocked out nucleon 
in general is of minor importance in inclusive scattering; the electron 
carries information only about the FSI that takes place within a distance of
order $1/\magq$ from the scattering vertex. Subsequent interactions of the recoil
nucleon on its way out of the nucleus (which are much more important for 
{\em e.g.} $(e,e^\prime p)$) do not influence the scattered electron. 

At low energy loss FSI does, however, play a role, as discussed in
Sect.~\ref{sec:FSI}.  While the distribution of the spectral
function $S(\magk,E)$ in $E$ leads to a convergence of $F(y,\magq)$ from below with 
increasing $\magq$, the FSI leads to a convergence from {\em above}. 
Fig.~\ref{fy1}, which presents the scaling function $F(y,Q^2)$ for Iron for
fixed values of $y$, shows this convergence from above and that, with 
the large momentum transfers recently made
available at  Jefferson Lab \cite{Arrington99}, convergence of
$F(y,Q^2)$ can be demonstrated for values of $-y$ as large as  0.5 GeV/c.  
\begin{figure}[htb]
\begin{center}
\includegraphics[scale=0.45,clip]{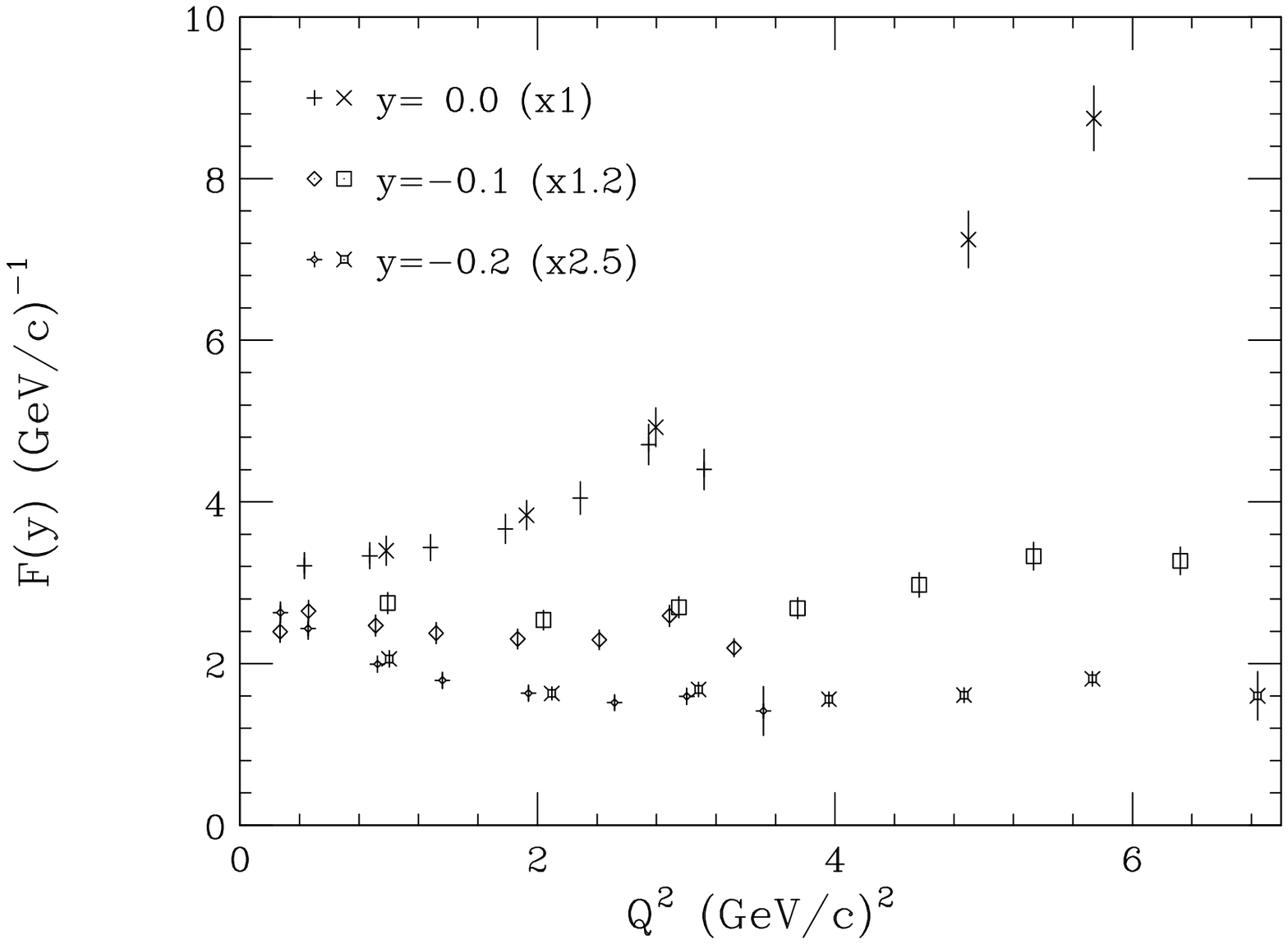}
\includegraphics[scale=0.45,clip]{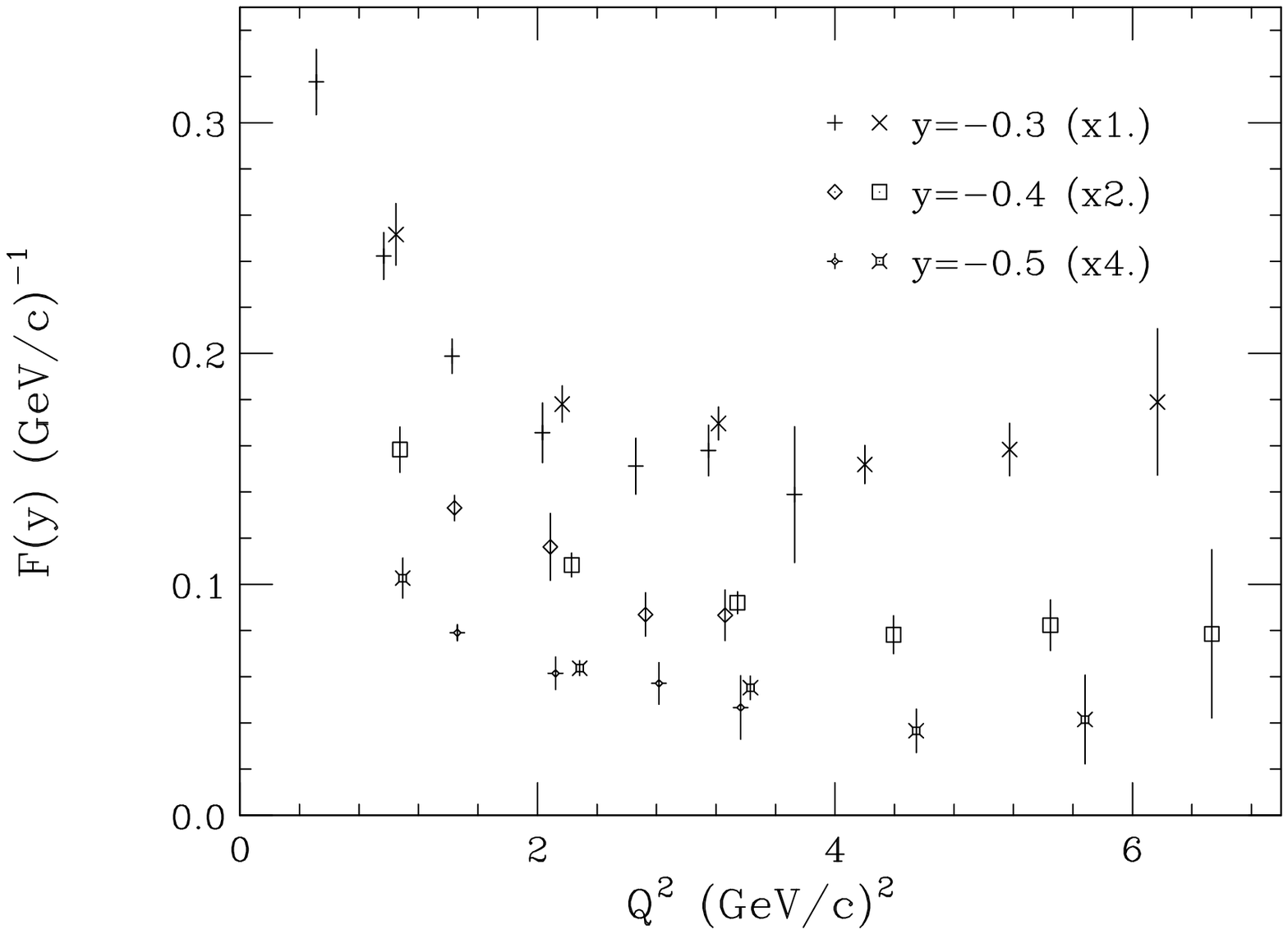}
{ \caption{ 
\label{fy1} Convergence of $F(y,Q^2)$ with increasing $Q^2$, for different 
values of $y$ \protect{\cite{Arrington99}}. 
 The rise of $F(y,Q^2)$ for $y=0$ and increasing $Q^2$ is due to
the $\Delta$-contribution.} }  
\end{center}
\end{figure}

Drawing conclusions on FSI starting from scaling should be done with caution.
 Scaling is usually
derived assuming the IA, so  the  experimental observation of scaling 
seemingly would suggest that FSI are unimportant. This, however, is not correct.
It has been shown \cite{Benhar99} how scaling in the presence of
FSI can come about: for \qes~ the width of the \qep~ becomes constant at large
$|{\bf q}|$ (due to relativistic kinematics). If at the same time the folding function
that accounts for FSI  becomes $q$-independent --- which is the case for NN
scattering where the total cross section is essentially independent of momentum
--- then the folded function also is independent of $|{\bf q}|$, and will scale.   The
same observation has been made for deep inelastic lepton-nucleon scattering
\cite{Paris02}. 

 It also has been
pointed out \cite{Weinstein82} that for a hard-core interaction
the scaling function is not directly related to the momentum distribution, due
to the effect of FSI.
However, for  less singular interactions, such as the Paris potential, it has
been shown \cite{Butler88}  via a calculation in  Brueckner-Goldstone theory
that the full response does converge to the IA result at large $|{\bf q}|$. 

{\bf Spectral function at large momentum.}  The  properties of $S({\kv},E)$ at large
$\magk$ are closely connected to  the behavior of the inclusive cross section 
at large $\magq$ (several GeV/c) and
comparatively low $\omega$ (several hundred MeV). This is qualitatively obvious 
when considering the limit of $\omega$=0. It is only possible (working in the
PWIA) to transfer a large momentum $\magq$ to the nucleon with the result that the
nucleon in the final state have both small energy, hence small momentum
$\bf{k+q}$, if the
initial nucleon had momentum $\bf{k} \sim -\bf{q}$ before the scattering. 

 The region of the low-$\omega$ tail of the quasi-elastic 
response is best studied by considering the ratio of the nuclear and deuteron
response (the latter being well known experimentally and accurately calculable
for any NN potential).
\begin{figure}[bht]
\begin{center}
\includegraphics[scale=0.45,clip]{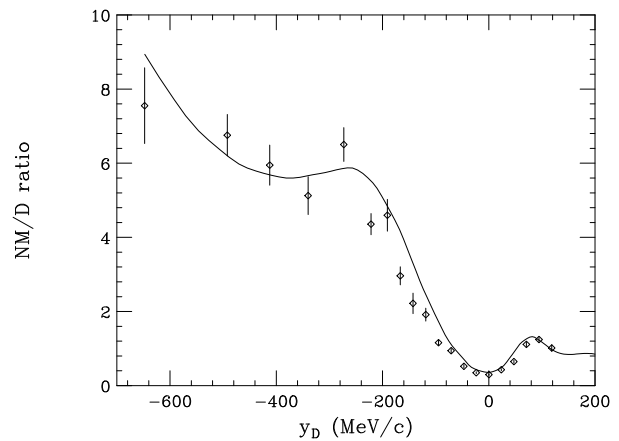}
\vspace*{-0.2cm}
{\caption[]{\label{rnm3630y}
Per nucleon 
cross section ratio of \nm\ and the deuteron taken at  3.6GeV and 30$^\circ$, 
as a function of the scaling
variable $y$. The result of the CBF calculation is shown as a solid line
\protect{\cite{Benhar94b}}.
 } } \end{center} 
\end{figure}
Figure \ref{rnm3630y} shows the ratio for one of the kinematics at large 
$\magq$, small $\omega$ where data are available. Data and theory are plotted 
as a function of the variable $y_D$, which is basically the component of the 
nucleon momentum $\kv$ parallel to $\qv$, calculated for the deuteron. 
The dip at $y=0$ results from the fact that the per-nucleon momentum space 
density at low momenta is higher for the deuteron (see Fig.~\ref{nm:Pke}),  the 
''plateau'' in the region of 
$y<-300$ MeV/c results from the fact that the deuteron and nuclear momentum
distributions have a similar fall-off at large momenta and essentially differ by
an overall factor.

Figure \ref{rnm3630y} shows that the particular \nm\ spectral function used 
in this calculation \cite{Benhar89} agrees well 
with the data.  Tests carried out by renormalizing the spectral function
 at $\magk > k_F$ have shown that the cross section ratio 
at $y<-$300 MeV/c is essentially proportional to $S(\magk,E)$ at $\magk>k_F$. One needs
to realize, however, that for a  quantitative study of $n(\magk)$  it is important 
to include FSI, as has been done in Fig.~\ref{rnm3630y}, although it was
suggested 
\cite{Frankfurt93} that the FSI could cancel in this ratio (see also 
Fig.~\ref{FSI:2}). Also, when dealing with
nuclei with A$>$2 it is important to use a spectral function $S(k,E)$ rather than
just a momentum distribution (which ignores the $E$-dependence of $S$), see
Fig.~\ref{FSI:2}.   

In terms of the scaling variable $x$ the inclusive cross sections now reach up
to $x \sim 3$. It is of course tempting to interpret the strength near $x=2$
($x=3$) as originating from scattering off 2 (3)-correlated nucleon systems
having 2$m$ (3$m$) mass. This type of interpretation \cite{Egiyan03}, however,
ignores the fact that the data exhibit clear $y$-scaling, hereby proving that
the electron scatters from constituents with nucleonic mass and nucleonic
form factor.
The interpretation of cross section ratios between nuclei as ratios of correlated
strength is also hindered by the fact that $x$, unlike $y$, is not simply
related to the momentum carried by the struck nucleon \cite{Liuti93}. In 
addition, the strength at very negative $y$ (large $x$) is strongly affected 
by A-dependent FSI, see Fig.~\ref{FSI:2}. 
 
{\bf Superscaling. } Recently,  the scaling idea has been pushed one step
further \cite{Donnelly99a}.
 Motivated by the Fermi-gas model, in which all momentum distributions
only differ by an overall scale factor --- the Fermi momentum --- Donnelly and
Sick have investigated whether the scaling functions of different
nuclei also can be related to each other, by adjusting one overall scale factor.
It turns out that this is possible for nuclei with $A \geq 12$; for lighter
nuclei deviations near the top of the quasi-elastic peak are visible. 
Fig.~\ref{dfy} shows an example the scaling function $f(\psi ')$ plotted as a
function of $\psi '$, which corresponds to the variable $y$ scaled by a ''Fermi
momentum''. 

\begin{figure}[bht]
\begin{center}
\includegraphics[scale=0.49,clip]{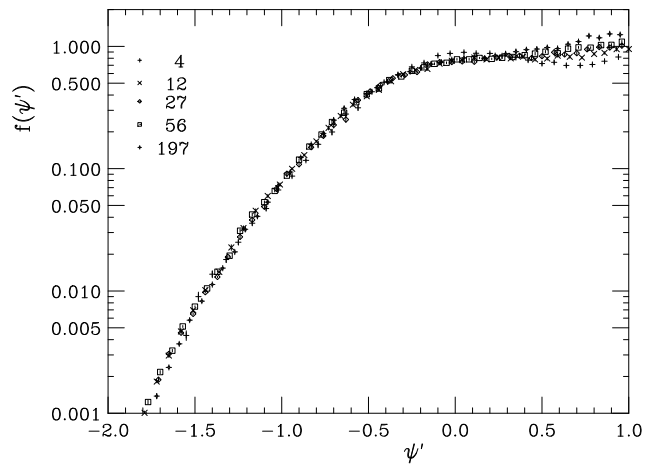}
\vspace*{-0.2cm}
{\caption[]{\label{dfy}
Scaling function for nuclei A = 4 -- 197 and fixed kinematics ($\magq \sim $1GeV/c)
as function of $\psi '$, which corresponds to the variable $y$ scaled by a 
''Fermi momentum''. 
 } } \end{center} 
\end{figure}

This scaling as a function of nuclear mass number actually is better realized 
than the normal scaling
which is broken by the non-\qe~ contributions to the response; these non-\qe~
contributions are not too dissimilar for different nuclei at the same 
kinematics.  As the momentum
used to scale $y$ is a slow and smooth function of $A$, this superscaling
property is particularly useful to interpolate data on $F(y)$ in order to
predict $F(y)$  for nuclei not experimentally investigated.

For the {\em longitudinal} response, superscaling is particularly well realized,
even at large energy loss where the transverse quasi-elastic response is
obscured by inelastic e-N scattering. Fig.~\ref{fpsip}
\begin{figure}[bht]
\begin{center}
\includegraphics[scale=0.5,clip]{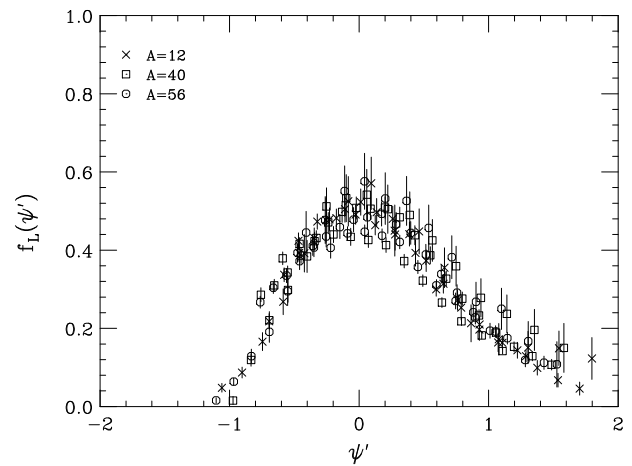}
\vspace*{-0.2cm}
{\caption[]{\label{fpsip}
Longitudinal superscaling function extracted  \cite{Donnelly99a} from the 
available data on $C$,
$Ca$ and $Fe$ for momentum transfers between 300 and 570MeV/c. 
 } } \end{center} 
\end{figure}
shows the longitudinal response function determined via superscaling from the
available separated data on $C$, $Ca$, $Fe$. This function, incidentally, also
displays a tail towards larger $\omega$ discussed in Sect.~\ref{sec:IA}.

This superscaling feature is particularly useful when one realizes that $(e,e^\prime)$
and $\nu$-induced quasi-elastic processes such as $(\nu,e)$ or $(\nu,\mu)$ 
differ only by the elementary vertex, the underlying nuclear physics being the
same. Cross sections for  $(e,e^\prime)$ together with superscaling then allow one to accurately
predict the cross section for neutrino-induced  reactions \cite{Amaro05}, 
currently an area of intense experimental activity. 

\section{Light nuclei} \label{light} 
Quasi-elastic scattering from light nuclei ($A \leq 4$) occupies a special
place. For these systems, several methods to calculate the inclusive response
have been used, often treating both the initial and the final state in  less 
approximate ways than imposed
by the complexities of the many-body system for heavier nuclei. For A=3, for
example, the first calculation of a realistic spectral function was
performed 30 years ago \cite{Dieperink76}. A variety of approaches to describe \qes~ 
from light nuclei is available; we address some of their results below.

 The longitudinal response for the A=3
nuclei has been calculated \cite{Efros04} using the Lorentz integral transform technique 
(see also
Sec.~\ref{eucl}). They use ground state wave functions expanded in terms of
correlated sums of hyper-spherical harmonics, calculated using different
NN forces ({\em e.g.} Argonne 18, Bonn-A) and three-body force (3BF) models. As
their calculation is essentially nonrelativistic, they restrict it to
$\magq <$500 MeV/c, and consider the longitudinal response to avoid the difficulty of
MEC. Fig.~\ref{orlan} shows  their results for $^3\!H$ and $^3\!He$ compared to
data.
\begin{figure}[hbt]
\begin{center}
\hspace*{-3mm}\includegraphics[scale=0.55,clip]{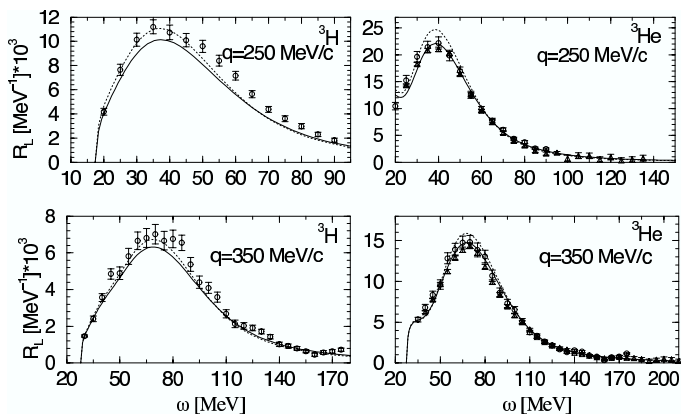}
{\caption[]{Comparison of experimental and theoretical response \cite{Efros04} 
for Argonne 18 potential (dotted) and Argonne V18+3BF (solid). Data are shown as 
open circles \cite{Dow88} and squares \cite{Marchand85}.  \label{orlan}}}
\end{center} 
\end{figure}

This calculation finds rather small differences between the responses calculated
using different modern NN forces. The three-body force leads to a systematic reduction
in the height of the \qep , presumably as a consequence of the tighter binding
which leads to a more compressed (extended) distribution in radial
(momentum)-space. While for $^3\!He$  ~the 3BF is helpful in explaining the
data, this does not seem to be the case for the \hyd~ nucleus.
 In a very recent paper \cite{Efros05}  the range of applicability of
this nonrelativistic calculation has been extended. 

The inclusive cross section has also been calculated  separately for
the 2-body and 3-body breakup \cite{Golak95}. For the ground state, 
Golak {\em et al.}  use a solution of the
34-channel Faddeev equation for the Bonn-B NN-potential. For the final state,
the authors separate the contribution from the (symmetrized) plane-wave approach
and the one from rescattering processes, summed to all orders. This calculation
again is non-relativistic, and does not include MEC. 

Fig.~\ref{golak} shows  the results  of Golak \et~  for $^3\!He$ and $^3\!H$, 
with the
contributions of 2- and 3-body breakup separately displayed. 
\begin{figure}[hbt]
\begin{center}
\includegraphics[scale=0.38,clip]{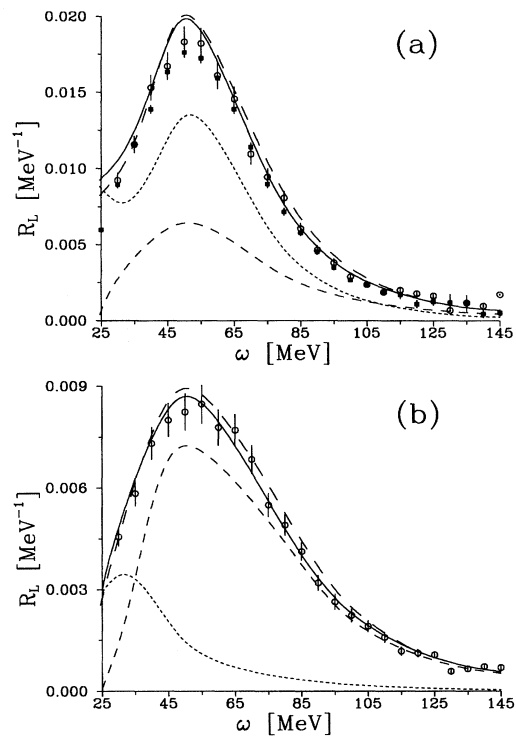}
{\caption[]{Comparison of theoretical \cite{Golak95} and experimental response 
at \magq=300MeV/c for $^3\!He$ (a) and $^3\!H$ (b).  Dotted: 2-body breakup, dashed: three-body 
breakup. Data are shown as open circles \cite{Dow88} and squares 
\cite{Marchand85}. \label{golak}}}
\end{center} 
\end{figure}

The inclusive cross section has also been calculated
using a realistic coupled-channel potential with single
$\Delta$-isobar excitation for the initial and final hadronic states, with the
corresponding e.m. current with two-baryon contributions \cite{Deltuva04}. 
The potential is
an extension of the  CD-Bonn NN-potential. The main $\Delta$-isobar effects on
observables then result from the effective three-nucleon force and the
corresponding effective two-and three-nucleon exchange currents. Both initial
bound state and final continuum state are exact solutions of the 3-particle
scattering equations. 

In particular at low momentum transfer Deltuva \et~ find large effects of the
$\Delta$ in the threshold region of the transverse response. At the larger
momentum transfers, {\em e.g.} at 500 MeV/c shown in Fig.~\ref{sauer} for
 $^3\!He$, the
contribution of $\Delta$ degrees of freedom is smaller, in agreement
with the observation made in Sec.~\ref{eucl}. The small shift between data and
calculation has been assigned to the use of non-relativistic kinematics
(necessary for consistency). Deltuva \et~ also have calculated
other e.m. observables such as elastic form factors and exclusive quasi-elastic
cross sections, and find  rather large changes when allowing for the coupled
channel nucleon-$\Delta$ case. 
\begin{figure}[hbt]
\begin{center}
\includegraphics[scale=1.2,clip]{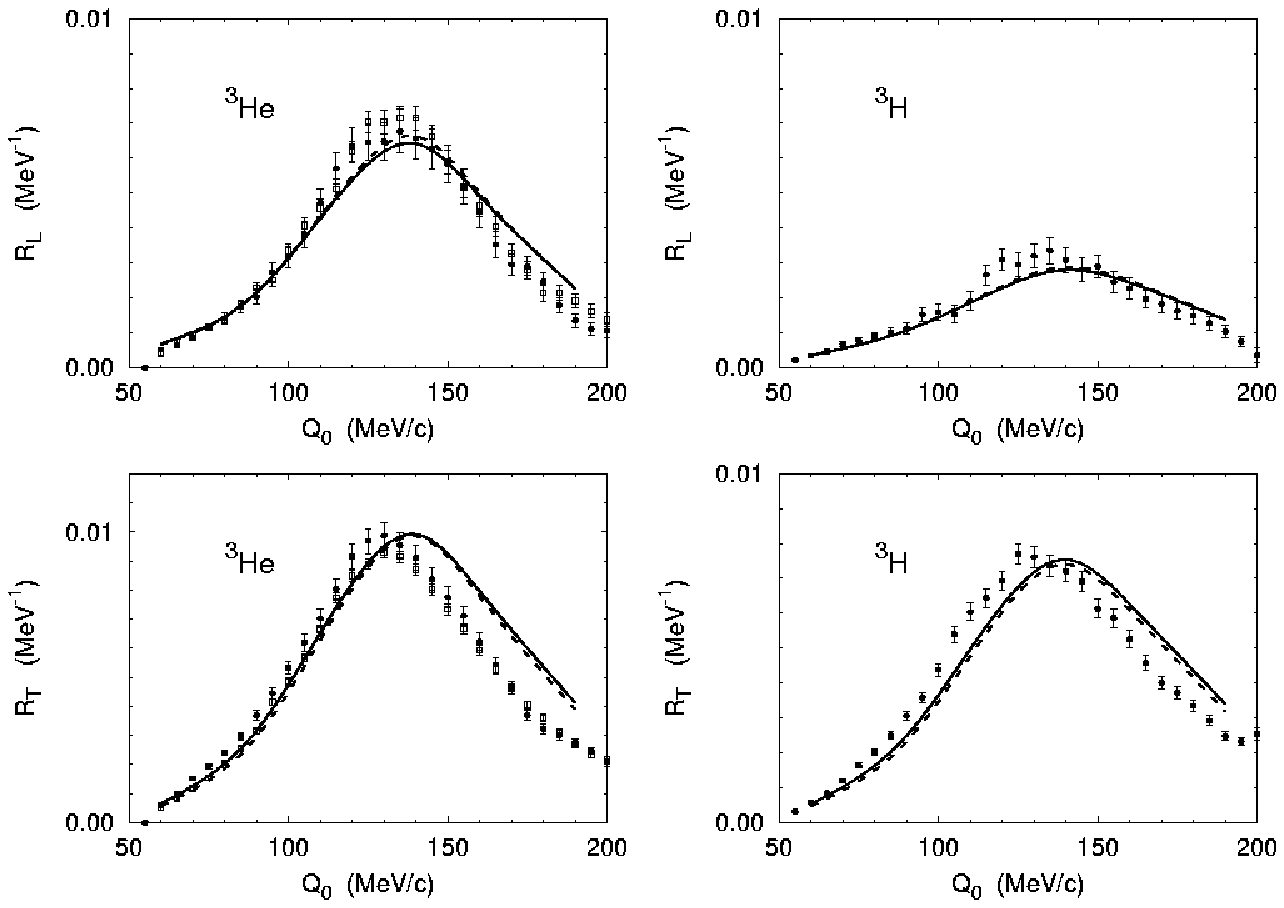}
{\caption[]{Comparison of theoretical \cite{Deltuva04} and experimental 
response at $\magq$=500MeV/c. 
Solid: coupled-channel nucleon-$\Delta$, dashed: purely nucleonic case. 
 Data are shown as circles  \cite{Dow88}  and squares \cite{Marchand85}.
 \label{sauer}}}
\end{center} 
\end{figure}

As pointed out in Sec.~\ref{scale} the scaling function  $F(y)$  is closely
related to the spectral function and momentum distribution. In the limit of very
large momentum transfer, and when neglecting the effects due to finite $E$ and
FSI, the scaling function is given by 
\beq
F(y)= 2 \pi \int_{|y|}^\infty n(\magk)~ d\magk 
\eeq
in which case the momentum distribution can be obtained from 
\beq
n(\magk) = \frac{1}{2\pi y} \frac{dF}{dy}, ~~~~ \mbox{with}~~~ k=|y|
\eeq
For the deuteron,  neglecting the effects of finite $E$ and FSI is not
unreasonable as
the data reach very high $q$. The  $n(k)$ derived via scaling \cite{Ciofi87b} 
is
compared in Fig.~\ref{ciofi} to the ones obtained from $(e,e^\prime p)$ reactions 
and theory. 

For heavier nuclei, both FSI and the effects of the distribution in  $E$ are 
no longer
negligible, and can only be incorporated in predictions of $F(y,Q^2)$ starting
from $S({\kv},E)$.  
\begin{figure}[hbt]
\begin{center}
\includegraphics[scale=0.45,clip]{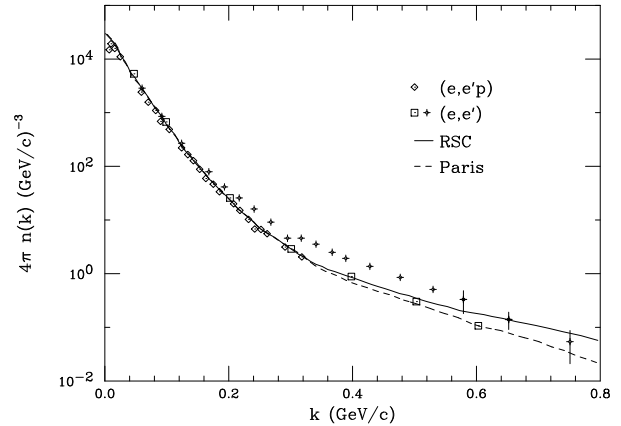}
{\caption[]{Comparison of deuteron momentum distribution obtained from (e,e')
(squares), (e,e'p) (triangles) and theory (curves) \cite{Ciofi87b}.  
\label{ciofi}}}
\end{center} 
\end{figure}

\section{Euclidean response } \label{eucl}
In previous sections, we have discussed \qes~ in terms of PWIA, with corrections
for FSI. This type of  theoretical description often is the only practical one,
as a less approximate treatment of the final, relativistic, continuum state is hard
to come by. 

For comparatively low momentum transfers,  alternative and more reliable
treatments are possible. For $A=3$ and $A=\infty$ the response in the
non-relativistic region can be calculated \cite{Golak95,Fabrocini89}.  For
nuclei in between, integrals  over the \qe~ response  can be studied. These
integrals  can be expressed as expectation values of the ground-state 
wave function, without need for an explicit
treatment of the continuum state. This approach has been followed via both 
 the Lorentz transform technique \cite{Efros94,Leidemann97a}, and the use of the
Laplace transform \cite{Carlson92}.  The most extensive results, 
including one- and two-body currents,  are
available for the latter approach \cite{Carlson02} which we discuss in more
detail below.

The Euclidean response is defined as an integral over the \qe~ response 
\beq
\tilde{E}_{T,L} (\magq,\tau) \ = \ \int_{\omega_{\rm th}}^\infty \ 
 \exp[- \omega \tau ] \  \ R_{T,L} (\magq,\omega) ~ d\omega\>\>
\eeq
with $\tau = Q^2/4m$. In the Lorentz transform technique an additional Lorentzian factor which
enhances the integrand at a given $\omega$ is used. The Lorentz transform technique
has the advantage that the response as a function of $\omega$ can be
reconstructed by the inverse transformation, something that is not practical for
the Euclidean response.

The longitudinal and transverse Euclidean response functions represent
weight\-ed sums of the corresponding $R_L(\magq,\omega)$ and $R_T(\magq,\omega)$:
at $\tau$=0 they correspond to the Coulomb and transverse sum rules,
respectively, while their derivatives with respect to $\tau$ evaluated
at $\tau$=0 correspond to the energy-weighted sum rules.  Larger values of
$\tau$ correspond to integrals over progressively lower energy loss regions
of the response.

In a non-relativistic picture, the $\tilde{E}_{T,L}$ can be simply obtained from:
\be \nonumber
\tilde{E}_L (\magq,\tau) \! &=&\!  \langle 0 | \rho^\dagger ({\bf q})
\exp [ -(H - E_0) \tau] \rho({\bf q}) | 0 \rangle \\
 &-& \exp\left(-\frac{q^2 \tau}{2 A m}\right) |\langle 0({\bf q}) 
|\rho({\bf q})|0\rangle |^2,
\ee
and similarly for $\tilde{E}_T (\magq,\tau)$, with the charge operator 
$\rho({\bf q})$ replaced by the current operator ${\bf j}_T({\bf q})$. 
The elastic contributions have been explicitly subtracted, and $|0({\bf q})\rangle$
represents the ground state recoiling with momentum ${\bf q}$.

The study of the Euclidean response has the outstanding advantage that 
$\tilde{E}(\magq,\tau)$ can be calculated
 from the ground state properties alone; no explicit treatment of the final
continuum state is required. For the A=3,4 ground states, very precise wave functions are
available, and the effects of MEC can be included using the two-body
operators well 
established in elastic and inelastic electron scattering from light nuclei
(for a review see Sick, 2001 \nocite{Sick01}). 

The Euclidean response has the disadvantage that we  lack an intuitive
interpretation of  this integrated quantity. Model studies \cite{Carlson02}
show the sensitivity of $\tilde{E}(\magq,\tau)$ to properties of $R(\magq,\omega)$, 
see Fig.~\ref{model}.
\begin{figure}[hbt]
\begin{center}
\includegraphics[scale=0.5,clip]{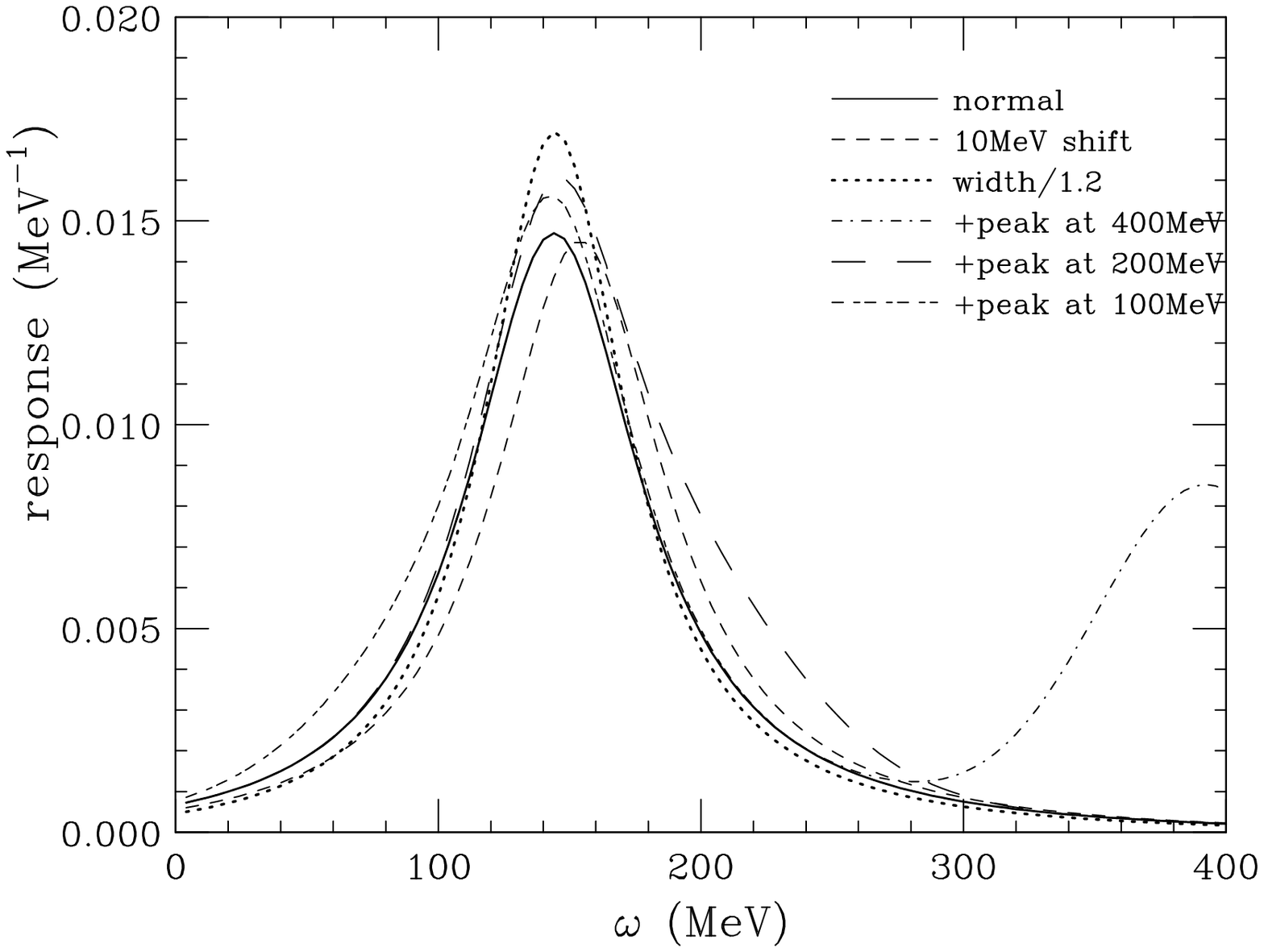}
\hspace*{2mm} \includegraphics[scale=0.5,clip]{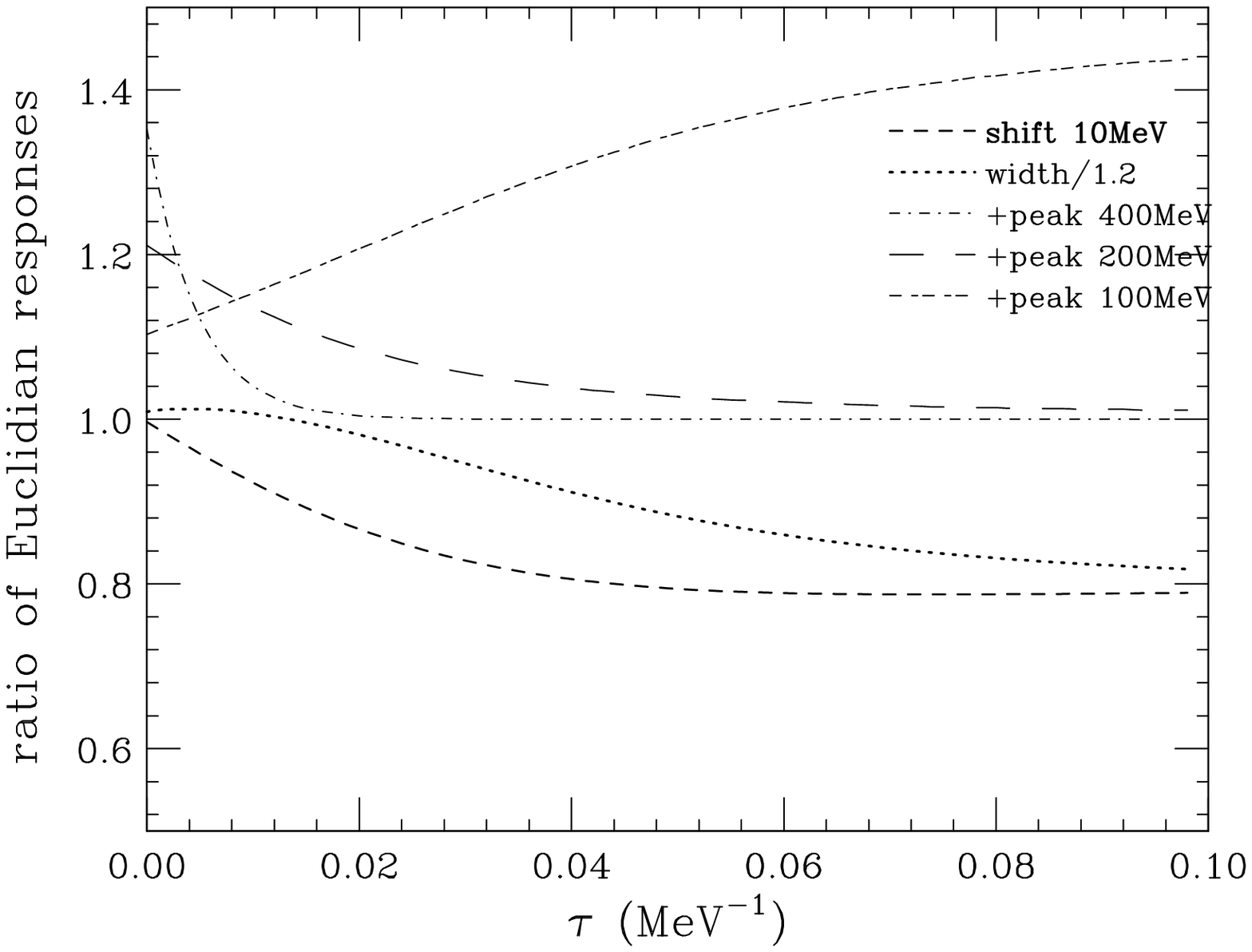}
\caption[]{ Model responses derived from the ''normal'' response
plus various modifications (top). The bottom figure gives the corresponding 
Euclidean response in terms of the ratio to the one from the ''normal'' one.
 \label{model}}
\end{center} 
\end{figure}
 The top figure shows various modifications of the 'normal' response, the bottom
 figure shows the effect upon the ratios of the resulting Euclidean responses to the 
'normal' one.
These studies show that, for the responses that can be extracted from
 the data, the region $0.01 \leq \tau \leq 0.05$ MeV$^{-1}$ is the most relevant one for a 
comparison with theory. Below $\tau = 0.01$ MeV$^{-1}$ the contribution of the tail of the
$\Delta$ in the experimental response is too important, above $\tau = 0.05$ MeV$^{-1}$
the response  is totally dominated by the contribution of very small values of $\omega$. 

The calculations discussed here \cite{Carlson02} have used the
standard expressions for the one-body 
electromagnetic operators, obtained from a relativistic reduction of the
covariant single-nucleon current. The two-body current operator consists of 
\lq\lq model-independent\rq\rq and
\lq\lq model-dependent\rq\rq \  components, in the standard classification scheme
\cite{Riska89}.  The model-independent terms
are obtained~\cite{Schiavilla90} from the nucleon-nucleon interaction.
For the model-de\-pen\-dent pieces, the calculation includes the isoscalar 
$\rho \pi \gamma$ and
isovector $\omega \pi \gamma$ transition currents as well as the isovector
current associated with excitation of intermediate $\Delta$-isobar resonances.
 The two-body charge operators \cite{Schiavilla90} include the $\pi$-, $\rho$-, and
$\omega$-meson exchange charge operators,  the (isoscalar) $\rho \pi \gamma$ and (isovector)
$\omega \pi \gamma$  couplings and the
single-nucleon Darwin-Foldy and spin-orbit relativistic corrections.

We show in Fig.~\ref{resp} the results obtained for $^3\!He$ and $^4\!He$ at one
value of $\magq$.  
 The ground-state wave functions used in this study
were obtained with Variational Monte Carlo approach \cite{Carlson02} and the Argonne 
 $v_8$  N-N interaction plus the UIX three-nucleon interaction. 

The helium nuclei studied by Carlson {\em et al.} are of particular interest as for
$^4 \! He$ the excess in the transverse strength is maximal among all nuclei, 
and grows by a factor of two between A=3 and A=4. 
 This excess ---  presumably due to MEC --- had
not been understood in the past; the many calculations of MEC 
for a multitude of nuclei gave results that
were rather discordant, and always much too small.  The data for the Helium nuclei
also show, that this excess covers the entire \qep , and not only the ''dip''
region between the \qe - and $\Delta$-peak.

\begin{figure}[hbt]
\begin{center}
\includegraphics[scale=0.5,clip]{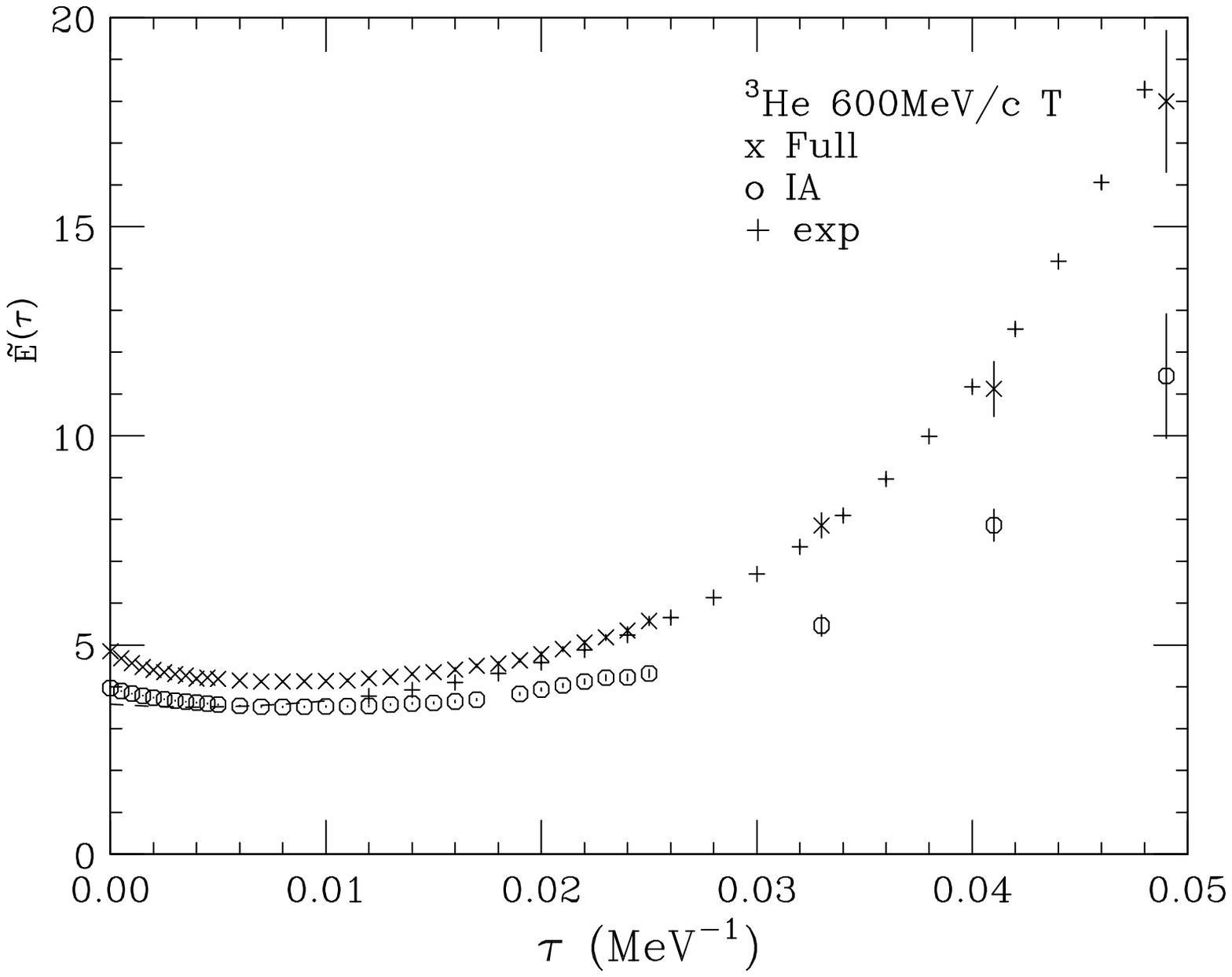}
\includegraphics[scale=0.5,clip]{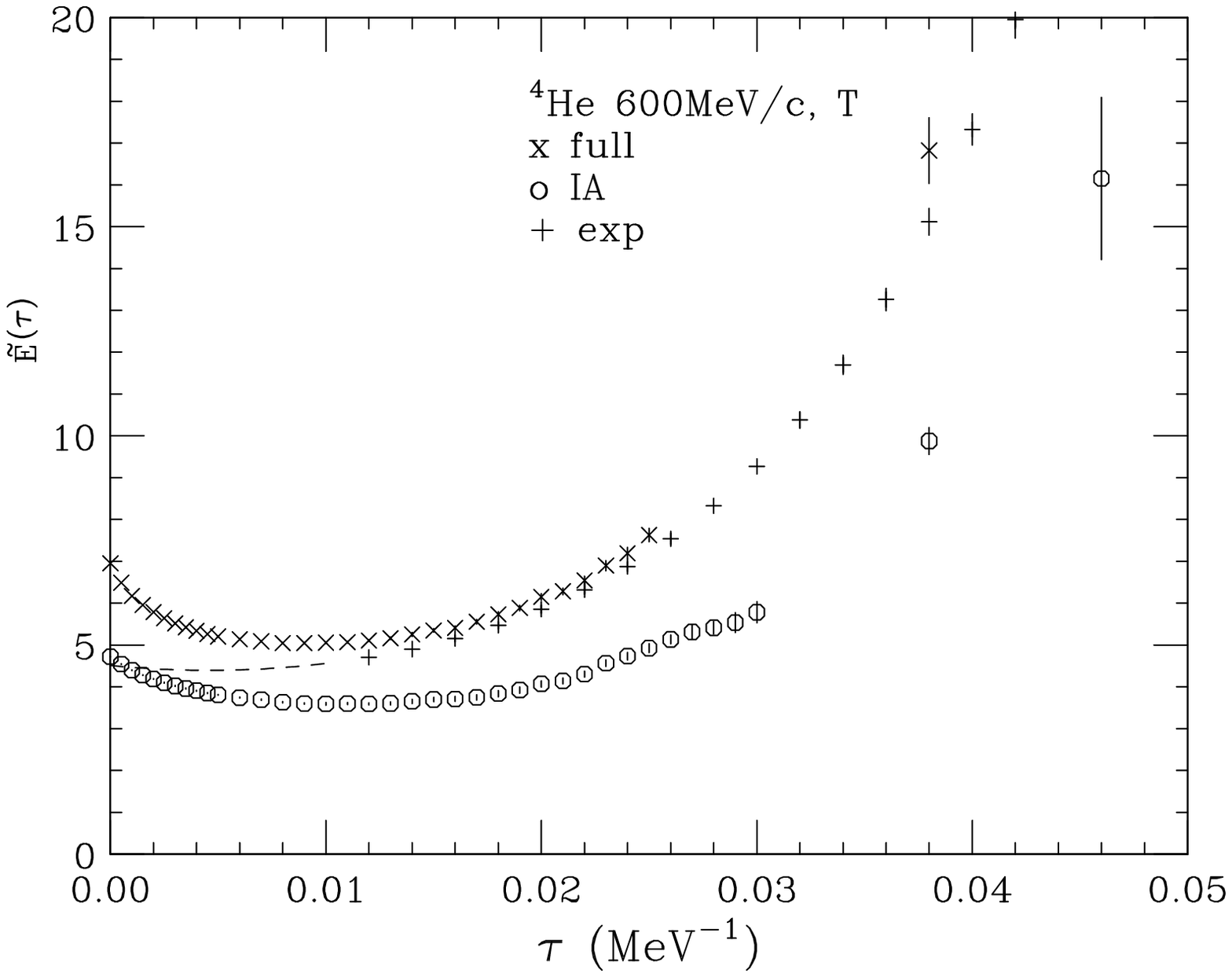}
{\caption[]{Transverse Euclidean response divided by the proton magnetic form
factor  for $^3 \! He$  
(top) and $^4 \! He$ at 
600 MeV/c momentum transfer, scaled by $e^{\tau \omega_{qe}}$. 
 Data ($+$), IA ($\circ$) and full calculation ($\times$). \label{resp}}} 
\end{center} 
\end{figure}

Fig.~\ref{resp} shows that  the full calculation, which includes MEC, is in good
agreement with the data (we pointed out above why the region $\tau < 0.01$ MeV$^{-1}$
should be ignored). The calculation predicts rather accurately the
enhancement of the transverse strength due to MEC and the doubling between A=3,4,
 and it also does quite well in predicting the $Q^2$-dependence (not shown).

The good reproduction of the transverse strength at first sight comes somewhat as
a surprise, given the  lack of success of previous MEC-calculations
\cite{Donnelly78,VanOrden81,Kohno81,Blunden89,Leidemann90,Dekker94,Amaro92,
Alberico84,Carlson94,Vandersluys95,Anguiano96,Fabrocini97a,Gadiyak98}. 
Carlson \et~have therefore investigated more in detail the reason for the large MEC
contribution. They have found that, in agreement with previous studies that have
received too little attention \cite{Leidemann90,Fabrocini97}, MEC only produce
large effects in combination with ground-state wave functions calculated {\em
including} the short-range n-p correlations. As most previous calculations were
based on independent-particle type wave functions, the smallness of the 
resulting MEC
contributions thus is  understood.  To verify this point further, Carlson \et~
have repeated their calculation using the same operators, but with a Fermi-gas 
wave function. Instead of a
 enhancement factor of 1.47 coming from MEC at $\magq=$600 MeV/c they find a 
factor of 1.06 only,
{\em i.e.} an eight times smaller MEC effect.   

The results of Carlson {\em et al.} also show, somewhat surprisingly, that the
MEC contribution is large at {\em low} momentum transfer. It decreases towards
the larger $Q^2$, in agreement with the expectation that at very large $Q^2$ it 
falls \cite{Sargsian01} like $Q^{-4}$ relative to \qes . 

From the above discussion it becomes clear that the Euclidean response, despite
inherent drawbacks,  is a very valuable quantity. Since the final continuum
state does not have to be treated explicitly, calculations of much higher
quality can be performed than for the response, and the role of two-body 
currents can be treated
quantitatively. The comparison between data and calculation has shown
in particular that for a successful prediction of MEC {\em correlated} wave
functions for the ground state are needed; such wave functions today are
available  up to $A \sim 12$ and for $A = \infty$. Unfortunately, the usage
of the Euclidean response for the time being is restricted to a regime where
relativistic effects are not too large, such that they can be included as
corrections.     

\section{L/T-separation and Coulomb sum-rule \label{incllt}}
In impulse approximation, and when neglecting the (small) contribution from
nucleonic convection currents, the longitudinal and transverse response functions $R_L$ and
$R_T$ contain the same information and have the same size. 
This has sometimes been called ''scaling of
the 0$^{th}$ kind'' (see Sec.\ref{scale}). It has been realized early on,
however, that the transverse response receives significant contributions from
meson exchange currents and $\Delta$-excitation (which are of largely transverse
nature). 
It therefore is clear that there is a high premium on separating the L-
and T-responses, both because the L-response is easier to interpret and because
of the additional information contained in the T-response.

\begin{figure}[hbt]
\begin{center}
\includegraphics[scale=0.5,clip]{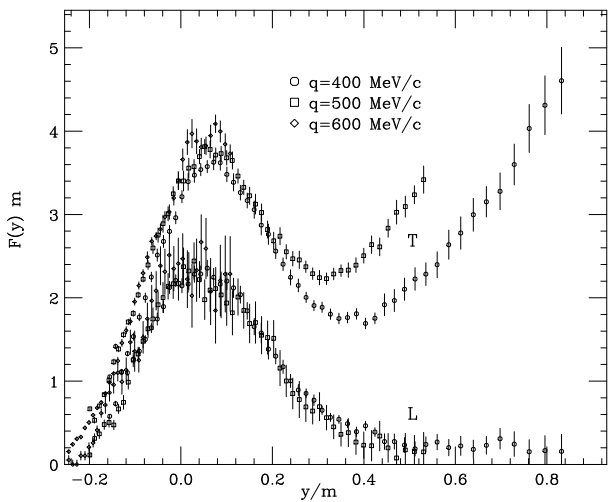}
{\caption[]{Longitudinal (lower data set) and transverse responses
 of $^{12}C$  \cite{Finn84}, plotted in terms of the scaling function $F(y)$.
 }\label{finn}} 
\end{center} 
\end{figure}

The separation of the L- and T-responses is performed using the 
Rosenbluth technique, which is justified only in the single-photon exchange
approximation. The cross section, divided by a number of kinematical
factors 
\beq
\frac{d\sigma}{d\Omega d\omega}~\frac{\epsilon}{\sigma_{Mott}}~\frac{\bf{|q|}^4}{Q^4} 
 = \epsilon~ R_L(\magq,\omega)+\frac{\bf{|q|}^2}{2Q^2}~ R_T(\magq,\omega)=\Sigma
\label{rosen} \eeq
is a linear function of the virtual photon polarization 
\beq
\epsilon = \big(1+\frac{2\bf{|q|}^2}{Q^2}~\tan^2\frac{\theta}{2}\big)^{-1}
\label{eps} \eeq
with $\qv$ ($Q$) being the 3-(4-)momentum transfer and $\epsilon$ varying from 0
to 1 for scattering angles $\theta$ between 180$^\circ$ and 0$^\circ$. The slope of the linear
function yields $R_L$, the intercept at $\epsilon = 0$ yields $R_T$.
Fig.~\ref{finn} shows an early example for an L/T-separation, and demonstrates
the excess observed for the transverse strength. 

While conceptually very straightforward, this L/T-separation is difficult in practice. It
involves data taking at the same $\magq$, but varying $\epsilon$, {\em i.e.}
varying beam energy. For an accurate separation of $R_L$ and $R_T$ obviously the
largest possible range in $\epsilon$, hence beam energy, is required. As data
are usually not taken at constant $\magq$, but  at a given beam energy 
and variable energy loss, obtaining the responses at  constant $\magq$ 
involves  interpolations of the data.
We show in Fig.~\ref{ltsep} two
examples for a Rosenbluth separation, performed on the low- and large-$\omega$
side of the \qep , which also illustrate the importance of the forward angle
(high energy) data for the determination of $R_L$, {\em i.e.}  the slope of the
fit.

\begin{figure}[hbt]
\begin{center}
\includegraphics[scale=0.55,clip]{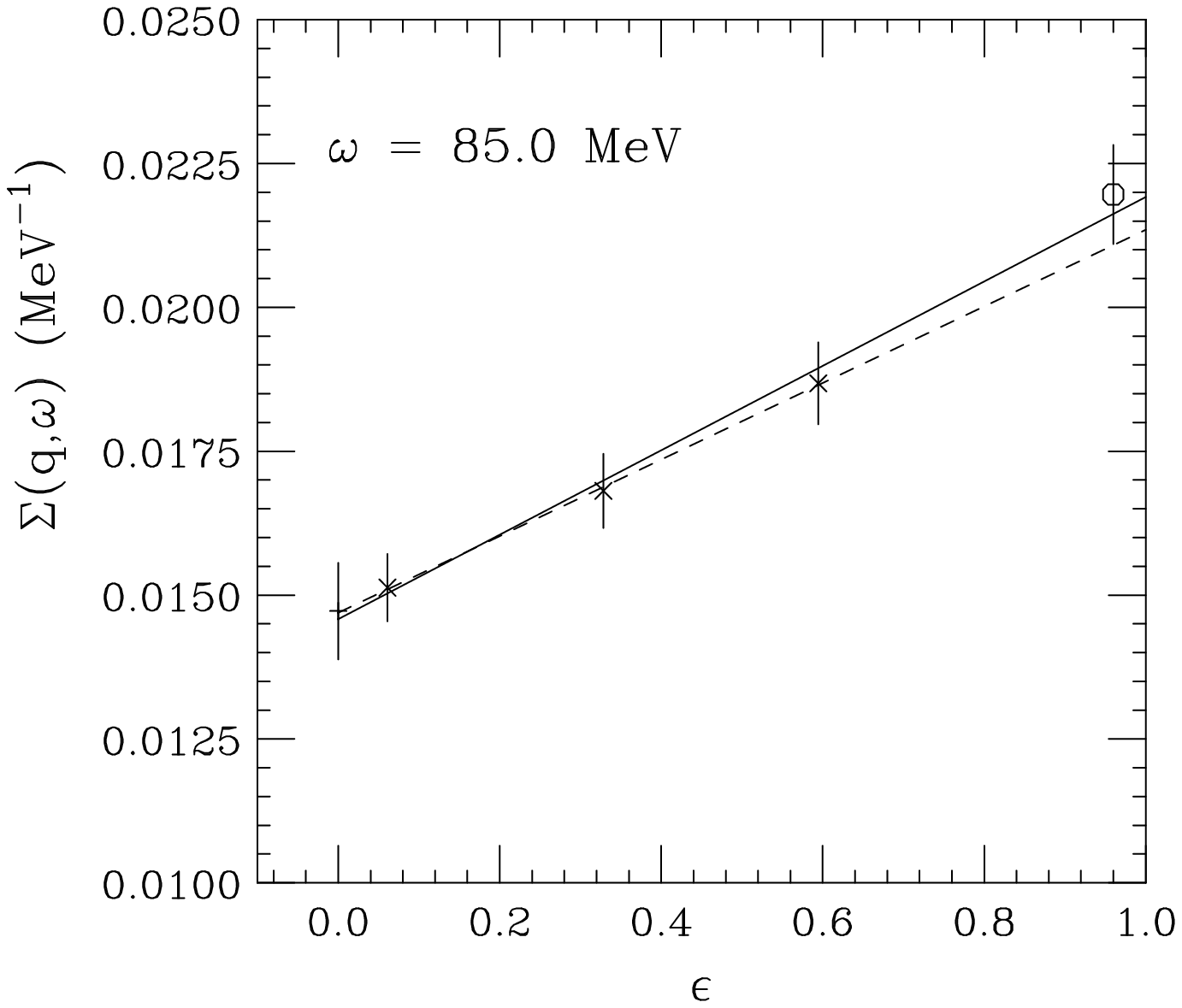}
\hspace*{1mm}\includegraphics[scale=0.55,clip]{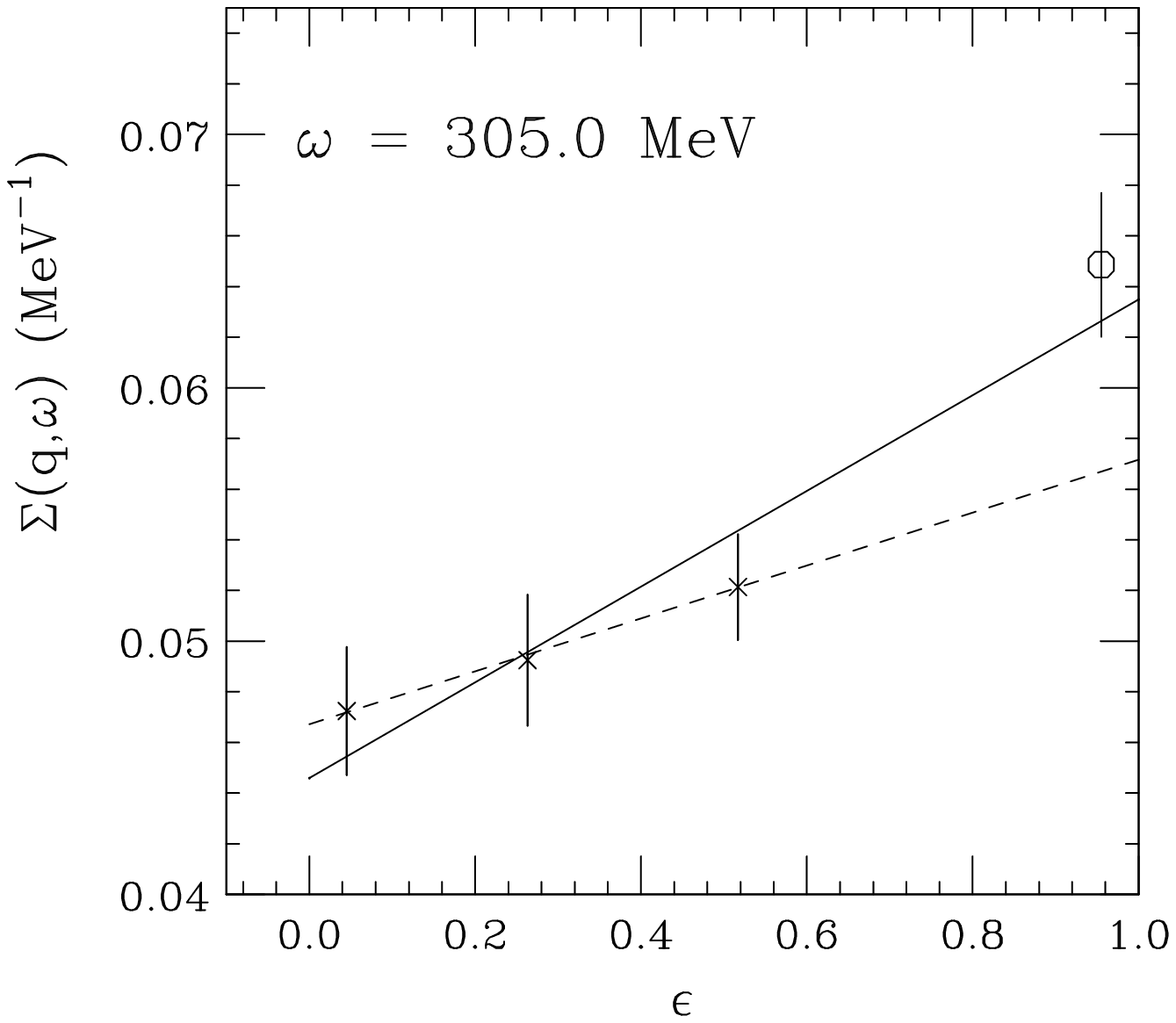}
{\caption[]{Rosenbluth separation for $^{56}\!Fe$ and $\magq=$570 MeV/c \cite{Jourdan96a}.
The dashed lines are fits to the Saclay data alone, the solid lines are fits to
the world data which include the forward-angle SLAC data and emphasizes the
importance of a large range in $\epsilon$.
}\label{ltsep}} 
\end{center} 
\end{figure}

The   Rosenbluth technique is applicable in Plane Wave Born Approximation
PWBA, and fails once Coulomb distortion of the electron waves is present.
Neglect of distortion 
is justified for the lightest nuclei alone, and only if $R_T$ is not
much bigger (or much smaller) than $R_L$. When one of the two contributions gets
too small, even minor corrections due to Coulomb distortion can have large
effects. At large $\magq$, for instance, even the determination of the proton charge
form factor via the Rosenbluth technique 
is significantly affected by Coulomb corrections \cite{Arrington04}.
In order to extract $R_L$ and $R_T$ in the presence of Coulomb distortion, the
data have first to be corrected for these effects; this is discussed in
Sec.~\ref{coul}.

Here,  we concentrate on the discussion of the longitudinal response.  
For practical reasons, the determination of the longitudinal response
is possible only in a  $q$-range that is 
somewhat limited. In addition, at low $\magq$, typically below twice the Fermi momentum, the
response is affected by Pauli blocking, which in many of the approaches
used to describe inclusive scattering is not properly treated. At large $\magq$,
typically above 0.8 GeV/c, the transverse response dominates the cross section, 
both due to the $|{\bf q}|^2/Q^2$ factor in Eq.~(\ref{rosen})
 and the increasing $\Delta$ contribution (see Fig.~\ref{finn}), such
that an accurate determination of $R_L$ becomes very difficult. Energy-dependent
experimental systematic errors must be handled with great care.

One particular use of the longitudinal response has received much attention: the
determination of the Coulomb Sum Rule (CSR). In the non-relativistic regime, when
 short-range correlations between nucleons and the effect of Pauli
blocking is neglected, the  CSR takes the simple form
\beq
S_L(\magq) = \frac{1}{Z} \int_{\omega^+}^\infty \frac{R_L(\magq,\omega)}{\tilde{G_e}^2}
d\omega
\label{csr} \eeq
where $\tilde{G_e} ^2= (G_{ep}^2 +  G_{en}^2 N/Z)$ and $\omega^+$ is the
threshold for particle emission. In the limit of  large $\magq$
$S_L$ should be one.  To say it in words: when neglecting the small
contribution from the neutron charge form factor $G_{en}$, the integral over the 
longitudinal response counts the number of
protons times the square of the proton charge form factor $G_{ep}^2$. 

The history of the CSR is a very checkered one. 
Early work proposed the CSR as a tool to study short-range correlations
SRC between nucleons \cite{Czyz63,Gottfried63}. These correlations move strength 
to large  energy loss, and
partly out of the physical region; at very large $\magq$, these correlation 
contributions go to zero in the sum rule.  
The series of L/T-separations performed at Bates \cite{Altemus80} 
and Saclay \cite{Meziani84} found effects in $S_L$  that were much bigger than could be
expected from SRC: in the region of $\magq$=350-550 MeV/c $S_L$ was up to 50\% lower
than expected, the deficit increasing with increasing $\magq$ and increasing
nuclear mass number $A$. These observations have widely been interpreted as a
medium-modification of the proton charge form factor
\cite{Noble81,Celenza86,Mulders86}.

 The experiments dealig with ther CSR have absorbed much of the attention. It 
was
 not generally known that some experiments \cite{Altemus80} had suffered from
 rescattering of electrons on the ''snout'' connecting scattering chamber and
 spectrometer \cite{Deady86} (see also Sec.~\ref{exp}). It also has taken a long time 
before  a  reanalysis \cite{Jourdan96a} removed a  number of
 deficiencies in the  Saclay analysis of the data (see below).

\begin{figure}[hbt]
\begin{center}
\includegraphics[scale=0.55,clip]{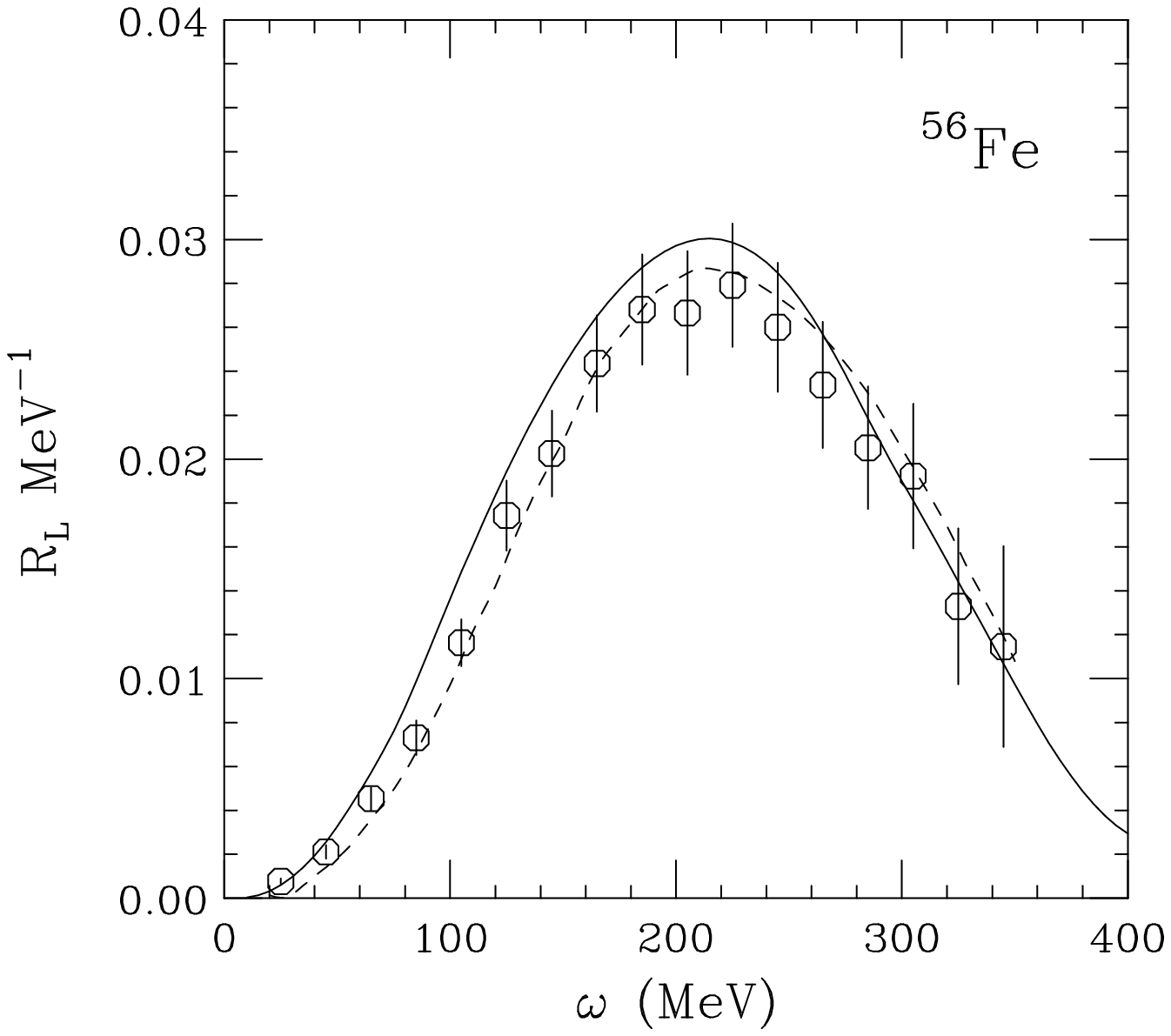}
\includegraphics[scale=0.55,clip]{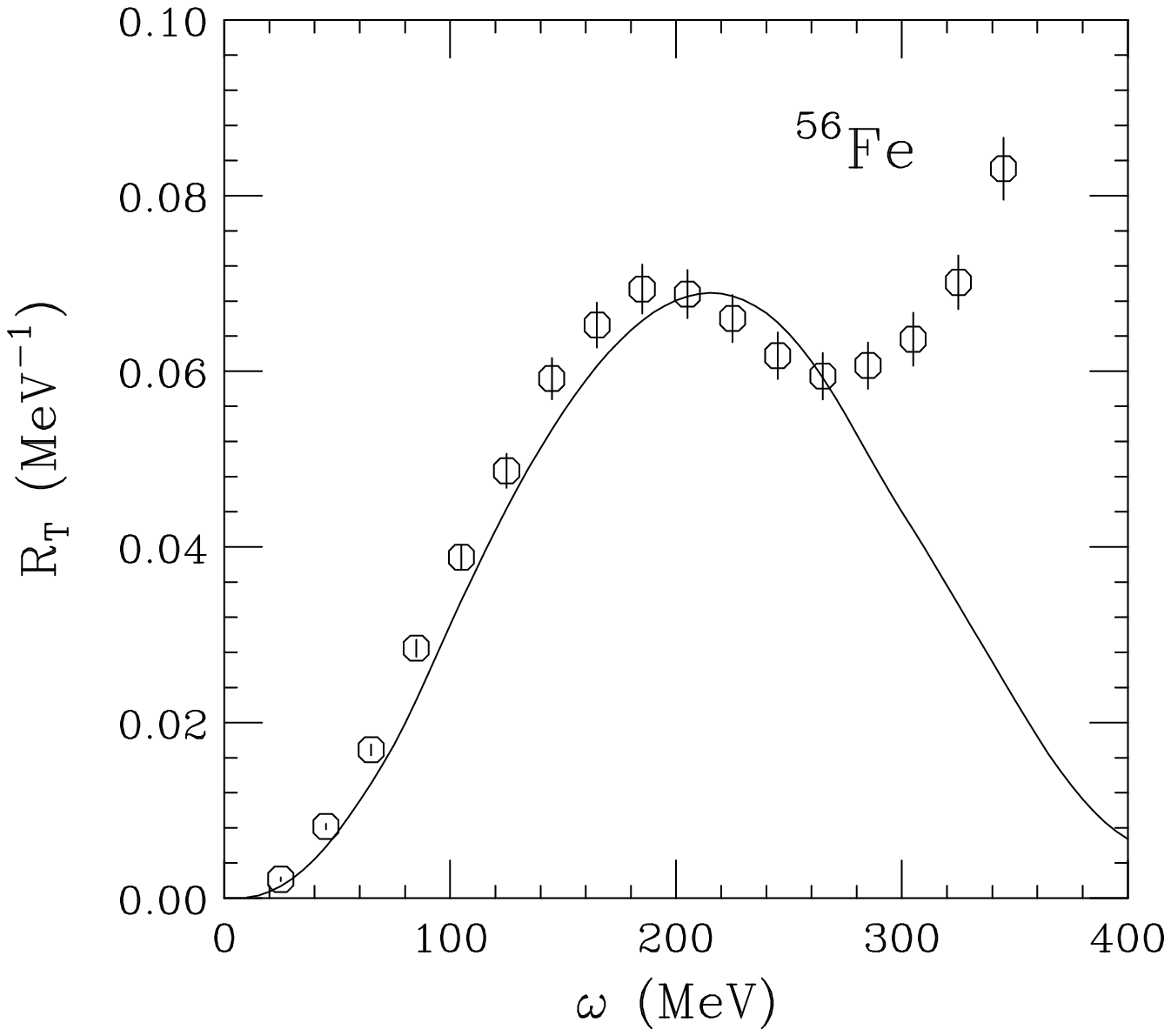}
{\caption[]{Separated response functions for $^{56}\!Fe$ and $\magq$=570MeV/c. 
The solid curve corresponds to the CBF calculation \cite{Fabrocini89}.
}\label{rlrt}} 
\end{center} 
\end{figure}

 L/T separations have been have been performed for a number of nuclei and
momentum transfers. 
\cite{Zimmerman69,Altemus80,Deady83,Barreau83,Finn84,Deady86,Blatchley86,Reden90,Chen91,Meziani92,Yates93,Williamson97}.
 Often (but not always, see {\em e.g.} Williamson, 1997), 
these separations were quite limited in the
$\epsilon$-range as data from one facility only were included; from the
determination of the proton form factors $G_{ep}$ and $G_{mp}$ it is well known
that the more reliable results are obtained from an analysis of the {\em world}
data spanning the largest possible range in $\epsilon$ (electron energy). 
 In almost all
cases, only approximate Coulomb corrections were included, using the EMA 
version of 
the effective-momentum transfer approach (see Sec.~\ref{coul}). Particularly 
for data producing low $S_L$ the  longitudinal response functions 
show an  unphysical behavior at large $\omega$: they dive steeply
towards zero, and would, for responses without discontinuities,  obviously be 
 negative  just beyond the range of  
$\omega$ shown (see {\em e.g.} Jourdan,1996).
From microscopic calculations \cite{Dieperink76,Dellafiore85,Fabrocini89} 
 using realistic nuclear spectral functions we
do know, however, that the response at large $\omega$ should approach zero
slowly, much more slowly  than the response at low $\omega$, as a 
consequence of the components of large removal energy $E$ present in realistic
 spectral functions $S({\kv},E)$ (see {\em e.g.} Fig.~\ref{orlan}). 
  
 The analysis of the data performed by \cite{Jourdan96a,Carlson03} included the 
{\em world} data to employ the largest $\epsilon$-range and improved a number
of aspects  of the Saclay analysis. In particular, the integral in
 Eq.~(\ref{csr}) was divided  by the correct $G_{ep}$, the well-known 
 relativistic corrections were included \cite{Forest84}, and Coulomb 
 corrections  in DWBA were made; in addition,  the
 contribution of $R_L(\omega)$ above the upper integration limit --- which 
 experimentally is far from  $\infty$ due to
 the limited range of data on $S_L(\omega)$ --- was added.  These 
L/T-separations extend to $\magq \sim$ 600 MeV/c,  only for
$^4\!He$, where the overlap of \qe~ and $\Delta$ strength is a lesser problem,
the L-strength is known up to 1 GeV/c.   The corrections to the 1984 analysis, 
which all happen to go in the same direction, increase the CSR 
 by a substantial amount. For example, the CSR
 for the largest $\magq$ and heaviest nucleus, $^{56}\!Fe$ and 570 MeV/c
analyzed in \cite{Jourdan96a}, amounts to  0.98$\pm$0.15. Fig.~\ref{rlrt} 
shows the corresponding L- and T-responses.   

In a subsequent reanalysis of the (e,e') data  \cite{Morgenstern01}
resulted in a less extreme reduction in the CSR result, for Iron the CSR was
found to be 82\% of the expected value. This result for the
sumrule is still significantly smaller than the one of \cite{Jourdan96a}, the
difference being largely due to the use of EMA for the Coulomb corrections. This
emphasizes that the Coulomb corrections are very important for the determination
of the CSR; particularly for the largest q-value of 570MeV/c  and the
large $\omega$, where the difference of \cite{Jourdan96,Morgenstern01} is 
largest,
 the backward angle data go down to scattered-electron energies as low as 
130MeV, where approximations like EMA fail \cite{Aste05}.

The issue of medium modifications of proton electromagnetic structure
has been recently revived by the results of
a polarization transfer $^4$He$(\vec{e},e^\prime\vec{p}\,)$$^3$H
measurement carried out at Jefferson Lab \cite{Strauch03}. However,
the interpretation of the experimental data in terms of medium-modified
form factors is challenged by the results of a theoretical calculation,
carried out
using accurate three- and four-nucleon bound-state wave functions,
a realistic model for the nuclear electromagnetic current
operator and a treatment of final-state-interactions with an optical
potential \cite{Schiavilla05}. In Ref. \cite{Schiavilla05}
no significant discrepancies are found between theory and experiment, both
for the ratio of transverse to longitudinal polarization transfers and for
the induced polarization, when free-nucleon electromagnetic form factors
are used in the current operator.

 To close the discussion of the Coulomb sum,  we add  as a caveat that even 
for the longitudinal response the contribution of
MEC is not entirely negligible.  In the $\magq \sim$1 GeV/c region it has been shown
\cite{Carlson03} that for $^4\!He$ the contributions are of the 
order of  10\%.

\section{Coulomb Corrections} \label{coul}
The effects of the static Coulomb field of nuclei upon \qes~ has posed a
persistent problem. The presence of the Coulomb potential --- for Lead of order
of 25-30 MeV in the nuclear interior --- has a major effect on  \qe~
cross section measured in the several-hundred MeV energy region. It 
invalidates the linear relation (\ref{rosen}) used to separate the longitudinal and
transverse responses using the Rosenbluth technique. 

While the treatment of the Coulomb distortion of the electron waves in 
the \qe~ region  presents no conceptual
problems,  the practical application has been difficult. Reliable
 calculations of Coulomb distortion have not been easily accessible
to analyses of experimental data. As a consequence, most experiments have been
analyzed {\em without} considering Coulomb distortion effects, or by using 
relatively simple recipes. 

The Coulomb distortion is more important for the lower-energy data and the
heavier nuclei. Its effects are visible in the separations of longitudinal
and transverse strength, where the small contribution --- in general the
longitudinal one --- is most affected. 
It is likely  that one of the
main problems with the longitudinal strength --- the ''diving to negative
responses'' mentioned in Sec.~\ref{incllt} ---  is  related to this aspect
 \cite{Traini88}. 

The Coulomb distortion can be treated in Distorted Wave Born Approximation (DWBA) 
by using electron wave functions
calculated as solutions of the Dirac equation for the known nuclear charge 
distribution; different programs have been employed
 \cite{Co87, Kim96,Udias93}, in particular by the group of Onley and Wright
\cite{Jin92,Zamani86}. In these
calculations, high-energy approximations \cite{Knoll74,Lenz71,Rosenfelder80} can
be employed,  as in general the electron energy is much higher than the Coulomb
potential. Such calculations have been successfully carried out, and can serve
as a benchmark. For a systematic analysis of experimental data, the numerical
effort is usually too big and un-practical. This is true in particular as often
computer codes have been developed for $(e,e'p)$, in which case the $(e,e')$
cross section has to be generated by summing over all possible initial and 
final states of the knocked-out  nucleon.

Simplifications are possible when going to the Eikonal approximation
\cite{Yennie65,Traini88,Giusti87}. When using the lowest order expansion in
$Z \alpha$, one finds two dominating effects: \\
~~~~1. As a consequence of the attractive electron-nucleus Coulomb interaction, the
effective energy  of the incident and scattered electron  at the moment of the 
scattering is  increased by the Coulomb potential $V$. This has the consequence
 that the effective momentum transfer $q_{eff}$ squared, on which the  
 response functions 
$R$ depend, is increased by a factor $(E_e+V)(E_{e'}+V)/E_e E_{e'}$.   \\
~~~~2. Due to the attractive Coulomb interaction, the electron ''plane'' waves are
focused onto the nucleus, hereby increasing the wave function  at the location 
of the nucleus, with a corresponding increase in the scattering cross section.

In the calculation of the inclusive cross section, the overall effect of these
corrections is accounted for by using the Mott cross section with the 
{\em un}modified 
electron energy, but with $q_{eff}$ as an argument of the response function. 
This approximation has
been baptized {\em Effective Momentum Approximation} (EMA). Versions in the
literature differ by the choice of  $V$. 
The results of EMA have been compared to the ones from DWBA calculations
\cite{Jourdan96,Kim96}.
 
Higher order terms have been included  
\cite{Traini88}. These terms have been calculated using   severe
approximations in the expansion around $r=0$, and are not recommended
\cite{Traini01}. Due to the approximations, the second-order effects found 
are about as large as the first order, which is indicative of problems of the
expansion. 

The term ''EMA'' is often confused in the literature because of the fact that
it is used  for two different
choices of the nuclear Coulomb potential $V$. This parameter $V$ is often
evaluated for the nuclear center $r=0$ assuming a homogeneous nuclear charge 
density, in which case $V=V_0=3 Z \alpha /2 R_{eq}$, with $R_{eq}$ being the
equivalent radius (a good approximation being $[1.1 A^{1/3} + 0.86 A^{-1/3}] fm
$ \cite{Kim96}). It has been recognized early on, however, that a better choice
would be an appropriate {\em average} Coulomb potential \cite{Rosenfelder80}. For
this reason, many applications of EMA use for $V$ the value at the nuclear
surface, $V_s=(2/3) V_0$, where most nucleons are located.   

In order to improve upon the quality of the Coulomb corrections without
resorting to the full solution of the Dirac equation, one can employ
 the Eikonal Distorted Wave Born Approximation \mbox{eDWBA},
where the electron waves, in the DWBA approach solutions of the Dirac equation,
are calculated using the Eikonal approximation \cite{Aste04a}. 
In this case the electron
current is modified by an additional Eikonal phase and a change in
amplitude.  This type of calculation 
%
%
can be carried out for realistic shapes of
the nuclear Coulomb potential, and eDWBA, contrary to the full solutions of the
Dirac equation, can be extended more easily  to the larger energies of 
interest for modern experiments.  


 
While eDWBA is a fairly practical approach that can be employed on a routine
basis, it still is much more involved than EMA. Aste and Jourdan 
have identified one problem of EMA, and have introduced an EMA-like
approach called \emap ~~\cite{Aste04b}. 
In EMA, one tries to treat two distinct effects mentioned
above: the increase of the electron momentum due to Coulomb interaction with the
nucleus, and the focusing of the electron waves. To handle  the increase of the
electron momentum (and momentum transfer) it clearly makes sense to use a 
potential
$V$ that corresponds to the  Coulomb potential {\em averaged} over all nucleons;
here the use of the potential at the nuclear surface is a good approximation.
For the focusing effect, on the other side, the value for the {\em nuclear
center} is a better approximation, as the focusing takes place all along the trajectory of
the electron through the nucleus. As the electron approaches (leaves)  the 
nuclear center the focusing is
smaller (larger). Hence the value at the center is a good compromise.  



Subsequent studies using solutions of the Dirac equation in the Coulomb field of
the nucleus \cite{Aste05} indicate that, in addition to the enhanced focusing in
the longitudinal direction accounted for by \emap, there is a reduction for
non-central electron trajectories. The two effects together give an overall
focusing which is not far from the one obtained with EMA.

Recent studies by Tjon and Wallace \cite{Tjon06}, performed using the Eikonal
expansion for the electron wave function, indicate that EMA somewhat
overestimates the Coulomb effects. The authors give a recipe for correcting the
deficiencies, which however is not very practical for analyses of data.

The recent work of \cite{Kim05}, who performed calculations of (e,e) in both
full DWBA and also in an approximate version DW which was found to agree well
with DWBA, indicates that EMA is a good approximation for the transverse part,
but that it works poorly for a longitudinal cross section. The authors, 
however, offer no explanation for the difference.

It would clearly be  desirable to certify, via exact calculations, the validity of
some EMA-type approach, as only such an approach could also be applied to the
important region of the large-$\omega$-side of the quasi-elastic peak, where one
has to deal with the overlap with a large $\Delta$-contribution, the Coulomb
corrections for which also need to be dealt with properly.

      The ideal way to experimentally check the Coulomb corrections is a comparison of
electron and positron scattering. Unfortunately, positron beams are hard to 
come
by, and experiments with the secondary positron beams  are much more difficult 
than with electrons. One such experiment has been carried out \cite{Gueye99}.
  The data, unfortunately, suffer from  normalization
problems  \cite{Aste04a}; from the {\em position} of the \qep~
 one can, however, deduce that it is appropriate to use $V_s$ for the 
calculation of $q_{eff}$.

\section{Nuclear matter}

For \nm\ the Schr\"odinger equation for nucleons bound by the nucleon -- nucleon
interaction (deduced from NN scattering) can be solved
with very few approximations. The translationally invariant nature of the 
medium, where 
solutions can all be written in terms of plane waves, simplifies the 
calculation very much. As a consequence, the quality of \nm\ wave functions
is comparable to the one for the A=2,3,4--nuclei.  Due to the ''exact''
nature of the \nm\ single particle wave function, {\em both} the long-range 
and short range 
properties are well under control. This is in contrast to finite nuclei,
 where calculations, that are designed to do well on the long-range properties 
(mean-field calculations), usually do badly on the short-range aspects. 

Unfortunately, only some integral properties such as density and binding energy
are known experimentally; little is known on the {\em short-range} properties of
\nm . Quasi-elastic scattering here provides valuable information.

It is important to realize that inclusive 
electron scattering at large $Q^2$ is sensitive only to rather
''local'' properties of the medium. The spatial resolution of $(e,e')$
is of order 1/$\magq$, which at large transfer is small. In particular, the scattered 
electron is not sensitive to the interactions of the recoiling nucleon
{outside} this range. This allows for an extrapolation from finite-nucleus data
using the LDA.

The extrapolation procedure \cite{Day89} starts  from the
consideration that the  nuclear response  is essentially  the incoherent sum 
of contributions from individual nucleons. As the average value of the 
density in the nuclear interior  and the shape of the density distribution in
the nuclear surface are approximately  $A$-independent,
the response can be divided into a volume component, proportional to the mass 
number $A$, and a nuclear surface component  proportional to $A^{2/3}$. 
It is the  former one that is of 
interest when discussing nuclear matter. The ratio of the
surface to volume contributions is thus proportional to $A^{-1/3}$.
Extrapolation of the nuclear response per nucleon to $A^{-1/3}= 0$ 
($A \rightarrow \infty$)
 as a linear function of $A^{-1/3}$  yields the nuclear matter response.

In order to illustrate this approach we reproduce  in Fig.~\ref{fig:nm}
  one example for an extrapolation
as a function of $A^{-1/3}$ \cite{Day89}. 
Ignoring  $^4 \!He$ (for which the properties of the density cited above  are 
not valid), the nuclear response
$s(q,\omega )$ for A = 12-197 is well fit by a linear function of
$A^{-1/3}$.  The plot on the bottom of Fig.~\ref{fig:nm} gives the same
extrapolation as a function of A. This figure reveals that the 
extrapolation as  a function of A is unwieldy though the curve  better
imparts the saturation of the response. Even heavy nuclei  significantly differ from nuclear matter
due to the large fraction of surface-nucleons. 

\begin{figure}[htb]
\begin{center}
\hspace*{1mm}\includegraphics[scale=0.45,clip]{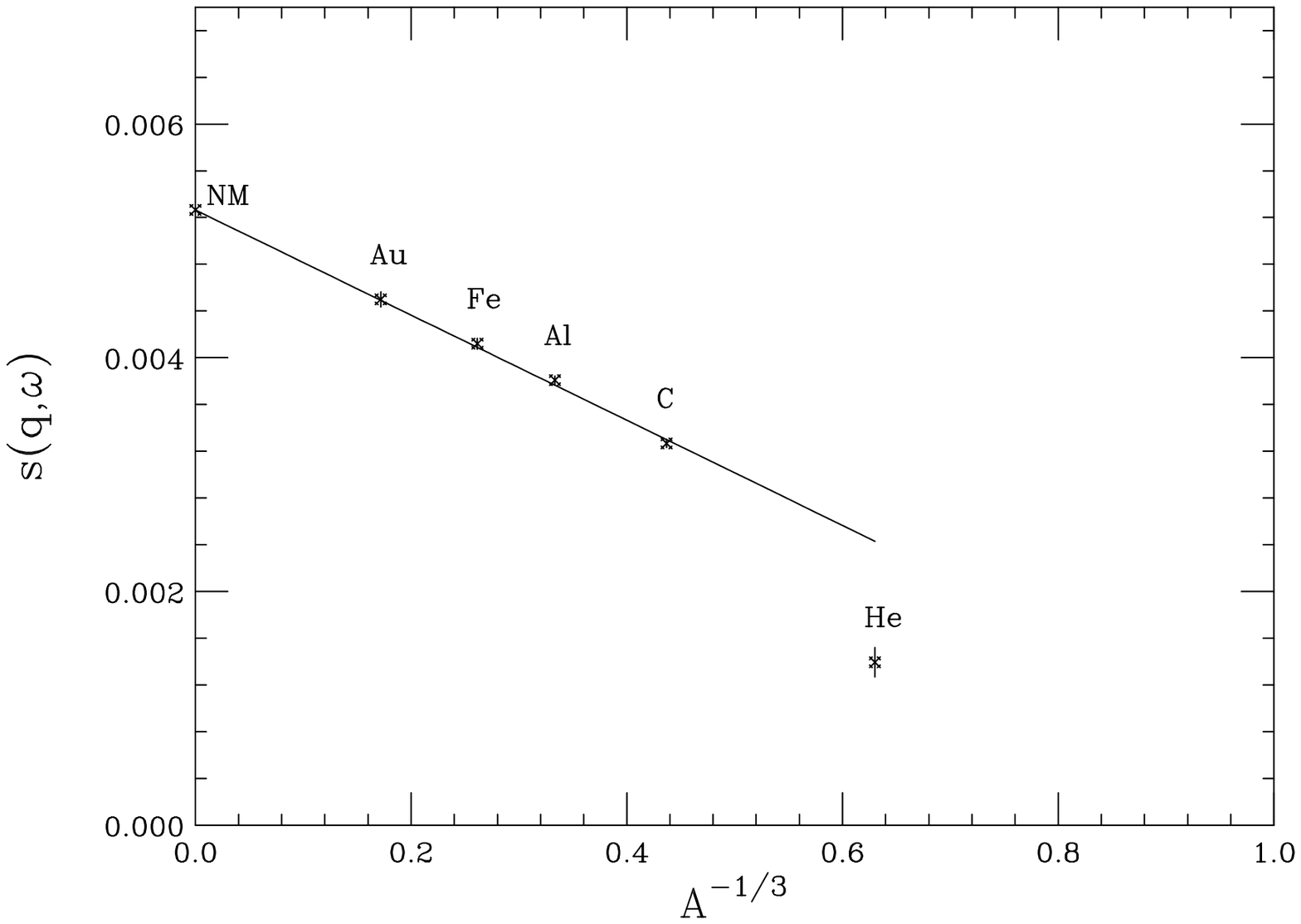}
\includegraphics[scale=0.45,clip]{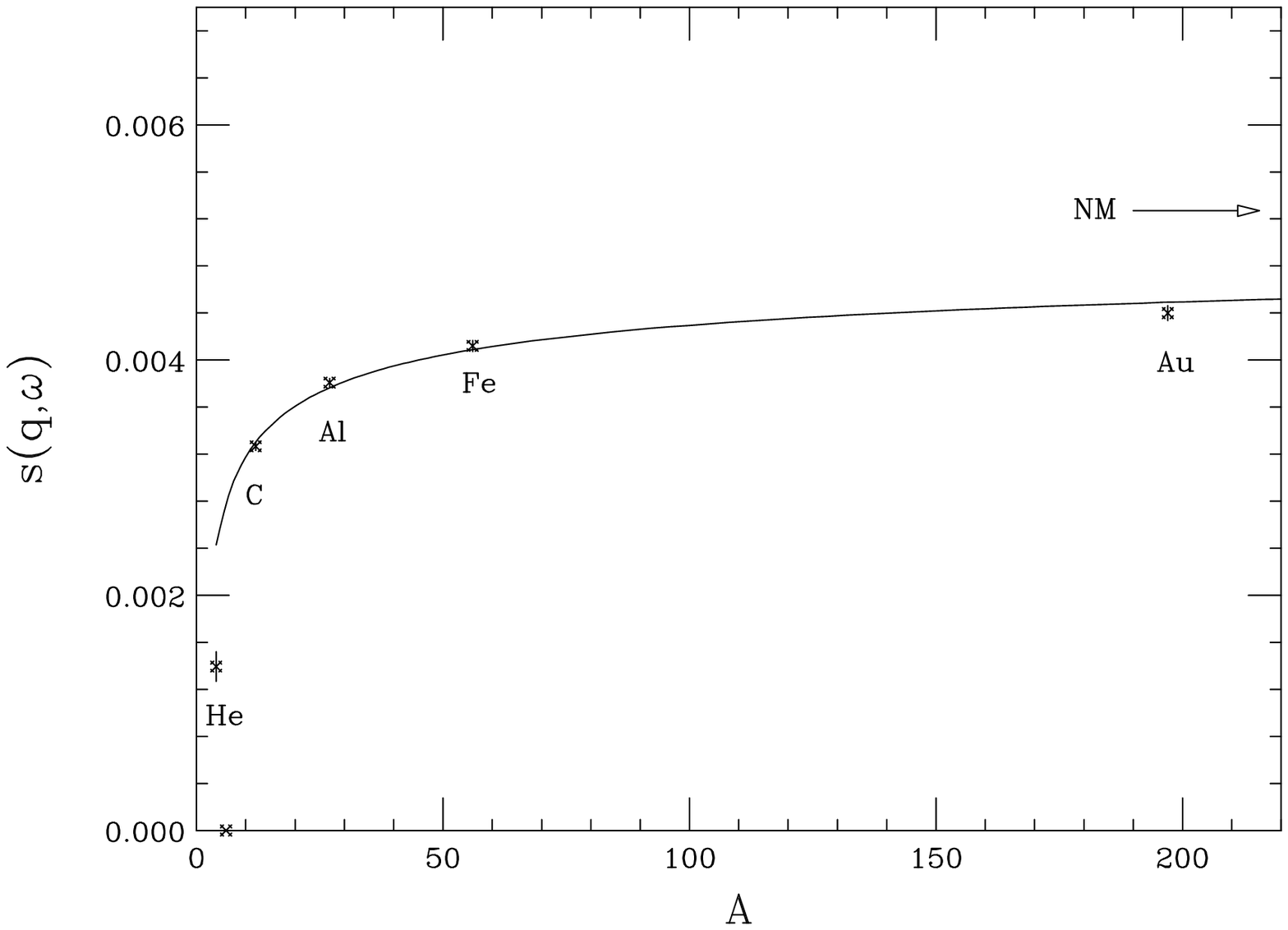}
\caption{Extrapolation of nuclear response per nucleon 
at fixed $q$ and $\omega$
($E=3.6$ GeV, $\theta = 16^\circ$, and $\omega = 180$ MeV) as a function
of $A^{-1/3}$ (top)  and as a function of $A$ (bottom) where the
extrapolated value of the nuclear matter response is indicated by the
arrow.} \label{fig:nm}\end{center}
\end{figure}

\begin{figure}[ht]
\centerline{\includegraphics[scale=0.5]{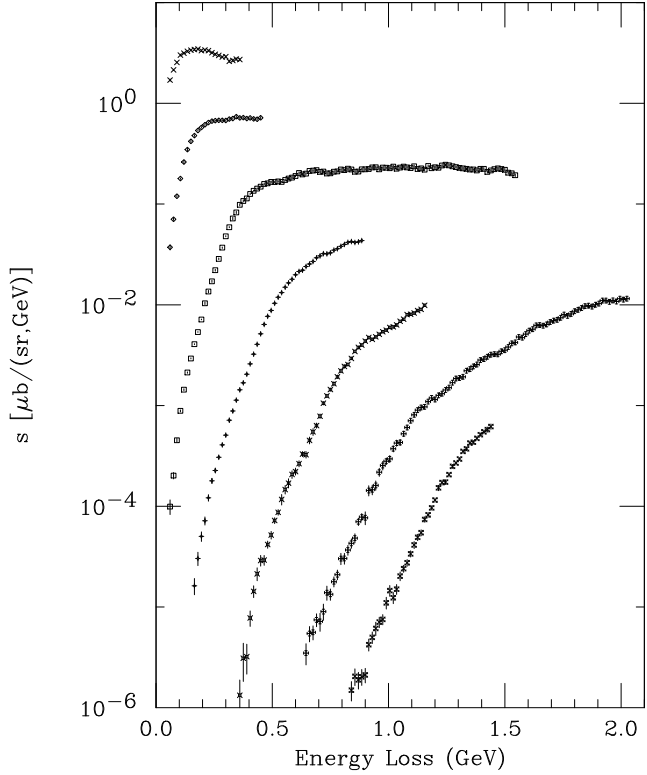}}
\begin{center}
\caption{Nuclear matter response at momentum transfers up to 3.5(GeV/c)$^2$
\cite{Day89} \label{fig:nmcom}}\end{center}
\end{figure}

In order to obtain the  response function for symmetric \nm , one makes use of
additional knowledge. For quasielastic
scattering, the relative contribution of protons and neutrons changes as
a function of A. Although the protons dominate due to the larger
electron-proton cross section, the contribution of neutrons is not
negligible. For the extrapolation, one  assumes that the
response functions for protons and neutrons are the same, and  the
trivial dependence on N,Z is removed by extrapolating the quantity
\begin{equation} s(q,\omega ) = \sigma (q,\omega ) / (Z \sigma_{ep} + N
\sigma_{en}) \end{equation} 
The nuclear matter response has been extrapolated \cite{Day89}  from 
the nuclear response measured for finite nuclei over a large
region of $q$ and $\omega$, see Fig.~\ref{fig:nmcom}. This was made possible by the availability
of data for nuclei with $A$ = 4, 12, 27, 56  and 197 \cite{Day93} 
taken at the same
incident energies and angles. Recent data taken at Jefferson Lab with $A$
= 3, 4, 9, 12, 63 and 197 will allow this  to be extended over an even
larger range of $q$ and $\omega$.

\section{Related areas} \label{other}
Inclusive scattering from composite systems is used as a tool in a number of
areas \cite{Silver89a}. The corresponding processes have many aspects that are closely related to
\qes~ from nuclei, but they also exhibit significant differences. We now 
address those areas where inclusive scattering has been harnessed to study
diverse composite systems.

These areas differ not only by the nature of the composite system investigated, 
but also by
the probe used: photons, low-energy electrons, neutrons, high energy electrons,
muons and neutrinos. The energies of the probes cover many orders of magnitude,
 from meV to GeV, scaling with  the
dimensions relevant for  the composite  systems.
 
Historically, the first area where inclusive scattering became prominent was the
measurement of the Compton profile, {\em i.e.} \qes~ of photons or X-rays from
electrons bound in atoms \cite{Williams77,Cooper85}.
  The Compton effect actually played a substantial role
in the early validation of quantum ideas,  and  with the experimental work
of DuMond \cite{DuMond47} became a practical tool for the investigation of electron  momentum 
distributions. Modern experiments involve energies in the region of 10 to 
hundreds
of keV, {\em i.e.} energies that in general are very large as compared to atomic
Fermi energies (eV).

The observable in Compton scattering, the so-called Compton profile, is the
longitudinal momentum distribution of the initially bound electrons, in direct
correspondence with the scaling function $F(y)$ determined in \qes~ from nuclei, see
Sec.~\ref{scale}. The momentum distribution of the electrons exhibits a
more complex structure, as not only the bound-electron $n(k)$ play a role,
but also --- and often more prominently --- the conduction electrons for
metallic targets.   

As an example we show in Fig.~\ref{comp} the Compton profile of Sodium. The
contribution of the core electrons has been removed; the parabola-like part
below a momentum of 0.5~$a.u.$ is due to the free-electron Fermi gas and
 the tail at
larger momenta is due to inter-electron-interactions \cite{Eisenberger72}.
\begin{figure}[hbt]
\begin{center}
\includegraphics[scale=0.4,clip]{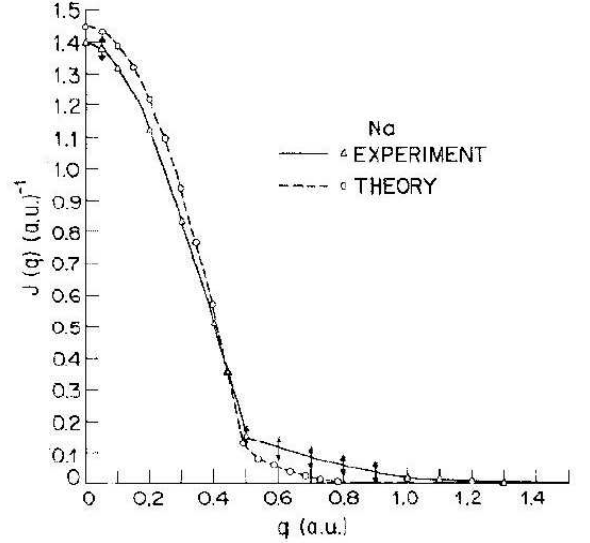}
{\caption[]{Compton profile of $Na$ as a function of electron
momentum, with the contribution of the
core electrons removed and finite-resolution effects unfolded
\cite{Eisenberger72}.
}\label{comp}} 
\end{center} 
\end{figure}

When compared to \qes~ on nuclei, Compton scattering possesses an additional
feature: for crystalline samples, the Compton profile can be measured for many
different orientations of the probe. This allows one to derive the reciprocal form factor
$B(\vec{r})$, from  which one can extract  the autocorrelation function in 
$r$-space, a quantity
that helps to deduce the spatial structure of the molecules in the crystal.

A variant of Compton scattering is electron Compton scattering, where the photon
is replaced by an electron, typically in the 50keV energy region.  While the
production of a good beam and the detection of the scattered particle is much 
easier, the requirement of  very thin targets, unfortunately, partially offsets
these  advantages.

A second area where inclusive scattering represents a popular tool is the  scattering
of low-energy neutrons from condensed matter systems such as quantum liquids. 
Neutrons with energies ranging from  10meV to 10keV, from reactors
or spallation neutron sources, are employed. The  energies are again very 
high as compared to energy scales of the system, of order
meV for liquid Helium for example.

Similar to \qes~ from nuclei, the measured structure function $S(q,\omega)$  yields, in
the impulse approximation, an integral over the momentum distribution. In the
limit of large $\magq$, it can  be written as a function of {\em one} variable 
$Y$, the  longitudinal momentum component. 

In  \qe~ neutron scattering, emphasis has been placed on the
understanding of the role of  FSI, for two reasons. 
First, the
interaction between two atoms is fairly singular at short inter-atom distances
$r$;
the typical Lennard-Jones potential rises very steeply ($r^{-12}$) at small $r$.
This strong FSI has a pronounced effect upon the observables, to the extent that
 the longitudinal momentum distribution can only be extracted after correction
for FSI. 

\begin{figure}[htb]
\begin{center}
\includegraphics[scale=1.0,clip]{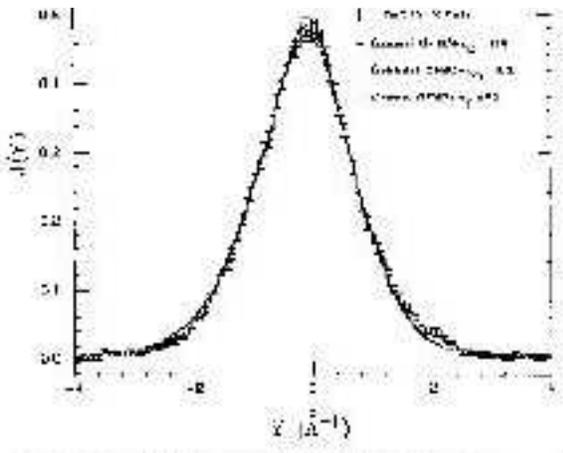}
{\caption[]{Profile for neutron scattering from liquid Helium at
$0.35K$. The curves, calculated using GFMC and including FSI effects, are 
shown for different fractions of the Bose condensate, 10\% giving the best fit
\cite{Sosnik90}.
}\label{sokol}} 
\end{center} 
\end{figure}

 FSI was studied in great detail for a second reason. Much of the emphasis in the
field was placed on the measurement of the fraction of Bose condensate in
superfluid $^4 \! He$. This Bose condensate was expected to produce a
$\delta$-function like spike in the momentum distribution $n(k)$ at $k=0$, 
which would lead to  a spike in the response at $Y=0$. This feature was not seen, 
a fact that  now has been understood as a consequence of FSI. 

The understanding of FSI in neutron scattering 
 has many aspects that are parallel to the discussion
given in Sec.~\ref{sec:FSI}. In particular, it has been found that the main effect of
FSI is a folding of the IA response. The  folding function  has a width
governed by the atom-atom total cross section. It has also been found, that for a
quantitative understanding of FSI, it is imperative to include in the description
of the initial state the atom-atom correlation function
$g(|\vec{r_i}-\vec{r_j}|)$. The treatment of the atom-atom potential 
is  comparatively difficult, yet is possible with  hard-core perturbation 
theory \cite{Silver88}. 

As an example, we show in Fig.~\ref{sokol} the scaling function measured at a
momentum transfer of 23\AA~ on superfluid Helium at $0.35K$ \cite{Sosnik90}. At 
this temperature a Bose condensate is predicted to occur. The $\delta$(Y=0)-function due to
the condensate, smeared with the FSI folding function, explains the data for a
condensate-fraction of 10\%.

The last related area we want to address concerns deep inelastic scattering
(DIS)  of
GeV electrons or muons from nucleons (for a review see {\em e.g.}
\cite{Ellis96}). Here, the energy spectrum of the
inclusively scattered lepton is used to derive the momentum distribution of
the quarks bound in the nucleon. This process  provided the first direct
evidence for the existence of pointlike constituents of fractional charge, 
subject to asymptotic freedom \cite{Politzer73,Gross73}, and
still is one of  the main sources of information on nucleon structure.    

\begin{figure}[hbt]
\begin{center}
\includegraphics[scale=0.41,clip]{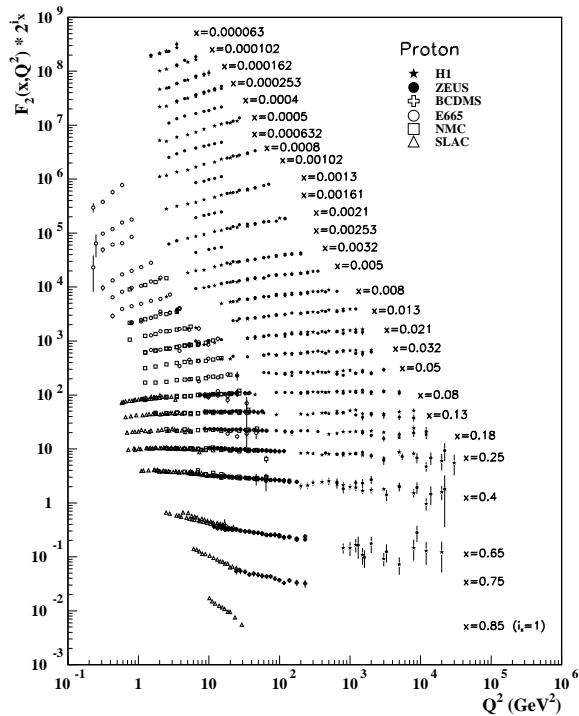}
{\caption[]{Structure function $F_2$ as a function of the momentum
transfer, for different values of the Bjorken scaling variable $x$
\cite{Eidelman04}.
}\label{struc}} 
\end{center} 
\end{figure}

DIS is generally analyzed in terms of the Bjorken scaling variable 
$x = Q^2/2m\omega$ 
\cite{Bjorken69} appropriate for  constituents of negligible 
rest mass. 
The variable $y$, which in the limit of small constituent 
mass equals the Nachtmann variable
$\xi$ (but for a trivial factor) is more appropriate 
for extending scaling to the lower momentum transfers
 \cite{Benhar00}. In terms of $\xi$, DIS 
and \qes~ from nuclei have many things in common. 

Apart from the parallel aspects (see Sec.~\ref{sec:IA}) DIS exhibits one special
feature: the evolution of the scaling function with increasing $Q^2$. This
evolution is a consequence of the fact that the lepton, at larger and larger
$Q^2$, resolves more and more of the nucleon structure and the other nucleon
constituents, the gluons. This evolution of the structure function has been
studied in great detail, and is well understood in terms of Quantum Chromo
Dynamics (QCD) \cite{Altarelli77,Gribov72}.  Fig.~\ref{struc} displays the evolution
of the proton structure function with momentum transfer. 

Unlike {\em e.g.}  the neutron-scattering mentioned above, DIS is always
analyzed in IA, neglecting the FSI of the recoiling quark; the structure
functions then are interpreted directly as the quark distribution functions. The
effects of FSI, although known to be present even in the $Q^2\rightarrow \infty$
limit \cite{Brodsky02}, are neglected  despite the fact that model calculations 
indicate that they are of substantial size \cite{Paris02}. 

Analogous experiments on DIS of neutrinos have also been performed; due to the
small rates, the data base in this area is much more
restricted. 

\section{Conclusions}
The field of inclusive quasi-elastic electron-nucleus scattering has seen
important progress during the last decade, both in terms of experimental results
and theoretical understanding.

Experiment has greatly benefited from the high-intensity GeV-energy facilities
and the high performance spectrometers and detectors that became available. This
has allowed one to extend the data to extreme values of momentum transfer and
energy loss. As a consequence, we now have, at least for selected nuclei, a
fairly complete data base (see the Web page
\url{http://faculty.virginia.edu/qes-archive}  that gives a rather complete
collection of the available cross sections).  Not yet satisfactory is the
situation for the longitudinal strength at large momentum transfer, where the
data base is very narrow. Improvements would also be desirable at low momentum
transfer, where much of the data base comes from experiments done in
spectrum-acquisition mode.

Theory has also made considerable progress. Nuclear many-body theory today 
provides reliable spectral functions, which are at the basis of any
quantitative understanding of quasi-elastic scattering, and especially important
in the region of high
momentum transfer and not too large energy loss.
In particular, the
nucleon-nucleon short-range correlations, which have long been known to play a 
major role, are now included in an adequate fashion. 

The $y$-scaling analysis of the data clearly shows that elastic scattering 
off a single nucleon is the dominant reaction mechanism at $x > 1$. This is the
region where quantitative information on nuclear properties can be extracted.
 
For the treatment of the recoil-nucleon final state interaction, 
various approaches have been developed, applicable in different regions of
momentum transfer and energy loss.
The results of calculations carried out within the scheme widely adopted 
at large momentum transfer (typically $|{\bf q}| > 1$ GeV), 
based on the eikonal approximation, indicate that final state 
interaction effects are large, indeed dominant, in the low energy loss tail 
of the inclusive cross section, and their inclusion leads to a quantitative 
account of the existing data up to $x \sim 2$.

 The role of meson exchange currents, which
surprisingly have a large effect at rather low momentum transfer and across the
entire quasi-elastic peak, also has been much better understood. Issues not yet
satisfactorily resolved concern the final state interaction at large momentum
transfer and very low energy loss, corresponding to $x >2$,  and the role of 
non-nucleonic degrees of
freedom at very large momentum transfer. 
 Recent studies of FSI in $(e,e^\prime p)$, carried out the within the formalism 
     described in Section \ref{sec:FSI} \cite{Schiavilla05}, show that the spin
     dependence of the NN scattering amplitude plays an important role. The 
     possible relevance of these effects in inclusive processes requires further
     investigations. The treatment of knock-out processes
accompanied by excitation of the nucleon also is not yet entirely satisfactory.

Recently, much consideration has been given to the connection between 
electron- and neutrino nucleus scattering \cite{Benhar05,Amaro05}. 
The generalization of the existing theoretical approaches and the exploitation
of the measured $(e,e^\prime)$ cross sections to predict the analogous neutrino-induced 
reactions, such as $(\nu,e)$ or $(\nu,\mu)$, will   be of great importance to 
reduce the systematic uncertainty in the interpretation of neutrino 
oscillation experiments.

Only partially  exploited is the relation to  quasi-elastic scattering of hadronic
probes ($p, \pi, K,..$) from which additional, and largely complementary,
information could be learned.
 
\section{Acknowledgments}
The authors want to thank Alex Dieperink, Bill Donnelly, Rocco Schiavilla and
Claude Williamson for helpful comments on the manuscript.

\bibliographystyle{apsrmp}
\bibliography{/usr/users/sick/sum2,/usr/users/sick/quasi/qesnff/rmpnnf}
\end{document}